\documentclass[aps,pra,twocolumn,superscriptaddress,nofootinbib,floatfix]{revtex4-2}

\usepackage{amsmath,amssymb,amstext}
\usepackage{graphicx}
\usepackage{braket}
\usepackage{bm}
\usepackage{placeins}
\usepackage[colorlinks=true,linkcolor=blue,citecolor=blue]{hyperref}

\newcommand{\C}{\mathcal{C}}
\newcommand{\A}{\mathcal{A}}
\newcommand{\Kry}{\mathcal{K}}
\newcommand{\dd}{\mathrm{d}}

\usepackage{xcolor}

\begin{document}

\hypersetup{urlcolor=blue}

\title{Krylov Complexity of Time-Dependent Optical Hamiltonians}

\author{Abhishek Chowdhury}
\email{achowdhury@iitbbs.ac.in}
\affiliation{Department of Physics, School of Basic Sciences, Indian Institute
of Technology Bhubaneswar, Jatni, Khurda, Odisha, 752050, India}
\author{Aryabrat Mahapatra}
\email{aryabratm23@iitk.ac.in}
\affiliation{Department of Physics, School of Basic Sciences, Indian Institute
of Technology Bhubaneswar, Jatni, Khurda, Odisha, 752050, India}
\affiliation{Department of Physics, Indian Institute of Technology Kanpur,
Kalyanpur, Kanpur, Uttar Pradesh, 208016, India}

\author{Ajit Prasad Mahapatra}
\email{a25ph09001@iitbbs.ac.in}
\affiliation{Department of Physics, School of Basic Sciences, Indian Institute
of Technology Bhubaneswar, Jatni, Khurda, Odisha, 752050, India}

\begin{abstract}
State Krylov complexity becomes experimentally concrete in driven optics
because its chain can be interrogated through excitation counting, parity,
return probability, and phase.  Its meaning, however, is fixed jointly by the
control history, the prepared state, and the reference used to follow explicit
time dependence.  We disentangle these ingredients for oscillator, $SU(2)$,
$SU(1,1)$, and three-level Hamiltonians and determine when preparation
dependence preserves exact solvability.  A fixed group displacement of an
extremal state can be absorbed into transformed controls, giving an
undo-and-count readout for arbitrary drives.  A polynomial preparation instead
Christoffel-transforms the spectral measure of a fixed operator; an affine
moving frame then transports its complete seed-adapted chain even when
laboratory Hamiltonians at different times do not commute.  We also resolve seed dependence in stroboscopic dynamics and driven
three-level geometry.  A periodically
forced mode started from any finite Fock state has cyclic dimension equal to
the denominator of a nonzero-displacement rational rotation and infinite
dimension for an irrational rotation.  At a half turn its complexity becomes
a Laguerre survival deficit and a magnitude-sensitive Wigner readout.  In a
driven $V$ atom, bright-state geometry interferes with excited-manifold
dynamics and controls bright--dark decoupling.  Preparation
therefore changes not only the initial state but also the chain and the optical
observable that realizes its complexity.

\end{abstract}

\maketitle

\section{Introduction}

Driven quantum-optical experiments already measure the quantities from which state
Krylov complexity is built.  Atomic populations, photon-number distributions,
parametric gain, field parity, and optical phase all resolve how a prepared
state explores its accessible Hilbert space.  The Krylov construction orders
that exploration relative to the preparation itself.  Starting from a
normalized seed $\ket{K_0}$, a time-independent Hamiltonian generates, through
the Lanczos algorithm, an ordered basis in which it is tridiagonal; the mean
Krylov-site index, denoted $\C(t)$, defines the spread complexity
\cite{Balasubramanian:2022tpr,Nandy:2024htc}.  The static construction is
therefore a property of the pair $(H,\ket{K_0})$, with a direct physical
meaning whenever the Krylov sites coincide with experimentally resolved
excitation sectors.  For an explicitly time-dependent Hamiltonian, the
prescribed reference path and its phase convention complete the defining data.

Optical control makes both entries of this pair nontrivial.  The Hamiltonian
changes during a pulse, while the seed may be an excited Dicke state, a Fock
state, a squeezed state, or a coherent superposition.  State-Krylov
constructions based on the extended generator $H(t)-i\partial_t$ and a
prescribed reference path were developed in
Refs.~\cite{Takahashi:2024hex,GrabaritsMedinaGuerraDelCampo2026}.
Complementary Floquet, Arnoldi, and Magnus approaches to driven complexity
appear in Refs.~\cite{Nizami,FarajiAstanehKafashi2026}.  Our aim is to turn
the joint dependence on control history and state preparation into optical
statements that can be read as populations, coherences, parity fringes, and
geometric phases.

We consider Hamiltonians that remain linear in a finite set of optical
generators,
\begin{equation}\label{eq:oH}
 H(t)=\sum_\alpha\lambda_\alpha(t)L_\alpha ,
\end{equation}
where the finite set of operators $\{L_\alpha\}$ closes on a
finite-dimensional Lie algebra and the $\lambda_\alpha(t)$ are prescribed
control functions.

The three rank-one,
single-ladder optical realizations are a collectively driven two-level ensemble
\cite{TavisCummings1968}, a forced cavity mode
\cite{WallsMilburn2008}, and nondegenerate parametric amplification
\cite{mollow1967quantum,YurkeMcCallKlauder1986}.  We also treat a coherently
driven three-level atom, where the Morris--Shore bright--dark decomposition
emerges from the recursion itself \cite{MorrisShore1983,DongTang2002}.
Algebraic closure organizes the propagator and the chain.  The exact status
depends jointly on the control history and the prepared seed, and is
summarized in Table~\ref{tab:seed-control-map}.  Our scope is analytic control
and observable identification in these algebraically closed models.
We retain static and rotating-frame constant-generator limits as reference
cases for distinguishing the effects of state preparation and algebra from
those of time-dependent control.  Chaos
classification requires separate spectral and dynamical diagnostics
\cite{Espanol:2022cqr}.

For the extremal optical preparations
used here, the first payoff is an experimental dictionary,
\begin{equation}\label{eq:dictionary}
 \C(t)=
 \begin{cases}
  \langle S_z\rangle+j, & \text{collective atomic excitation},\\[2pt]
  \langle a^\dagger a\rangle, & \text{cavity photons},\\[2pt]
  \langle n_a\rangle=\langle n_b\rangle, &
       \text{signal--idler pair number}.
 \end{cases}
\end{equation}
Here $S_z$ is the collective population-inversion operator for spin $j$,
$a$ is the cavity annihilation operator, and $n_a$ and $n_b$ are the
signal- and idler-mode number operators.  The third line is restricted to the
zero-number-difference sector.
Changing the preparation changes the experimental dictionary.

The control history and the preparation define two independent axes of the
problem.  A time-independent group displacement $W_{\rm s}$ maps an extremal
preparation to another point on its
coherent-state orbit and conjugates the control
coefficients without leaving the oscillator, $SU(2)$, or $SU(1,1)$ algebra.
Consequently, $W_{\rm s}\ket{\ell,0}$ retains the complete arbitrary-control
extremal-seed reduction.  The opposite $SU(2)$ extremal state is covered after
a fixed $\pi$ rotation and reversal of the ladder.  For an interior weight,
the same covariance fixes the first hopping exactly, while the subsequent
recursion contains new seed-dependent information.  The two-photon example in Sec.~\ref{sec:tipping} resolves that information
through coincidences, while its mean mode occupations remain unchanged.

A polynomial preparation addresses the seed axis in a different way.  For a
fixed self-adjoint operator $X$ and normalized cyclic seed $\ket\chi$, the
orthonormal polynomial $p_r$ is defined by
$\ket{K_r^{(X)}}=p_r(X)\ket\chi$.  An admissible polynomial $Q$ is more
general.  If $Q(E)=\sum_{n=0}^{d}q_n p_n(E)$, then, before finite-support
identifications, $Q(X)\ket\chi$ is a superposition of the first $d+1$ old
Krylov rungs.  Thus every $p_r$ is an allowed choice of $Q$, but $Q$ prepares
one old rung only when it is proportional to $p_r$.

The new seed has the
positive Christoffel measure $|Q(E)|^2\dd\mu(E)/\mathcal N_Q$, with $E$ the
spectral variable of $X$, $\dd\mu$ its seed measure, and $\mathcal N_Q$ the
normalization.
In this precise sense a finite-degree polynomial preparation is local in
the reference Krylov index. A generic group displacement is instead
nonlocal in the corresponding weight basis, spreading an oscillator or
$SU(1,1)$ extremal state over infinitely many weights and a spin extremal
state over its full finite multiplet.

For a prescribed moving reference $W(t)\ket\chi$, the first hopping is the
uncertainty of $W^\dagger H W-iW^\dagger\dot W$, equivalently the relative
Fubini--Study speed.  We evaluate this exact local response for oscillator
Fock weights, interior $SU(2)$ weights, and excited $SU(1,1)$ weights.

Complete evolution follows when the moving frame
transports one fixed $X$ and its generator is affine in $X$.  Every admissible
$Q(X)\ket\chi$ then evolves in its Christoffel-transformed chain at the
effective time $\Delta\tau=\tau(t)-\tau(0)$
\cite{GrabaritsMedinaGuerraDelCampo2026,ChowdhuryMahapatra2026}.  The
one-photon oscillator member gives
$\C=(\Delta\tau)^2+1-\exp[-2(\Delta\tau)^2]$, including noncommuting
laboratory Hamiltonians.  Periodic forcing provides a complementary unitary
problem.  For nonzero displacement, a Fock seed $\ket r$,
$r\in\mathbb N_0$, has Arnoldi dimension $q_F$ when the rotation number is the
reduced rational $p/q_F$ with $q_F\ge2$, and infinite dimension for an
irrational rotation.  At a half turn the chain has at most two sites, and its
complexity is the survival deficit shown in
Fig.~\ref{fig:floquet-halfturn}.

For resonant fixed-direction driving, time enters through the accumulated
pulse area alone.  The compact $SU(2)$, flat Heisenberg--Weyl, and noncompact
$SU(1,1)$ orbits then give bounded, quadratic, and hyperbolic
spread profiles. Their full site distributions remain binomial, Poisson, and negative
binomial, so the mean and variance obey one curvature-dependent relation.
The same orbit carries a phase variable.
If $\Phi$ is the coherent-state phase coordinate,
the Berry connection in the coherent-state gauge of the Gauss chart is
$\left(-\C\,\dd\Phi\right)$, and a closed trajectory acquires
\begin{equation}
 \gamma_{\rm g}=-\oint\C\,\dd\Phi \pmod{2\pi}.
\end{equation}
Thus population spread and interferometric phase are conjugate readouts of the
same optical orbit.  This connects the coherent-state geometry
developed for operator Krylov spread
\cite{Caputa:2021sib} with the standard geometric phase
\cite{Berry1984,AharonovAnandan1987}; recent time-dependent Krylov studies also
exhibit Berry-phase information in moving spin bases
\cite{ZhengHeLiuZhang2026}.

A second experimental bridge appears for cavity seeds of definite
photon-number parity.  Their odd-chain
probability is an affine displaced-parity readout of the seed Wigner function
along the drive-induced displacement.  Photon-added states
therefore display the sign structure of their Wigner function, while cat
states produce interference fringes controlled by branch separation and
width.  The required displaced-parity readout is the standard observable of
cavity Wigner tomography
\cite{LutterbachDavidovich1997,Deleglise2008}.

The three-level system separates pulse magnitude, pulse geometry, and spectral
mixing particularly cleanly.  For static couplings, a ground-state seed
generates the bright state at the first step and the dark direction at the
second whenever the detuning difference opens that rung.  Equal detunings
close the reachable chain at the bright state.  A uniform three-site chain
gives complete transfer, while the equal-coupling time-averaged spread has an
exact rational profile with computable flanking maxima.

For complex
time-dependent couplings, the first hopping is the
total coupling amplitude
$G(t)$ and the second is
$\|P_D(H_e-i\partial_t)\ket B\|$, with $P_D$ the projector on the dark
direction, $H_e$ the excited-manifold Hamiltonian, and $\ket B$ the bright
state of Sec.~\ref{sec:vchain}. The latter becomes the
Fubini--Study speed of the normalized coupling spinor for a degenerate excited
doublet and, more generally, contains gauge-invariant interference between
excited-manifold dynamics and bright-state motion.  Its vanishing gives an
exact bright--dark decoupling condition.  The same connection term controls
the dark-state leakage in stimulated Raman adiabatic passage
\cite{ZlatanovRangelovVitanov2022,Vitanov2017}.

Closed unitary evolution with prescribed classical drives is the exact setting
of the formulas below.  At each application we state the resonance, frame,
seed-preparation, and moving-basis tangency conditions and identify the
associated population, parity, or phase readout.  Weak-probe spectra, loss,
and dephasing can then be compared with these coherent reference profiles.

Section~\ref{sec:framework} develops the extended recursion, the
prepared-seed constructions, and the metric and symplectic geometry.
Sections~\ref{sec:atoms} and \ref{sec:photons} apply them to collective atoms
and driven modes, including the nonlowest-seed Floquet construction in
Sec.~\ref{sec:floquet-forced}.  Section~\ref{sec:three} treats the static and
time-dependent three-level chains, including adiabatic passage.
Section~\ref{sec:discussion} assembles the experimental consequences and the
remaining extensions.
The four appendices give the constant-coefficient Gauss decomposition
(Appendix~\ref{app:gauss}), the group identity used in the text
(Appendix~\ref{app:unitarity}), the parabosonic derivation of the photon-added
chain (Appendix~\ref{app:dunkl}), and the moving-seed and dressed-generator
derivations (Appendix~\ref{app:moving-seeds}).

\section{Spread complexity for driven optical systems}\label{sec:framework}
The construction requires two
choices that will be used repeatedly.
In the state-Krylov construction adopted here, explicit time dependence enters
the recursion through the extended generator rather than through a Hamiltonian
frozen at each instant, and the residual phase freedom is fixed by requiring
the Lanczos hoppings to be positive.
Once these are
settled a single geometric statement covers the extremal sectors of all three algebras, and those sectors are
distinguished by the curvature of their
coherent-state orbits.  We use units with $\hbar=1$, so Hamiltonian coefficients have units of
angular frequency.

\subsection{The construction and its seed}

Let the Hamiltonian $H$ be self-adjoint and
$\ket{K_0}$ a normalized state.
The Krylov space
is the cyclic subspace
$\Kry=\overline{\operatorname{span}}\{\ket{K_0},H\ket{K_0},H^2\ket{K_0},\dots\}$,
which may be a proper subspace of the full Hilbert space.\footnote{
For an unbounded operator, the seed is assumed to lie in the domains of the
powers required at each Lanczos step.  The Fock, coherent, squeezed, and
finite-dimensional seeds used below satisfy this requirement.  For an
infinite chain we also require the orthogonal polynomials to be complete in the
spectral cyclic subspace of the seed, equivalently that polynomials are dense in
$L^2$ of the seed spectral measure.  This is automatic for finite support and
holds for the Gaussian-type and exponentially decaying discrete measures used
below.} Gram--Schmidt
orthogonalization of that sequence is the Lanczos recursion
\cite{Lanczos1950}. With $b_0=0$ and $\ket{K_{-1}}=0$,
\begin{align}\label{eq:lanczos}
  a_n&=\braket{K_n|H|K_n}, \nonumber\\
  \ket{A_{n+1}}&=(H-a_n)\ket{K_n}-b_n\ket{K_{n-1}},\\
  b_{n+1}&=\|A_{n+1}\|,\qquad
  \ket{K_{n+1}}=\ket{A_{n+1}}/b_{n+1}.\nonumber
\end{align}
In this basis $H$ is tridiagonal, and expanding
$\ket{\psi(t)}=\sum_n\phi_n(t)\ket{K_n}$ turns the Schr\"odinger equation into
hopping on a one-dimensional chain. The spread complexity is the mean position
on that chain \cite{Balasubramanian:2022tpr},
\begin{equation}\label{eq:complexity}
  \C(t)=\sum_n n\,|\phi_n(t)|^2 .
\end{equation}

Two properties of Eq.~\eqref{eq:lanczos} will be used throughout.  A zero
$b_{n+1}$ closes the cyclic subspace at site $n$, so a finite chain bounds the
spread and an eigenstate seed gives $\C(t)=0$.  The normalized seed may be any
state in the Hilbert space.  Ground states and lowest-weight vectors form an
especially tractable class, but they are choices rather than prerequisites.
Changing the seed changes the chain and, consequently, the physical observable
represented by Eq.~\eqref{eq:complexity}.

\subsection{Explicit time dependence and the phase convention}
\label{sec:time-dependence}

For an explicitly time-dependent Hamiltonian, we construct a moving Krylov
basis using the differential expression
\begin{equation}\label{eq:extended}
  \mathcal G=H(t)-i\partial_t ,
\end{equation}
acting on differentiable state vectors. The enlarged-space $t$--$t'$
formalism motivates this expression \cite{Howland1974}; here the recursion
uses inner products in the physical Hilbert space at each time, as in
Ref.~\cite{Takahashi:2024hex}.  Indeed, substituting
$\ket{\psi(t)}=\sum_n\phi_n(t)\ket{K_n(t)}$ into the Schr\"odinger equation
gives the amplitude matrix
\begin{equation}\label{eq:Gmatrix}
  \mathcal G_{mn}(t)=\braket{K_m|H(t)|K_n}
  -i\braket{K_m|\partial_tK_n}.
\end{equation}
The second term is the connection of the moving basis.  For a differentiable orthonormal frame in the domain of the Hermitian
$H(t)$, differentiating $\braket{K_m(t)|K_n(t)}=\delta_{mn}$ shows that the
amplitude matrix is Hermitian. The pointwise three-term recursion therefore
defines real $a_n(t)$ and, with the site phases fixed, nonnegative hoppings
$b_n(t)$.

For the three dynamical algebras, denote the coefficient of the raising generator $L_+$ by
$q(t)=\rho(t)e^{-i\varphi(t)}$, with $\rho\ge0$, and
denote the weight states by $\ket{\ell,n}$.  Here $\ell=j$ for
$\mathfrak{su}(2)$, $\ell=h$ for the positive discrete series of
$\mathfrak{su}(1,1)$, and $\ell=\tfrac12$ for the
Heisenberg--Weyl normalization used below.  The moving basis, built on the
extremal-weight seed $\ket{K_0}=\ket{\ell,0}$,
\begin{equation}\label{eq:gauge}
  \ket{K_n(t)}=e^{-in\varphi(t)}\ket{\ell,n}
\end{equation}
absorbs the raising phase.  Its off-diagonal elements are
$b_{n+1}(t)=\rho(t)\beta_{n+1}$, where $\beta_{n+1}>0$ is the representation
matrix element, and its connection shifts the $n^{\text{th}}$ diagonal entry by
$\left(-n\dot\varphi\right)$.

The same phase construction extends to any coupling graph with vanishing loop
flux modulo $\pi$.  Vertex phases remove all edge phases on a tree.  On a graph
with cycles, the sum of edge phases around each loop is gauge invariant, and
a basis gauge with real edge couplings exists precisely when every such sum is
$0$ modulo $\pi$. Here the gauge freedom is the independent vertex rephasing
$\ket n\mapsto e^{i\theta_n^{\rm g}}\ket n$.  Fixing phases along a spanning
tree makes all tree edges real; each remaining chord is real exactly when the
flux of its fundamental loop is $0$ modulo $\pi$.  This is the graph version
of the Peierls phase construction \cite{Peierls1933}. The condition modulo $\pi$ permits signed real edges; making every
nonzero edge positive requires vanishing loop flux modulo $2\pi$.

The one-generator ladders in this paper are chains and
satisfy the condition automatically.  Time-dependent vertex phases
$\theta_i^{\rm g}(t)$, one per weight state, add their
$\dot\theta_i^{\rm g}$
  connection shifts to the diagonal.
For the driven-spin ensemble introduced in Sec.~\ref{sec:atoms}, let $\lambda_0(t)$ denote the coefficient of $S_z$ in
Eq.~\eqref{eq:oH}. Then
\begin{align}\label{eq:gaugedcoeffs}
  a_n(t)&=\lambda_0(t)(n-j)-n\dot\varphi(t),\nonumber\\
  b_{n+1}(t)&=\rho(t)\sqrt{(n+1)(2j-n)} \,.
\end{align}
A common shift of the $a_n$ contributes only an overall phase.  For a
monochromatic drive, $\lambda_0=\omega_0$ and $\dot\varphi=\omega$, so the diagonal retains
only the detuning $\Delta=\omega_0-\omega$. At an isolated zero of
$\rho(t)$, the angle $\varphi$ marks a coordinate-patch boundary and the basis
continues from either neighboring interval.  Global chain termination occurs
when the coupling vanishes throughout the protocol
or the reachable cyclic subspace closes.

\emph{Which problem is solved.}  Time dependence and state preparation are
independent choices.  A fixed group displacement changes the preparation and
conjugates the control coefficients.  For an extremal seed this covariance
preserves the arbitrary-control solution, but for an interior seed it does not
determine the later Lanczos hoppings.

A degree-$d$ polynomial $Q(X)$ changes the spectral measure of a fixed
reference operator $X$ and prepares a superposition of only its first $d+1$
old Krylov rungs, up to termination of the original chain
\cite{ChowdhuryMahapatra2026}. Its action has finite range in that reference
chain, since a degree-$d$ polynomial in a tridiagonal matrix connects sites
at most $d$ steps apart. A nonzero oscillator displacement or $SU(1,1)$
squeeze instead gives an extremal state an infinite weight tail; a generic
$SU(2)$ rotation occupies its finite multiplet. The distinction is between
finite-range polynomial preparation and generally nonlocal group dressing
in the specified basis, not spatial locality. Rebuilding the orthonormal
chain after a finite-degree preparation can still change Lanczos
coefficients at arbitrarily high index.
Table~\ref{tab:seed-control-map} separates the two
axes and records the exact status of each construction; $X$ in the table is
not another name for a generic laboratory Hamiltonian $H(t)$.

\begin{table*}[t]

\caption{\label{tab:seed-control-map}Exact status of the control and
seed combinations used in this paper.  ``Complete'' means that the
seed-adapted chain amplitudes, and hence its spread complexity, are determined
for the stated class.}
\small
\begin{ruledtabular}
\newcommand{\tabrr}{\raggedright}
\begin{tabular}{p{0.19\textwidth}p{0.22\textwidth}p{0.49\textwidth}}
\tabrr Control class & \tabrr Prepared first site & \tabrr Exact result \tabularnewline
\hline
\tabrr Constant $X$, or $f(t)X$ & \tabrr Admissible polynomial seed
$Q(X)\ket\chi/\sqrt{\mathcal N_Q}$ & \tabrr Complete fixed-chain solution with
measure $|Q(E)|^2\dd\mu(E)/\mathcal N_Q$.  The old-rung choice
$Q=p_r$ is a distinguished special case. \tabularnewline
\tabrr Arbitrary controls in one oscillator, $SU(2)$, or $SU(1,1)$ algebra &
\tabrr $\ket{\ell,0}$ & \tabrr Complete reduction to one linear or Riccati equation for the
Gauss coordinate $x(t)$, followed by Eq.~\eqref{eq:threesums}. \tabularnewline
\tabrr The same arbitrary controls & \tabrr Fixed group-displaced extremal seed
$W_{\rm s}\ket{\ell,0}$; also $W_{\rm s}\ket{j,j}$ for $SU(2)$ & \tabrr Complete by
time-independent conjugation.  The opposite $SU(2)$ pole uses a fixed $\pi$
rotation and reversed ladder.  The
transformed controls, chain data, and readout are given in
Eqs.~\eqref{eq:displaced-extremal-controls}--\eqref{eq:displaced-extremal-readout}. \tabularnewline
\tabrr The same arbitrary controls & \tabrr $W_{\rm s}\ket{\ell,r}$ with $r>0$ for the
oscillator or $SU(1,1)$, and $0<r<2j$ for $SU(2)$ & \tabrr
Conjugation and the first hopping are exact.  The spin-one middle weight has a complete arbitrary-control chain
(Sec.~\ref{sec:tipping}). The fixed-$X$, affine and Floquet reductions
give further complete classes. \tabularnewline
\tabrr Affine moving frame, Eq.~\eqref{eq:dressed-fixed-generator} &
\tabrr $W(0)Q(X)\ket\chi/\sqrt{\mathcal N_Q}$ & \tabrr Complete transported chain with
Christoffel measure, Eqs.~\eqref{eq:dressed-lanczos-coefficients}--
\eqref{eq:time-dependent-christoffel}, even when $[H(t),H(t')]\ne0$. \tabularnewline
\tabrr Periodic forced oscillator & \tabrr Fock seed $\ket r$, $r\in\mathbb N_0$ & \tabrr Exact monodromy and
cyclic dimension.  The full complexity is closed at the half turn; for a
higher rational fold it follows from a finite $q_F$-site Arnoldi problem. \tabularnewline
\tabrr Generic $H(t)$ and prescribed $W(t)\ket\chi$ & \tabrr Any seed in the stated domain &
\tabrr The first hopping is exactly the relative Fubini--Study speed in
Eq.~\eqref{eq:moving-seed-b1}; subsequent hoppings require the extended
recursion unless one of the preceding reductions applies.
\end{tabular}
\end{ruledtabular}
\end{table*}

For the arbitrary-control rows, ``complete reduction'' does not assert a
waveform-independent time expression.  It means that the propagator and
complexity are fixed once the indicated scalar or fundamental-representation evolution problem is
solved.  A full nonextremal chain is called complete only in the rows that
also determine its amplitudes or finite Arnoldi recursion.  Periodic forcing
uses the one-cycle unitary and therefore defines a stroboscopic Arnoldi chain,
distinct from the instantaneous chain of the extended generator.

The first site is part of the physical definition of a time-dependent Krylov
problem.  In the construction of Ref.~\cite{Takahashi:2024hex}, it may be any
normalized path $\ket{K_0(t)}$ satisfying
$\ket{K_0(0)}=\ket{\psi(0)}$.  A fixed path makes
$|\phi_0(t)|^2$ the survival probability.  An instantaneous eigenstate makes
it the adiabatic fidelity.  In both cases the chain quantifies departure from
a reference fixed independently of the exact solution.  We use a fixed first site for the static
and phase-gauged ladders and for the ground-seeded $V$ system.  The dressed
family of Sec.~\ref{sec:dictionary} and STIRAP use explicitly declared moving
references; the tipped-spin result is stated in its rotating frame.

The first hopping coefficient gives an exact local measure of that departure.
Let $\ket{K_0(t)}=W(t)\ket{\chi}$, where $\ket\chi$ is fixed and normalized and $W(t)$ is unitary.  Define
\begin{align}\label{eq:moving-seed-generator}
  G_W(t)&=W^\dagger H W-iW^\dagger\dot W,\\
  P_0(t)&=\ket{K_0(t)}\!\bra{K_0(t)} \nonumber.
\end{align}
We assume that the displayed products are defined on a common invariant
core containing $\ket\chi$, that $G_W$ has the self-adjoint realization used
below, and that $\ket\chi\in\operatorname{Dom}(G_W^2)$.

The first Lanczos residual then obeys
\begin{align}\label{eq:moving-seed-b1}
 b_1^2(t)
 &=\left\|[1-P_0(t)]\,[H(t)-i\partial_t]\ket{K_0(t)}\right\|^2
 \nonumber\\
 &=\braket{\chi|G_W^2|\chi}-\braket{\chi|G_W|\chi}^2 .
\end{align}
\noindent Thus $b_1$ is the uncertainty of the Hamiltonian seen in the moving frame.
Equivalently, it is the Fubini--Study speed of the reference ray relative to
the Schr\"odinger flow.  Suppose the ray belongs to a group orbit with real
  coordinates $u^\mu$ and quantum metric $g_{\mu\nu}$.  Whenever the
  Hamiltonian flow is tangent to this orbit, define its coordinate velocity by
  $-i[H,\varrho(u)]=v_H^\mu\partial_\mu\varrho(u)$, where
  $\varrho(u)$ is the rank-one projector onto the ray.  Then
Eq.~\eqref{eq:moving-seed-b1} becomes
\begin{equation}\label{eq:qgt-b1}
 b_1^2=g_{\mu\nu}(\dot u^\mu-v_H^\mu)
       (\dot u^\nu-v_H^\nu).
\end{equation}
For a fixed reference this reduces to the energy variance.  For an instantaneous eigenstate it
receives contributions only from the part of the moving-frame connection that
leaves the chosen eigenstate.  These are the matrix elements cancelled by
counterdiabatic driving.  Thus $b_1$ is the instantaneous
nonadiabatic-coupling scale; accumulated leakage also depends on spectral gaps
and on the protocol history.

Equation~\eqref{eq:moving-seed-b1} gives the exact local response of a
nonextremal preparation.  Write the transverse part of $G_W$ as
$q_W\,a^\dagger+\bar q_W\,a$ for the one-mode Heisenberg--Weyl group $H(1)$, as
$\Omega_xJ_x+\Omega_yJ_y$ for $\mathfrak{su}(2)$, and as
$\xi_1K_1+\xi_2K_2$ for $\mathfrak{su}(1,1)$, where
$K_1=(K_++K_-)/2$ and $K_2=(K_+-K_-)/(2i)$.  Here $q_W$ may be complex,
$\Omega_{x,y}$ and $\xi_{1,2}$ are real, and the representation labels obey
$m,n\in\mathbb N_0$, $-j\leq m_z\leq j$, and $h>0$.  The corresponding
weight-state seeds are $\ket m$, $\ket{j,m_z}$, and $\ket{h,n}$,
respectively, with $G_W$ the moving-frame generator of
Eq.~\eqref{eq:moving-seed-generator}.  Longitudinal Cartan terms act
diagonally on these seeds and drop out of the variance.  Direct evaluation
gives
\begin{align}\label{eq:nonextremal-first-hop}
 b_{1,H(1)}^2
   &=(2m+1)|q_W|^2,\nonumber\\
 b_{1,SU(2)}^2
   &=\frac{j(j+1)-m_z^2}{2}
     \left(\Omega_x^2+\Omega_y^2\right),\\
 b_{1,SU(1,1)}^2
   &=\frac{n^2+2hn+h}{2}
     \left(\xi_1^2+\xi_2^2\right).\nonumber
\end{align}
The transverse susceptibility grows as $2m+1$ for a displaced Fock seed,
peaks near the equator of a spin multiplet, and grows quadratically with
excitation number in the positive discrete series.  These exact first-hop
factors quantify the immediate departure from each chosen reference.  The
subsequent recursion determines whether the full chain retains a weight-ladder
form.

\subsection{Gauss coordinates and the geometry of the spread}
\label{sec:gauss}

Because $H(t)$ stays inside a finite-dimensional algebra, its propagator can
be factorized. We use $[J_0,J_\pm]=\pm J_\pm$ and
$[J_+,J_-]=2J_0$; a self-adjoint compact generator has
$C=\bar A$ and real $B$. For an $\mathfrak{su}(2)$-type generator
$H=A(t)J_++B(t)J_0+C(t)J_-$ we use the Wei--Norman/Gauss factorization
\cite{WeiNorman1963,Ban:93}
\begin{equation}\label{eq:gaussfact}
  U(t)=e^{x(t)J_+}e^{y(t)J_0}e^{z(t)J_-},
\end{equation}
and $i\dot U=HU$ reduces to a system for the three coordinates.
Differentiating Eq.~\eqref{eq:gaussfact}
and multiplying by $U^{-1}$,
reducing the three
adjoint actions by Baker--Campbell--Hausdorff, and matching the coefficients of
$J_+$, $J_0$ and $J_-$, gives
\begin{equation}\label{eq:weinorman}
  i\dot x=A+Bx-Cx^2,\quad
  i\dot y=B-2Cx,\quad
  i\dot z=Ce^{y},
\end{equation}
with $x(0)=y(0)=z(0)=0$. Only the Riccati equation for $x$ must be solved,
after which $y$ and $z$ follow by quadrature. These are local coordinates, so a
pole of $x$ marks a change of chart and not a singularity of $U$.
Appendix~\ref{app:gauss} gives the constant-coefficient solution and the
$\mathfrak{su}(1,1)$ sign changes.

Acting with Eq.~\eqref{eq:gaussfact} on an extremal seed produces a generalized
coherent state \cite{perelomov1972coherent,Glauber1963}, whose expansion
coefficients are the chain amplitudes. For the three families these are
\begin{align}\label{eq:amplitudes}
  |\phi_n|^2=&\binom{2j}{n}\frac{|x|^{2n}}{(1+|x|^2)^{2j}},\qquad
  \frac{|x|^{2n}}{n!}\,e^{-|x|^2},\nonumber\\[2pt]
  {~}& \binom{n+2h-1}{n}|x|^{2n}\bigl(1-|x|^2\bigr)^{2h}.
\end{align}
These are, in order, binomial, Poisson, and negative-binomial
distributions.  Their means are the three complexities
defined by Eq.~\eqref{eq:complexity},
\begin{equation}\label{eq:threesums}
  \C=\frac{2j\,|x|^2}{1+|x|^2}\,,\qquad
  \C=|x|^2,\qquad
  \C=\frac{2h\,|x|^2}{1-|x|^2},
\end{equation}
for the spin-$j$ representation of $SU(2)$, the Fock
representation of $H(1)$, and the positive discrete-series representation
$D_h^+$ of $\mathfrak{su}(1,1)$, respectively.  Only $|x|$ enters. Unitarity fixes
$\operatorname{Re}\,y=\ln(1\pm|x|^2)$ for the two curved families and
$\operatorname{Re}\,y_H=-|x|^2/2$ for the oscillator's central coordinate
$y_H$, defined in Appendix~\ref{app:unitarity}.
Here $0\le n\le 2j$ in the compact case, $n\in\mathbb N_0$ in the other
two cases, and $|x|<1$ for $D_h^+$.

The three means share a normalized geometric coordinate.
Introduce the representation-independent orbit metric
\begin{equation}\label{eq:normalized-orbit-metric}
 \dd\Sigma_\kappa^2=
 \frac{4\,\dd x\,\dd\bar x}{(1+\kappa|x|^2)^2},
 \qquad \kappa=+1,0,-1 .
\end{equation}
Let $d$ be the shortest
geodesic separation between the seed and the endpoint on the unit Bloch
sphere, the flat phase plane, or the unit hyperboloid.  In the corresponding
stereographic charts,
\begin{equation}\label{eq:stereo}
 |x|=\tan\frac d2,\qquad |x|=\frac d2,\qquad
 |x|=\tanh\frac d2
\end{equation}
for normalized curvature sign
$\kappa=+1,0,-1$, respectively, with $x$ the Gauss coordinate of
Eq.~\eqref{eq:gaussfact}.  Here $d$ is the endpoint separation in
Eq.~\eqref{eq:normalized-orbit-metric}.  The representation's physical
Fubini--Study metric is
$\dd s_{\rm FS}^2=(\ell/2)\dd\Sigma_\kappa^2$, and hence has Gaussian
curvature $2\kappa/\ell$.  Substitution into
Eq.~\eqref{eq:threesums} gives the common distance relation
\begin{equation}\label{eq:distancelaw}
 \C(t)=2\ell\,S_\kappa^2\!\left(\frac{d(t)}{2}\right),\qquad
 S_\kappa=\sin,\ \mathrm{id},\ \sinh ,
\end{equation}
where $\ell=j,\tfrac12,h$ in the same order.  The coherent-state geometry
for operator Krylov spread is developed in
Ref.~\cite{Caputa:2021sib}, and direct Bloch-sphere treatments of state-Krylov qubit dynamics
appear in Refs.~\cite{SeetharamanSinghNath2025,CafaroClementsAnuboyina2026}.
Equation~\eqref{eq:distancelaw} makes the optical trichotomy immediate.  At fixed
endpoint separation, the compact spread is bounded, the flat spread is
quadratic, and the noncompact spread is hyperbolic.  The drive
carries the
remaining dynamical information through $d(t)$.\footnote{On the sphere, the
shortest separation has domain $0\leq d\leq\pi$, and
$2j\sin^2(d/2)$ increases monotonically from $0$ to $2j$.  If the physical
trajectory continues past the antipode, its shortest endpoint distance
decreases and the finite chain recurs.  Thus the recurrence belongs to
$d(t)$, rather than to a turnover of $\C$ as a function of shortest distance.}
This relation applies to these extremal
coherent-state orbits.  The
photon-added, cat, three-level, and moving-reference sectors acquire their own
seed-adapted recurrences below.

The same group orbit fixes the fluctuations.  At every point reached by a
unitary drive within the algebra, the site distribution remains binomial,
Poisson, or negative binomial.  Writing
$V_{\rm K}=\sum_n(n-\C)^2|\phi_n|^2$ for the variance of the Krylov-site
index, one obtains
\begin{equation}\label{eq:eos}
 V_{\rm K}=\C\left(1-\kappa\frac{\C}{2\ell}\right),
 \qquad \kappa=+1,\,0,\,-1 .
\end{equation}
For the compact, flat, and noncompact families, Eq.~\eqref{eq:eos} uses the
same curvature sign $\kappa$ as Eq.~\eqref{eq:distancelaw}.  It holds for an
extremal seed and for any seed obtained from it by a time-independent group
displacement, under every envelope and detuning that keeps the evolution
within the same algebra.  Equation~\eqref{eq:eos} therefore collects all three
curvatures in a single relation and shows that pulse shaping moves the state
along a fixed mean--variance curve.  The two curved cases at fixed Hamiltonian
were obtained in Ref.~\cite{FuKimPalPal2024}.  Related phase-space fluctuations
are analyzed in Ref.~\cite{SChowdhury2026}, while coherent-state chains are
treated in Ref.~\cite{HaqueJacobi2024}.

The relation is directly testable because the Krylov-site distribution is the
population histogram for the extremal optical seeds.  In terms of the Mandel
parameter \cite{Mandel1979},
\begin{equation}\label{eq:mandel}
 Q_{\rm M}\equiv\frac{V_{\rm K}}{\C}-1
   =-\kappa\,\frac{\C}{2\ell},\qquad \C>0 .
\end{equation}
\noindent The sign distinguishes sub-Poissonian compact statistics, Poissonian oscillator
statistics, and super-Poissonian noncompact statistics.  In either curved
family, $|Q_{\rm M}|/\C=1/(2\ell)$ also returns the representation scale, namely the
atom number $2j$ or the pair-sector index $2h$.  Within the stated coherent
unitary setting, changing the drive preserves Eq.~\eqref{eq:eos}.
A measured deviation signals departure from this coherent unitary orbit, for
example through a changed seed recurrence as in Sec.~\ref{sec:photonadded}, a
nonlinear generator, or mixed and dissipative dynamics.

The metric controls the spread magnitude, while the symplectic connection
controls the phase.  Write the normalized coherent state as
$\ket{x}=\mathcal N\sum_n c_nx^n\ket n$, with $c_n$ and $\mathcal N$ real,
and let $\Phi=\arg x$.  Normalization cancels the radial part of
$\braket{x|\dd x}$, whereas differentiating $x^n$ contributes
$in\,\dd\Phi$.  The remaining sum is exactly the mean chain index,
\begin{equation}\label{eq:actionangle}
 i\bra{x}\dd\ket{x}=-\C\,\dd\Phi .
\end{equation}
Thus $\C$ and $\Phi$ form a local canonical (Darboux) pair on the Gauss chart, with
curvature two-form
$\mathcal F=-\dd\C\wedge\dd\Phi$.  On the compact orbit a second chart
is required at the antipode, and the two Berry potentials differ by a gauge
transformation on their overlap.

The angle is a laboratory quantity,
the phase of the transverse
lowering coherence $\braket{S_-}$ (equivalently, minus the conventional
Bloch-vector azimuth), the field phase, or the phase of the pair coherence
$\braket{ab}$.  Related complexity--connection pairings occur
in Krylov constructions \cite{Chakraborty2026}; the optical setting
makes both coordinates directly observable.

For any cyclic, possibly nonadiabatic evolution confined to the orbit, the
Aharonov--Anandan geometric phase is
\begin{align}\label{eq:complexity-geometric-phase}
 \gamma_{\rm g}
 &=\arg\braket{\psi(0)|\psi(T)}
   +i\int_0^T\!\dd t\,\braket{\psi(t)|\dot\psi(t)}\nonumber\\
 &\equiv-\oint\C\,\dd\Phi \pmod{2\pi}.
\end{align}
The second line uses a single-valued coherent-state section in one chart; a
loop crossing a chart boundary includes its transition phase.  It is the
symplectic area enclosed in the measured $(\C,\Phi)$ plane
\cite{Berry1984,AharonovAnandan1987}.  On the precession cone of
Sec.~\ref{sec:tipping}, this becomes
$\gamma_{\rm g}=-j\Omega_{\rm c}$, where $\Omega_{\rm c}$ is the enclosed
solid angle.  The oscillation amplitude in Eq.~\eqref{eq:tipped} and the
interferometric phase per period therefore determine the same cone.
Adiabatic geometric phases of parametric amplifiers and their
photon-statistical signatures provide the historical optical context
\cite{Gerry1989,VegaChorenoOjedaGuillenMota2024}.

The compact orbit also exposes the topology cleanly.  With the orientation
used above, $(2\pi)^{-1}\int_{S^2}\mathcal F=-2j$; its magnitude is the Chern
number $2j=N$ of the spin coherent-state line bundle.  The phase plane and
hyperboloid are noncompact, so the corresponding full-orbit curvature
integrals are unbounded rather than quantized.  This topological distinction
constrains the global orbit, while the driven trajectories below probe its
local metric and symplectic geometry.

Displaced Fock orbits, the Heisenberg--Weyl family of
Sec.~\ref{sec:photons}, separate the symplectic and metric responses of the
optical state.  For $D(\alpha)\equiv
\exp(\alpha a^\dagger-\bar\alpha a)$ and
$\ket{\psi_m(\alpha)}=D(\alpha)\ket m$, the Weyl identity for
$D^\dagger\dd D$ gives
\begin{align}\label{eq:displaced-fock-geometry}
 \mathcal B_m
 &=\frac{i}{2}
   (\bar\alpha\,\dd\alpha-\alpha\,\dd\bar\alpha),\nonumber\\
 \mathcal F_m&=i\,\dd\bar\alpha\wedge\dd\alpha,\\
 \dd s_m^2&=(2m+1)\,\dd\bar\alpha\,\dd\alpha . \nonumber
\end{align}
The connection and curvature are independent of the photon number in the
seed, whereas the quantum metric grows by the factor $2m+1$.  A closed
displacement loop therefore accumulates the same geometric phase for every
$m$, fixed by its signed phase-space area, while the first Krylov hopping of a
moving displaced reference is enhanced by the seed excitation, in agreement
with Eq.~\eqref{eq:nonextremal-first-hop}.

 This distinction is
visible in the resonantly forced mode. Writing $\alpha=|\alpha|e^{i\Phi}$
gives $\mathcal B_m=-|\alpha|^2\dd\Phi$ for every $m$. The vacuum chain
has $\C=|\alpha|^2$, so Eq.~\eqref{eq:actionangle} follows.  For the fixed-quadrature resonant protocol of
Sec.~\ref{sec:photonadded}, let $\Omega(t)/2$ be the real rotating-frame
forcing envelope and define
$\A(t)=\tfrac12\int_0^t\Omega(t')\,\dd t'$.
The displacement is $\alpha(t)=-i\A(t)$.
The one-photon result from our companion paper
\cite{ChowdhuryMahapatra2026} is
$\C=\A^2+1-e^{-2\A^2}$, reproduced in Eq.~\eqref{eq:photonaddedC}.
The seed-independent curvature in Eq.~\eqref{eq:displaced-fock-geometry}
fixes the phase of any closed displacement loop, while the complexity
depends on the seed-adapted chain of the chosen control history.
 Recent connections between Krylov complexity
and geometric phase are discussed in Refs.~\cite{Chakraborty2026,ZhengHeLiuZhang2026};
here both quantities are tied to independently measurable phase-space data.

\subsection{What the complexity measures in the laboratory}
\label{sec:dictionary}

For an extremal-weight seed, the phase convention in Eq.~\eqref{eq:gauge}
leaves the Krylov basis aligned with the weight basis.  The chain-number
operator $\hat N_{\rm K}=\sum_n n\ket{K_n}\bra{K_n}$ is then a shifted Cartan
generator.  For a collective-spin weight $\ket{j,m_z}$, the site label is
$n=m_z+j$ and $\hat N_{\rm K}=S_z+j$.  For the forced mode,
$\hat N_{\rm K}=a^\dagger a$.  For a signal--idler pair seeded in the
zero-number-difference sector, $\hat N_{\rm K}=\hat n_a=\hat n_b$.  These
group-theoretic identifications are standard
\cite{Caputa:2021sib,GrabaritsMedinaGuerraDelCampo2026}; in the optical
realizations they turn spread complexity into an excited-atom, photon, or pair
count measured by existing population readout.

\begin{figure}[tb]
\includegraphics[width=\columnwidth]{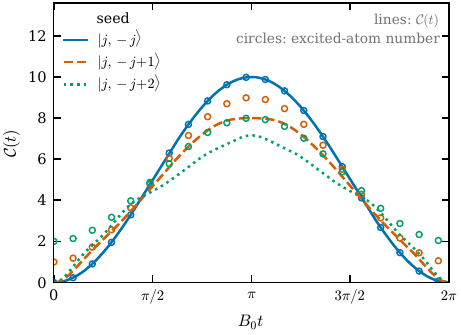}
\caption{\label{fig:seeds} Spread complexity for a resonantly driven $j=5$ ensemble.  Lines show
$\C(t)$ and circles show the measured excitation $\langle S_z\rangle+j$.
They coincide for the extremal seed.  For the interior seeds $m=1,2$, the
excitation remains above $\C$, with maximum gaps of one and two units because
the Krylov vectors mix level states.}
\end{figure}

An interior seed defines a different observable.  If the ensemble starts in
$\ket{j,-j+m}$, the Lanczos vectors are superpositions of Dicke levels and
$\hat N_{\rm K}$ develops off-diagonal matrix elements in the level basis.
Its expectation is then a seed-adapted coherent combination of populations and
coherences rather than the excited-atom
population alone.  Figure~\ref{fig:seeds} shows this
separation for the resonantly driven $j=5$ example.  The extremal curve equals
$\langle S_z\rangle+j$ at every time, whereas the two interior preparations
produce signed gaps of up to one and two units in the plotted protocol.
Related extremal and interior-seed behavior in an $SU(2)$ spin model is analysed
in Ref.~\cite{AguilarGutierrez2025}.  Thus preparation selects the operator
whose chain expectation is called complexity; it does more than rescale a
fixed population.

The distinction between an old rung and a new seed is essential.  Let
$\{\ket{K_n^{(X)}}\}$ be the chain generated by one fixed self-adjoint operator
$X$ from $\ket{K_0^{(X)}}$, with orthonormal polynomials $p_n$ and spectral
measure $\dd\mu$.  Its $r^{\text{th}}$ rung obeys
\begin{equation}\label{eq:old-rung-seed}
 \ket{K_r^{(X)}}=p_r(X)\ket{K_0^{(X)}},\quad
 \dd\mu_r(E)=|p_r(E)|^2\dd\mu(E).
\end{equation}
This spectral transformation is the Christoffel construction
\cite{Szego1939,ChowdhuryMahapatra2026}.

The polynomial $p_r$ is the unique normalized degree-$r$ member of the
orthogonal family associated with $(X,\ket{K_0^{(X)}})$, up to its fixed
leading-sign convention.  It should be distinguished from a general
admissible preparation polynomial $Q$.  Expanding
$Q(E)=\sum_{n=0}^{d}q_n p_n(E)$ gives
$Q(X)\ket{K_0^{(X)}}=\sum_{n=0}^{d}q_n\ket{K_n^{(X)}}$ before any
finite-support identifications.  Hence $Q=p_r$ prepares one old rung, whereas
a generic $Q$ prepares a superposition of old rungs and induces the normalized
measure $|Q(E)|^2\dd\mu(E)/\mathcal N_Q$.  Overall rescaling is removed by
$\mathcal N_Q$; for complex coefficients the positive factor is
$Q^\sharp Q$, as specified below.  On finite spectral support, polynomials
that agree at every support point define the same seed.

Choosing an old rung as the initial state creates a new chain whose first site
is $\ket{\widetilde K_0}=\ket{K_r^{(X)}}$.  It does not relabel the old chain.
Because $X\ket{K_r^{(X)}}$ generally contains both
$\ket{K_{r-1}^{(X)}}$ and $\ket{K_{r+1}^{(X)}}$, one normally has
$\ket{\widetilde K_n}\ne\ket{K_{r+n}^{(X)}}$.  Equation~\eqref{eq:old-rung-seed}
instead rebuilds the recurrence from the Christoffel-weighted measure.

For a fixed one-ladder generator, every Fock, Dicke, or pair-number state
reached before termination is an old rung of the vacuum, lowest-spin, or
lowest-pair chain, respectively.  The same reading applies to the static $V$
system.  Here $\ket B=\ket{K_1}$ and,
when $b_2>0$, the dark ray is the second rung $\ket{K_2}$, whose representative
is fixed by the positive-hopping convention; at $b_2=0$
the ground-seeded chain terminates at $\ket B$ and the orthogonal dark vector
belongs to an excluded one-site sector.

A factor $e^{-ir\varphi(t)}$ attached
to one weight-state seed is only a time-dependent global phase.  It changes a
diagonal gauge convention, not the seed ray, its spectral measure, or any
chain probability.  For a superposition of weight states, the basis vectors
acquire weight-dependent phases and the expansion coefficients transform with
the inverse phases if the physical ray is to remain unchanged.

Equation~\eqref{eq:moving-seed-b1} determines $b_1(t)$ for every normalized
seed in the stated common invariant domain, with a finite second moment of
$G_W$. The following constructions retain different parts of the dynamics.
Constant conjugation transports a solved seed problem, affine transport
preserves a fixed spectral measure and its polynomial modifications, and
parity determines a sector probability even when the full first moment
remains open.

The first handle is \emph{conjugation}.  Let $W_{\rm s}$ be a
time-independent unitary group element and prepare
$\ket{\psi_0}=W_{\rm s}\ket{\ell,0}$.  The identity
  $W_{\rm s}^\dagger\mathcal T
  e^{-i\int_0^t H(s)\,\dd s}W_{\rm s}
  =\mathcal T e^{-i\int_0^t W_{\rm s}^\dagger H(s)W_{\rm s}\,\dd s}$ maps the Krylov chain
unitarily to the extremal-seed chain of
$H'=W_{\rm s}^\dagger H W_{\rm s}$
\cite{BalasubramanianCaputaSimon2026}.  When
$W_{\rm s}^\dagger L_\alpha W_{\rm s}$ remains in the same algebra,
Eq.~\eqref{eq:distancelaw} applies with the separation measured from the
prepared seed.  Tipped ensembles, displaced coherent states, and squeezed
vacua have this form.

For the three optical ladders discussed in this paper, the covariance can be made fully explicit for
arbitrary time-dependent controls.  For the oscillator and the two curved
families, write
\begin{align}
 H_{\rm o}(t)&=\epsilon_0+\nu\hat n+q a^\dagger+\bar q a,
 \qquad \hat n=a^\dagger a,\nonumber\\
 H_\kappa(t)&=\epsilon_0+B L_0+qL_++\bar qL_-,
 \qquad \kappa=+1,-1,
\end{align}
where all coefficients may depend on time, $\epsilon_0,B,\nu$ are real, and
$\kappa=+1$ and $-1$ denote $SU(2)$ and $SU(1,1)$, respectively.  We choose
$W_{\rm s}=D(\alpha_{\rm s})$ for the oscillator.  For the curved families,
fix the coset representative by
\begin{equation}
 W_{\rm s}(z_{\rm s})=
 e^{z_{\rm s}L_+}
 e^{\ln(1+\kappa|z_{\rm s}|^2)L_0}
 e^{-\bar z_{\rm s}L_-},
\end{equation}
so that
\begin{equation}
 W_{\rm s}\ket{\ell,0}
 =(1+\kappa|z_{\rm s}|^2)^{-\kappa\ell}
 e^{z_{\rm s}L_+}\ket{\ell,0},
\end{equation}
with $z_{\rm s}\in\mathbb C$ for $SU(2)$ and
$|z_{\rm s}|<1$ for $SU(1,1)$.  Then
$H_{\rm s}'=W_{\rm s}^\dagger H W_{\rm s}$ has the same form, with
\begin{align}
 \nu_{\rm s}'&=\nu,\qquad
 q_{\rm s}'=q+\nu\alpha_{\rm s},\nonumber\\
 \epsilon_{\rm s}'&=\epsilon_0+\nu|\alpha_{\rm s}|^2
 +q\bar\alpha_{\rm s}+\bar q\alpha_{\rm s}
 \qquad \text{(oscillator)},\nonumber\\[2pt]
 q_{\rm s}'&=
 \frac{q+Bz_{\rm s}-\kappa\bar qz_{\rm s}^{\,2}}
 {1+\kappa|z_{\rm s}|^2},  \label{eq:displaced-extremal-controls} \\
 B_{\rm s}'&=
 \frac{B(1-\kappa|z_{\rm s}|^2)
 -2\kappa(q\bar z_{\rm s}+\bar qz_{\rm s})}
 {1+\kappa|z_{\rm s}|^2},\nonumber\\
 \epsilon_{\rm s}'&=\epsilon_0
 \qquad \text{($SU(2)$ or $SU(1,1)$)}. \nonumber
\end{align}
The oscillator formula includes the number rotation used for a forced mode;
for the strict Heisenberg--Weyl algebra, $\nu=0$ and a fixed displacement
changes only the central term.  A fixed stabilizer phase rotates
$q_{\rm s}'$ but leaves the positive hoppings and complexity unchanged.

On any interval where $q_{\rm s}'\ne0$, set
$q_{\rm s}'=\rho_{\rm s}e^{-i\varphi_{\rm s}}$.  The physical
positive-hopping chain and its recurrence data are
\begin{align}
 \ket{K_n^{({\rm s})}(t)}
 &=W_{\rm s}e^{-in\varphi_{\rm s}(t)}\ket{\ell,n},\nonumber\\
 a_{n,{\rm s}}(t)&=\epsilon_{\rm s}'(t)+B_{\rm s}'(t)w_n
 -n\dot\varphi_{\rm s}(t),\nonumber\\
 b_{n+1,{\rm s}}(t)&=\rho_{\rm s}(t)\beta_{n+1},
 \label{eq:displaced-extremal-chain}
\end{align}
where $B_{\rm s}'=\nu$ in the oscillator row.  At an isolated zero of
$q_{\rm s}'$, the phase gauge is continued from either neighboring interval.
The representation data are
\begin{equation}
\begin{array}{c|c|c}
 &w_n&\beta_{n+1}\\ \hline
oscillator&n&\sqrt{n+1}\\
SU(2)&n-j&\sqrt{(n+1)(2j-n)}\\
SU(1,1)&n+h&\sqrt{(n+1)(2h+n)} .
\end{array}
\end{equation}

Let $x_{\rm s}(0)=0$ be the Gauss or displacement coordinate of
$U_{\rm s}=W_{\rm s}^\dagger U W_{\rm s}$.  It obeys
\begin{align}
 i\dot x_{\rm s}&=\nu x_{\rm s}+q_{\rm s}'
 &&\text{(oscillator)},\nonumber\\
 i\dot x_{\rm s}&=q_{\rm s}'+B_{\rm s}'x_{\rm s}
 -\kappa\bar q_{\rm s}'x_{\rm s}^{\,2}
 &&\text{($SU(2)$ or $SU(1,1)$)},
\end{align}
and the complete displaced-seed complexities are
\begin{equation}\label{eq:displaced-extremal-complexity}
 \C_{\rm s}(t)=
 \begin{cases}
 |x_{\rm s}|^2, & \text{oscillator},\\[2pt]
 \dfrac{2j|x_{\rm s}|^2}{1+|x_{\rm s}|^2}, & SU(2),\\[7pt]
 \dfrac{2h|x_{\rm s}|^2}{1-|x_{\rm s}|^2}, & SU(1,1).
 \end{cases}
\end{equation}
Thus arbitrary control histories are reduced exactly to one scalar
initial-value problem.  A waveform-specific time expression requires that
linear or Riccati equation to be solved.  The seed-relative distribution has
the direct optical readout
\begin{align}\label{eq:displaced-extremal-readout}
 P_{n,{\rm s}}(t)=
 \left|\braket{\ell,n|U_{\rm s}(t)|\ell,0}\right|^2,
 \quad
 \C_{\rm s}(t)=\sum_n nP_{n,{\rm s}}(t).
\end{align}
Experimentally, this is obtained by undoing the fixed displacement, collective
rotation, or two-mode squeeze before photon, Dicke, or pair-number counting.

The same conjugation maps a displaced excited seed
$W_{\rm s}\ket{\ell,r}$ to $\ket{\ell,r}$ under $H_{\rm s}'$.  Every $r>0$
is nonextremal for the oscillator and $SU(1,1)$ ladders, whereas the interior
$SU(2)$ range is $0<r<2j$.  The endpoint $r=2j$ is the highest weight and is
already included in the exact extremal class by replacing $W_{\rm s}$ with
$W_{\rm s}R_\pi$, where $R_\pi\ket{j,-j}$ is proportional to $\ket{j,j}$;
the physical chain is then ordered downward from the highest weight.  For the
first jump, Eq.~\eqref{eq:nonextremal-first-hop} gives
\begin{equation}\label{eq:displaced-first-jump-b1}
 b_{1,{\rm s}}^2=
 \begin{cases}
 3|q_{\rm s}'|^2, & \text{oscillator},\\
 2(3j-1)|q_{\rm s}'|^2, & SU(2),\\
 2(3h+1)|q_{\rm s}'|^2, & SU(1,1),
 \end{cases}
 \qquad (r=1).
\end{equation}
For the oscillator and $SU(1,1)$, and for $SU(2)$ when $j>1/2$, these are exact
local responses.  When $j=1/2$, the $SU(2)$ seed with $r=1=2j$ is the opposite
extremal state and its two-site arbitrary-control chain is complete.
For $j=1$, the interior seed $r=1$ also closes explicitly for arbitrary
controls, as shown in Sec.~\ref{sec:tipping}; fixed conjugation transports
that complete solution to $W_{\rm s}\ket{j=1,m=0}$. For general interior
weights, the later recurrence needs further information, with the complete
classes treated here distinguished in Table~\ref{tab:seed-control-map}.

A fixed group-displaced Fock seed is likewise unitarily equivalent to the
corresponding primed nonextremal problem; that equivalence alone does not close
its chain.  A Christoffel solution follows only when the undressed seed is an
old rung, or more generally $Q(X)\ket\chi$, in the cyclic subspace of the same
fixed operator $X$.  These are precisely the hypotheses of
Eq.~\eqref{eq:dressed-polynomial-seed}.

The second exact mechanism is
\emph{dressing}, which combines a moving frame with a Christoffel-transformed
seed.
Let $X$ be self-adjoint, let $W(t)$ be unitary, and let
$\tau(t),\gamma(t)$ be real.  We assume that $W$ is strongly
differentiable on a common invariant core for $X$, $\dot W$, and the displayed
products, and that the corresponding self-adjoint realizations are used; these
conditions are automatic in the finite-dimensional atomic examples.
The explicitly time-dependent Hamiltonian
\begin{equation}\label{eq:dressed-fixed-generator}
 H(t)=i\dot W W^\dagger+\dot\tau\,WXW^\dagger+\dot\gamma\,\mathbf 1
\end{equation}
is obtained from
$H=i\dot UU^\dagger$ using the exact propagator
\begin{equation}\label{eq:dressed-propagator}
 U(t,0)=e^{-i\Delta\gamma}W(t)e^{-i\Delta\tau X}W^\dagger(0),
 \end{equation}
where $\Delta\tau=\tau(t)-\tau(0)$ and $\Delta\gamma$ is defined analogously.  The first term in
Eq.~\eqref{eq:dressed-fixed-generator} is the inertial contribution that
keeps the result valid for a genuinely moving frame.  Hamiltonians at
different times may therefore fail to commute even though the dressed
evolution is generated by the single fixed operator $X$.
The effective-time reduction follows when
$G_W=\dot\tau X+\dot\gamma\,\mathbf 1$ on the relevant invariant cyclic
subspace; a generic prescribed pair $H(t),W(t)$ need not have this affine
fixed-operator form.

Now prepare
\begin{align}\label{eq:dressed-polynomial-seed}
 \ket{\psi_Q(0)}
 &=\frac{W(0)Q(X)\ket\chi}{\sqrt{\mathcal N_Q}},\\
 \mathcal N_Q&=\braket{\chi|Q^\dagger(X)Q(X)|\chi} \nonumber .
\end{align}
Here $Q$ is any polynomial for which $0<\mathcal N_Q<\infty$.
For unbounded $X$, the vector $X^kQ(X)\ket\chi$ must exist to every
Lanczos order used; an infinite chain requires the corresponding moments to
remain finite.  Finite-step statements need only the associated finite set of
moments.  The all-level expansion below additionally requires polynomial
completeness in $L^2(\mu_Q)$, or direct invariance of the polynomial Krylov
closure under $e^{-isX}$.  The finite-support and Gaussian- or Poisson-type
transformed measures used here satisfy this condition.

Let
$\ket{K_n^{Q,X}}$ be the fixed Krylov basis generated by $X$ from
$Q(X)\ket\chi/\sqrt{\mathcal N_Q}$.
Denote its Lanczos coefficients by $a_n^{Q,X}$ and $b_{n+1}^{Q,X}$.  On an
interval where $\dot\tau\ne0$, the transported recurrence has
\begin{equation}\label{eq:dressed-lanczos-coefficients}
 a_n(t)=\dot\gamma(t)+\dot\tau(t)a_n^{Q,X},
 \qquad
 b_{n+1}(t)=|\dot\tau(t)|b_{n+1}^{Q,X} .
\end{equation}
For $\dot\tau>0$ the canonical basis is
$W(t)\ket{K_n^{Q,X}}$; for $\dot\tau<0$ it is the site-rephased basis
$(-1)^nW(t)\ket{K_n^{Q,X}}$.  An isolated zero of $\dot\tau$ is a patch
boundary at which the two positive-hopping gauges are joined; the basis rays
and all site probabilities continue continuously.  The continuous transported
gauge uses $W(t)\ket{K_n^{Q,X}}$ throughout.
If $\dot\tau=0$ on a finite interval, $G_W=\dot\gamma\,\mathbf 1$ there and the
instantaneous recursion terminates at the seed.  Keeping the higher transported
sites is then the basis convention inherited continuously from the preceding
interval, and every site probability is frozen across the plateau.

In that gauge, the amplitudes and complexity are
\begin{align}\label{eq:dressed-amplitudes}
 \phi_n^Q(t)&=\frac{e^{-i\Delta\gamma}}{\sqrt{\mathcal N_Q}}
   \braket{K_n^{Q,X}|e^{-i\Delta\tau X}Q(X)|\chi},\nonumber\\
 \C_Q(t)&=\C_Q^X(\Delta\tau).
\end{align}
Here $\C_Q^X(s)$ denotes the spread complexity generated by the fixed
operator $X$ from the normalized seed
$Q(X)\ket\chi/\sqrt{\mathcal N_Q}$ at evolution parameter $s$.
On an interval with $\dot\tau<0$, amplitudes in the canonical positive-hopping gauge
acquire $(-1)^n$; the probabilities and $\C_Q$ are unchanged.
All transported-chain probabilities
depend on the envelope through the effective time $\Delta\tau$, while $W(t)$
specifies their laboratory embedding.  The seed spectral measure is exactly
\begin{equation}\label{eq:time-dependent-christoffel}
 \dd\mu_Q(E)=\frac{|Q(E)|^2}{\mathcal N_Q}\,\dd\mu(E).
\end{equation}
For complex coefficients it is useful to write
$|Q(E)|^2=Q^\sharp(E)Q(E)$, with
$Q^\sharp(z)=\overline{Q(\bar z)}$; $Q(E)^2$ agrees with this Hermitian product only when $Q(E)$ is real for
$\mu$-almost every $E$ on the spectral support.
For a finite atomic measure, zeros of $Q$ can remove spectral atoms,
\begin{equation}\label{eq:christoffel-support}
 \operatorname{supp}\mu_Q
 =\{E\in\operatorname{supp}\mu:Q(E)\ne0\}.
\end{equation}
The transformed chain then has as many sites as there are surviving atoms and
may terminate earlier than the original chain.  An isolated zero inside a
continuous support changes the weight and recurrence without removing that
support.

Within the affine moving-frame row of Table~\ref{tab:seed-control-map}, $W(t)$
and $Q$ encode independent choices.  Setting $W=\mathbf1$ and $Q=p_r$
rebuilds the autonomous old-rung chain; taking $Q=1$ transports the extremal
seed; and combining a nonconstant $W(t)$ with $Q=p_r$ transports the rebuilt
nonextremal chain.  Each case is exact because the moving-frame generator is
affine in the same fixed $X$.  A moving algebra direction outside this form
requires the extended recursion rather than a reweighting of one fixed
measure.

An explicit displacement-dressed optical corollary follows by taking
$X=a+a^\dagger$ and $W(t)=D[\alpha(t)]$.  The Hamiltonian
\begin{equation}\label{eq:displacement-dressed-hamiltonian}
 H_{\rm dd}(t)=i\dot D[\alpha]D^\dagger[\alpha]
 +\dot\tau\,D[\alpha](a+a^\dagger)D^\dagger[\alpha]
 +\dot\gamma\,\mathbf 1
\end{equation}
is a forced mode with independently programmable quadratures and a scalar
phase.\footnote{The two operator identities are
$i\dot D D^\dagger=i\dot\alpha\,a^\dagger-i\dot{\bar\alpha}\,a
+\tfrac{i}{2}(\alpha\dot{\bar\alpha}-\bar\alpha\dot\alpha)$ and
$D(a+a^\dagger)D^\dagger=a+a^\dagger-\alpha-\bar\alpha$.
They make the Hermiticity and laboratory drive quadratures explicit.}
Its generators at different times are generically noncommuting, while
Eq.~\eqref{eq:dressed-propagator} remains exact.  For the nonextremal seed
$D[\alpha(0)]\ket1$, obtained from $Q(X)=X$ and $\ket\chi=\ket0$, the
transported-chain complexity is
\begin{equation}\label{eq:displacement-dressed-photon-added}
 \C_{\rm dd}(t)=(\Delta\tau)^2+1-e^{-2(\Delta\tau)^2}.
\end{equation}
The arbitrary displacement path changes the laboratory state and force
quadratures, whereas the chain populations depend only on the effective time
$\Delta\tau$, equivalently the accumulated fixed-generator area
when $\dot\tau$ is a pulse envelope.  The same construction applies to Dicke and fixed-number-difference pair
sectors when $W(t)$ is, respectively, an $SU(2)$ rotation or an $SU(1,1)$
squeeze, the seed is polynomial in one fixed generator $X$, and the physical
Hamiltonian has the affine moving-frame form in
Eq.~\eqref{eq:dressed-fixed-generator}.

The transported-frame construction is the fixed-generator mechanism of
Ref.~\cite{GrabaritsMedinaGuerraDelCampo2026}; the polynomial change of seed
and measure is the Christoffel construction of
Refs.~\cite{Szego1939,ChowdhuryMahapatra2026}.  Equations~
\eqref{eq:dressed-fixed-generator}--\eqref{eq:time-dependent-christoffel}
combine them into an exact time-dependent result for excited-weight and
polynomially jumped preparations.
Higher Fock seeds use $Q=p_m$, the normalized Hermite polynomial of the vacuum
quadrature measure, and give the confluent
Christoffel transforms discussed in Sec.~\ref{sec:photonadded}.
\footnote{For a seed consisting of one weight
state, the factor $e^{-im\varphi(t)}$ in Eq.~\eqref{eq:gauge} multiplies the
whole seed by a time-dependent global phase.  It shifts the diagonal
connection convention but leaves $|Q(E)|^2$, the off-diagonal Lanczos
coefficients, and the complexity unchanged.  For a superposition, this
componentwise rephasing is no longer global and defines a different moving
reference unless the expansion coefficients are transformed inversely.  A
physical change of the relative phases changes the complex polynomial $Q$; it
changes the normalized seed measure, and hence the recurrence data used here,
only through $Q^\sharp(E)Q(E)/\mathcal N_Q$ on the spectral support.}
Appendix~\ref{app:moving-seeds} gives the derivation.

The third is \emph{parity}. If the generator is odd under an involution that
the seed respects, the chain inherits a grading, and one exactly computable
observable survives even when the full complexity does not close. This is what
makes cat seeds tractable in Sec.~\ref{sec:parity}.

The zero-spread condition also has a compact exact form.  For a fixed
reference, an instantaneous eigenvector gives $b_1(t)=0$ at that instant.  The
spread vanishes throughout a protocol precisely when the seed is a common
eigenvector,
\begin{equation}\label{eq:darkcrit}
 H(t)\ket{\psi_0}=\lambda(t)\ket{\psi_0}\qquad\text{for every }t.
\end{equation}
For a moving reference $W(t)\ket\chi$, the equivalent condition is
$G_W(t)\ket\chi=g(t)\ket\chi$ throughout the protocol.  An isolated zero of
$b_1(t)$ is a local closure and the chain may reopen.  The fixed-direction
family $H(t)=f(t)X$, with $X\ket{\psi_0}=\lambda\ket{\psi_0}$,
gives the
examples below.
For the fixed group-displaced extremal seeds above, the common-eigenray condition is
simply $q_{\rm s}'(t)=0$ throughout the protocol.  It becomes
$q(t)+\nu(t)\alpha_{\rm s}=0$ for a coherent oscillator seed.  In $SU(2)$ it
requires every control vector to share the fixed Bloch axis selected by the
seed.  In $SU(1,1)$ a fixed $|z_{\rm s}|<1$ must solve
$q+Bz_{\rm s}+\bar qz_{\rm s}^{\,2}=0$ for every $t$.  Thus a fixed coherent
input can lock an elliptic generator, whereas a nonzero parabolic or
hyperbolic generator has no normalizable coherent eigenray at finite
squeezing.
A constant generator and a resonant shaped envelope
preserve
the one-site chain, whereas an independently varying detuning rotates the
generator away from that eigendirection.

\section{Collective two-level atoms}\label{sec:atoms}
The compact optical family begins with an ensemble of two-level atoms.  Its
$SU(2)$ coherent-state manifold is a sphere, whose positive curvature and
finite representation bound the chain spread.  A fully de-excited ensemble is
both the standard experimental preparation and the extremal seed for which
Eq.~\eqref{eq:dictionary} identifies complexity with the excited-atom number.

\subsection{The driven ensemble}

Consider $N=2j$ identical two-level atoms with transition frequency $\omega_0$,
coupled uniformly to a classical field so that the dipole interaction is the
same for each. In the symmetric (Dicke) sector the atomic operators combine
into collective spin generators \cite{WallsMilburn2008}, and
\begin{equation}\label{eq:su2H}
  H(t)=\omega_0S_z+\frac{\Omega(t)}{2}
  \left(e^{-i\omega t}S_++e^{i\omega t}S_-\right),
\end{equation}
where $\Omega(t)\ge0$ is the real Rabi envelope and $\omega$ the carrier
frequency.\footnote{The optical effective Hamiltonians in this section and in
Eqs.~\eqref{eq:forcedH} and \eqref{eq:opaH} use the rotating-wave approximation.
For the atomic problem, $\Omega$ and $|\Delta|$ are small compared
with the carrier scale and the envelope varies slowly on the scale
$1/\omega_0$; the forced-mode and three-wave-mixing models use the analogous
separation of coupling, mismatch, and carrier scales.} The same Hamiltonian
describes a spin in a rotating magnetic field, and a driven qubit. It also
arises from the fully quantized cavity problem. If the field is a large,
weakly depleted coherent state, replacing $a\mapsto\beta e^{-i\omega t}$ in the
Tavis--Cummings Hamiltonian \cite{TavisCummings1968} returns
Eq.~\eqref{eq:su2H} with
$\Omega=2g|\beta|$.  We use this semiclassical reduction
in the weak-depletion, weak-back-action regime.  Dissipative cavity dynamics
require the open-system extension discussed in Sec.~\ref{sec:discussion}.
The seed is the fully de-excited state $\ket{j,-j}$, an extremal weight, so the
dictionary applies and $\C$ is the number of excited atoms.

\subsection{Monochromatic driving}

\begin{figure}[tb]
\includegraphics[width=\columnwidth]{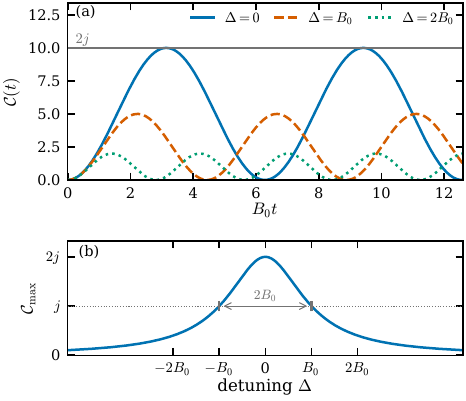}
\caption{\label{fig:rabi} Monochromatically driven ensemble with $j=5$. (a) Complexity oscillates at the
generalized Rabi frequency.  Resonance reaches the bound $2j$, while detuning
raises the frequency and reduces the amplitude. (b) The maximum over time is
the coherent-excitation Lorentzian in Eq.~\eqref{eq:lorentz}, with half-width
$B_0$.}
\end{figure}

For constant $\Omega=B_0$, the transformation to the frame rotating at
$\omega$ removes the time dependence entirely. Including the inertial term,
\begin{equation}\label{eq:rotframe}
  H_{\rm rot}=VHV^\dagger+i\dot VV^\dagger=\Delta S_z+B_0S_x,
\end{equation}
with $\Delta=\omega_0-\omega$ and $V=e^{i\omega tS_z}$.  This is the
constant-generator reduction of the phase transformation in
Sec.~\ref{sec:time-dependence}, Eq.~\eqref{eq:gaugedcoeffs}.
The chain amplitudes are binomial, with the single-atom Rabi probability
\begin{equation}\label{eq:rabip}
  p(t)=\frac{B_0^2}{B_0^2+\Delta^2}\,
  \sin^2\!\left(\frac{t}{2}\sqrt{B_0^2+\Delta^2}\right),
\end{equation}
so that $|\phi_n|^2=\binom{2j}{n}p^n(1-p)^{2j-n}$ and
\begin{equation}\label{eq:su2C}
  \C(t)=2j\,p(t).
\end{equation}
Each atom flops independently, and the ensemble complexity is the mean number
of atoms that have flopped. Equation~\eqref{eq:su2C} is in fact more general than
the monochromatic case that produced it. For \emph{any} $\mathfrak{su}(2)$
drive $H(t)=\bm\lambda(t)\!\cdot\!\bm S$ seeded at $\ket{j,-j}$ the collective
propagator is a single $SU(2)$ element applied to every atom, so
Eq.~\eqref{eq:su2C} holds exactly, with $p$ the single-atom excitation
probability, for every envelope, detuning and sequence. Three drives below give
$p$ in closed form, and Eq.~\eqref{eq:rabip} gives the first closed form.
On resonance $\C$ reaches its bound $2j$ at the inversion times;
off resonance the maximum is
\begin{equation}\label{eq:lorentz}
  \C_{\max}(\Delta)=\frac{2jB_0^2}{B_0^2+\Delta^2},
\end{equation}
a Lorentzian of half-width $B_0$ in a detuning scan.  It is measured from the
maximum excited-state population of the closed Rabi evolution
(Fig.~\ref{fig:rabi}).  Incorporating decay, dephasing, and a probe observable
would turn this coherent profile into a spectroscopic response.

\subsection{Shaped pulses and the accumulated area}\label{sec:arealaw}

\begin{figure}[tb]
\includegraphics[width=\columnwidth]{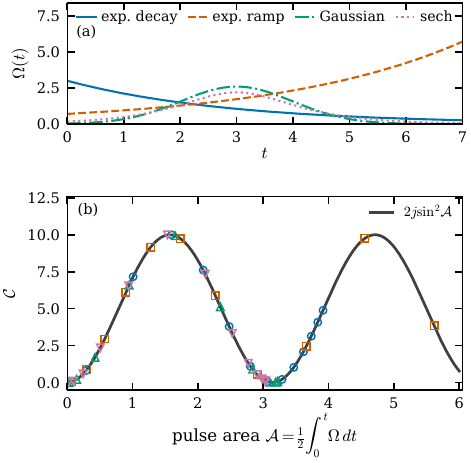}
\caption{\label{fig:arealaw} Pulse-area collapse for Eq.~\eqref{eq:arealawsu2} with $j=5$. (a) Exponentially
decaying, exponentially ramped, Gaussian, and hyperbolic-secant resonant
envelopes. (b) Direct Schr\"odinger evolution plotted against each pulse's
accumulated area.  All four results lie on the universal curve
$2j\sin^2\A$ (solid), so the envelope enters the complexity only through
$\A(t)$.}
\end{figure}

Now let the envelope be arbitrary. On resonance ($\omega=\omega_0$) the
rotating-frame generator is $\Omega(t)S_x$, a fixed operator carrying a
time-dependent coefficient,  the commuting $f(t)X$ control class in
Table~\ref{tab:seed-control-map}, so generators at different times commute and the
propagator is an ordinary exponential of the time integral.  With the pulse-area convention introduced in
Sec.~\ref{sec:gauss},
\begin{equation}\label{eq:area}
  \A(t)=\frac12\int_0^t\Omega(t')\,\dd t',
\end{equation}
the propagator is $\exp[-i\A\,(2S_x)]$. In the defining representation
this is $\exp(-i\A\sigma_x)$, a rotation of the Bloch sphere through the angle
$2\A$ about $\hat x$. The orbit length accumulated
along that great circle is $2|\A|$ for a one-way pulse, whereas the geodesic
separation of the endpoints is globally
fixed by $\cos d=\cos2\A$ with
$d\in[0,\pi]$; only before the first antipode does
$d=2|\A|$.  Globally,
$\sin^2(d/2)=\sin^2\A$. Hence
Eq.~\eqref{eq:distancelaw}
gives
\begin{equation}\label{eq:arealawsu2}
  \C(t)=2j\sin^2\A(t).
\end{equation}
This expression holds \emph{for every envelope}.\footnote{The constant-phase reduction underlying
Eq.~\eqref{eq:arealawsu2} is also
integrated, for an arbitrary real envelope, in
Ref.~\cite{GrabaritsMedinaGuerraDelCampo2026}, including its compact, noncompact, and $\mathfrak{su}(4)$ examples.}
Two pulses of equal area produce equal complexity
whatever their shapes, and a plot of $\C$ against $\A$ collapses them onto one
curve (Fig.~\ref{fig:arealaw}). The same argument in the other two families
gives $\C=\A^2$ for a resonantly forced mode and $\C=2h\sinh^2\A$ for a
resonantly pumped pair, the three cases differing only through $S_\kappa$ in
Eq.~\eqref{eq:distancelaw}.\footnote{The flat member is the mean photon number of a
displaced vacuum.  Ref.~\cite{GrabaritsMedinaGuerraDelCampo2026} derives the
corresponding result for an arbitrary complex drive envelope.}

The controlling quantity is the pulse area familiar from coherent optics
\cite{McCallHahn1969}. A decaying drive has finite
total area, so the complexity approaches the constant
$2j\sin^2[\Omega_0/(2\eta)]$ for $\Omega=\Omega_0e^{-\eta t}$, leaving the
ensemble frozen in a partially inverted state. Complete inversion
occurs whenever $\A=(2k+1)\pi/2$, with $k\in\mathbb Z$; it is the asymptotic
state only when the total area has one of these values. A ramped drive has
unbounded area and keeps cycling, ever faster.

The same commuting structure has a second consequence and provides a
speed-limit diagnostic of resonance for the rotating-frame pair used below.
Since $\C=\braket{\hat N_{\rm K}}$ with
$\hat N_{\rm K}=S_z+j$, its rate of change is
$\dot\C=i\braket{[H,\hat N_{\rm K}]}$.  The Robertson inequality
\cite{Robertson1929} applied to $(H,\hat N_{\rm K})$ gives the Krylov speed
limit \cite{MandelstamTamm1945,HornedalCarabbaMatsoukasDelCampo2022, GrabaritsMedinaGuerraDelCampo2026},
\begin{equation}\label{eq:qsl}
  |\dot\C|\le2\,\Delta H\,\Delta N_{\rm K} .
\end{equation}

Robertson's equality condition requires
$(H-\braket{H})\ket\psi$ and
$(\hat N_{\rm K}-\braket{\hat N_{\rm K}})\ket\psi$ to be proportional with a
purely imaginary constant.  It can be tested on one atom because
Eq.~\eqref{eq:su2C} reduces the ensemble dynamics to that problem.  On
resonance, a spin starting from $\ket\downarrow$ evolves as
$\ket\psi=\cos\A\ket\downarrow-i\sin\A\ket\uparrow$, with
$\braket H=0$ and $\braket{\hat N_{\rm K}}=\sin^2\A$.  Hence
\begin{align}
  (H-\braket{H})\ket\psi&=\tfrac{\Omega}{2}
     \bigl(\cos\A\ket\uparrow-i\sin\A\ket\downarrow\bigr),\label{eq:qslsat}\\
  i(\hat N_{\rm K}-\braket{\hat N_{\rm K}})\ket\psi
  &=\sin\A\cos\A
     \bigl(\cos\A\ket\uparrow-i\sin\A\ket\downarrow\bigr).\nonumber
\end{align}
The two vectors are parallel wherever their ratio is defined, and the
inequality is saturated throughout the nonzero-spread resonant evolution.
For the rotating-frame pair $(H_{\rm rot},\hat N_{\rm K})$, resonance realizes
phase-locked geodesic transport and saturates the speed limit throughout the
nonzero-spread evolution.  For $B_0\ne0$, detuned evolution is strictly below
the bound, apart from return instants where both $\Delta N_{\rm K}$ and
$|\dot\C|$ vanish.\footnote{For the purely longitudinal case $B_0=0$, the
seed is stationary and the zero-speed equality holds identically at all times.
Reference
\cite{GrabaritsMedinaGuerraDelCampo2026} obtains the corresponding criterion in
the interaction picture, where the uncertainty is $\Delta H_I$ and saturation
also occurs at half inversion, $|x|=1$.  The rotating-frame statement here
refers to the pair $(H_{\rm rot},\hat N_{\rm K})$.}

The rotating-frame map separates carrier oscillations from genuine envelope
dynamics.  Monochromatic resonance is equivalent to a static generator,
whereas a shaped pulse introduces the physically distinct choice between
finite and unbounded accumulated area.

\subsection{Detuned exponential driving and compact--noncompact continuation}\label{sec:bessel}

A detuned exponential envelope rotates the algebra direction and converts the Riccati
equation in
Eq.~\eqref{eq:weinorman} into a Bessel problem. The underlying
two-level evolution is the Demkov model \cite{Demkov1964}.  Its
Bessel solution determines the projective ratio and hence the extremal-seed
complexity. Keeping the independently normalized solution also reconstructs
the stabilizer coordinate and the full Gauss propagator.  The same
reduction controls a detuned exponentially varying
$\mathfrak{su}(1,1)$ pump after analytic continuation, so it closes the compact
and noncompact driven cases in parallel.
\par Let
$\Omega(t)=\Omega_0e^{-\eta t}$ describe an exponentially decaying drive at
fixed detuning $\Delta$, with $\Omega_0$ the peak Rabi frequency, and let
  $U_{ij}$ be the entries of the $SU(2)$ propagator in the
  spin-$\tfrac12$ representation of
Eq.~\eqref{eq:gaussfact}. Take $\Omega_0>0$ and $\eta>0$; for
$\Omega_0=0$ the uncoupled result $U_{22}=e^{i\Delta t/2}$ is immediate.
Set $s_0=\Omega_0/(2\eta)$ and
$u_{\rm B}(t)=e^{-i\Delta t/2}U_{22}(t)$.  When regarded as a function of
$s=s_0e^{-\eta t}$, this auxiliary amplitude obeys
\begin{equation}\label{eq:besselode}
  s^2u_{\rm B}''
  -\frac{i\Delta}{\eta}su_{\rm B}'
  +s^2u_{\rm B}=0 ,
\end{equation}
where primes denote derivatives with respect to $s$.  Its two independent
solutions are $s^\nu J_\nu(s)$ and $s^\nu J_{-\nu}(s)$, with
\begin{equation}\label{eq:besselp}
  \nu=\frac12+\frac{i\Delta}{2\eta}.
\end{equation}
Restoring the factored phase gives the propagator entry
\begin{equation}\label{eq:u22-bessel}
 U_{22}(t)=s_0^{\,i\Delta/(2\eta)}s^{1/2}
 \bigl[d_+J_\nu(s)
       +d_-J_{-\nu}(s)\bigr].
\end{equation}
The constants follow from $U_{22}(0)=1$ and
$\dot U_{22}(0)=i\Delta/2$, equivalently from the Bessel Wronskian at
$s=s_0$.\footnote{A direct elimination of $U_{12}$ gives
$s^2U_{22}''+[s^2+\Delta^2/(4\eta^2)-i\Delta/(2\eta)]U_{22}=0$.
Its solutions are precisely $s^{1/2}J_{\pm\nu}(s)$, providing an independent
check on the phase restoration in Eq.~\eqref{eq:u22-bessel}.}
\par For finite $t$ ($\Omega>0$), the companion equation gives
$U_{12}=(2i\dot U_{22}+\Delta U_{22})/\Omega$; unitarity fixes
$U_{11}=U_{22}^{*}$ and $U_{21}=-U_{12}^{*}$.  Hence all three Gauss
coordinates follow without an additional quadrature,
\begin{equation}\label{eq:lambda0closed}
 x(t)=\frac{U_{12}}{U_{22}},\quad
 y(t)=-2\ln U_{22},\quad
 z_{\rm G}(t)=\frac{U_{21}}{U_{22}},
\end{equation}
with the logarithm continued from $y(0)=0$.  These Gauss coordinates
are valid on intervals where $U_{22}\ne0$.  At a zero of $U_{22}$ the unitary
propagator remains regular and is continued in the complementary Gauss, or
Bruhat, chart.  For a detuned
$\mathfrak{su}(1,1)$ pump in Sec.~\ref{sec:opa} the same reduction changes the ordinary Bessel
functions to modified Bessel functions.

The useful structural step is to retain the normalization data together with
the projective ratio.  The ratio $x=U_{12}/U_{22}$ fixes the point on the
coherent-state orbit, while the independently normalized $U_{22}$ fixes the
stabilizer coordinate $y$; compact unitarity then fixes
  $z_{\rm G}$.  The
result is the full group element, including the phases of the chain amplitudes.

\subsection{Prepared seeds, tipping and spin locking}
\label{sec:tipping}

Suppose the ensemble is first tipped by a preparation pulse, so that the seed
is a spin-coherent state pointing along $\hat n_0$ rather than along $(-\hat z)$.
Conjugation (Sec.~\ref{sec:dictionary}) applies, and
Eq.~\eqref{eq:distancelaw}
gives the answer without further work. For the constant-envelope Hamiltonian
$H_{\rm rot}$ in Eq.~\eqref{eq:rotframe}, the Bloch vector
precesses about the Rabi vector $\bm\Omega=(B_0,0,\Delta)$ at rate
$|\bm\Omega|$.  For $|\bm\Omega|>0$, write
$\hat{\bm\Omega}=\bm\Omega/|\bm\Omega|$; the zero generator gives
$\C(t)=0$.  A vector making an angle $\vartheta$ with that axis sweeps a
cone, and after a time $t$ its geodesic separation $d$ from where it began
satisfies $\sin^2(d/2)=\sin^2\vartheta\,\sin^2(|\bm\Omega|t/2)$. Since
$\sin^2\vartheta=1-(\hat n_0\!\cdot\!\hat{\bm\Omega})^2$,
Eq.~\eqref{eq:distancelaw} gives
\begin{equation}\label{eq:tipped}
  \C(t)=2j\left[1-(\hat n_0\!\cdot\!\hat{\bm\Omega})^2\right]
  \sin^2\!\left(\frac{|\bm\Omega|t}{2}\right).
\end{equation}
The reachable complexity is multiplied by the square of the sine of the angle
between the preparation and the drive axis. In the limiting case
$\hat n_0=\pm\hat{\bm\Omega}$ the seed is an eigenstate of the generator, the
chain terminates at $b_1=0$, and $\C\equiv0$ for all time. The ensemble is
\emph{spin locked} \cite{Redfield1955}. On resonance this is the familiar
statement that a spin aligned with the drive does not evolve, and in Krylov
language it is a one-element chain.

Adiabatic rapid passage is a complementary control protocol.  The
detuning is swept linearly through zero at a fixed rate $\dot\Delta$, so the
 generator changes direction in the algebra,
as described by the Gauss-coordinate evolution in Sec.~\ref{sec:gauss},
Eq.~\eqref{eq:weinorman}. The single-atom probability is then the
Landau--Zener transition probability \cite{Zener1932}, whose consequences for
spread complexity across a transition were developed by Grabarits and
del~Campo \cite{GrabaritsDelCampo2025}.
For the ideal asymptotic sweep
$t_i\to-\infty$, $t_f\to+\infty$, and approximately for finite endpoints far
from resonance, the final complexity is
\begin{equation}\label{eq:arp}
  \C_\infty=2j\left[1-e^{-\pi B_0^2/2|\dot\Delta|}\right].
\end{equation}

The resonant population is periodic in accumulated area, returning
the ensemble to its ground state at $\A=\pi$.  A small area error obeys
$\delta\C=2j\sin(2\A)\delta\A+O(\delta\A^2)$, so the leading preparation
error is quadratic at return and inversion points.  Rapid passage instead
depends monotonically on $B_0^2/|\dot\Delta|$.  A sufficiently slow sweep
brings $\C_\infty$ close to $2j$ and leaves the population at the far end of
the chain after the pulse.  Its control parameter is the sweep adiabaticity
rather than a target final area.

A fixed nonlowest preparation can also retain complete arbitrary-control
dynamics. Take the middle weight $\ket{j=1,m=0}$ under
$H=\lambda_0(t)S_z+\rho(t)(e^{-i\varphi(t)}S_++e^{i\varphi(t)}S_-)$,
with $\rho>0$ on the phase patch. In the lossless two-mode realization
$S_+=a^\dagger b$, $S_z=(n_a-n_b)/2$, this is the two-photon input
$\ket{1,1}_{ab}$ \cite{CamposSalehTeich1989}. Define
\begin{align}\label{eq:two-photon-rungs}
 \ket{B_\varphi}&=\frac{e^{-i\varphi}\ket{2,0}_{ab}
                         +e^{i\varphi}\ket{0,2}_{ab}}{\sqrt2},\nonumber\\
 \ket{D_\varphi}&=\frac{e^{-i\varphi}\ket{2,0}_{ab}
                         -e^{i\varphi}\ket{0,2}_{ab}}{\sqrt2}.
\end{align}
On a subinterval with $\delta=\lambda_0-\dot\varphi\ne0$, the full chain is
$\ket{K_0}=\ket{1,1}_{ab}$, $\ket{K_1}=\ket{B_\varphi}$,
$\ket{K_2}=\operatorname{sgn}(\delta)\ket{D_\varphi}$, with
\begin{equation}\label{eq:two-photon-chain}
 a_0=a_1=a_2=0,\quad b_1=2\rho,\quad b_2=|\delta|,\quad b_3=0.
\end{equation}
Indeed, $H\ket{B_\varphi}=2\rho\ket{1,1}_{ab}+\lambda_0\ket{D_\varphi}$
and $-i\partial_t\ket{B_\varphi}=-\dot\varphi\ket{D_\varphi}$.
The second hopping therefore resolves interference between the level
splitting and the drive's phase motion, the same mechanism developed for
the $V$ system in Sec.~\ref{sec:vtime}.

The amplitudes follow from the fundamental propagator
$U_{1/2}=\left(\begin{smallmatrix}u&v\\-v^*&u^*\end{smallmatrix}\right)$
in the single-photon order $(\ket{1,0}_{ab},\ket{0,1}_{ab})$.
It solves $i\dot U_{1/2}=H_{1/2}U_{1/2}$, $U_{1/2}(0)=I$, with the
same controls in the spin-half representation. Setting
$\zeta_2=e^{i\varphi}uv$, its symmetric square gives
\begin{equation}\label{eq:two-photon-amplitudes}
 \begin{aligned}
 \phi_0&=|u|^2-|v|^2,\qquad \phi_1=2i\operatorname{Im}\zeta_2,\\
 \phi_2&=2\operatorname{sgn}(\delta)\operatorname{Re}\zeta_2.
 \end{aligned}
\end{equation}
This determines the full spread for any smooth control history once the
two-dimensional evolution is solved. At zeros of $\delta$, the signed frame
$(\ket{1,1}_{ab},\ket{B_\varphi},\ket{D_\varphi})$ continues the
probabilities. If $\delta$ vanishes from preparation onward, the dark rung
is unoccupied and the chain has two sites; a later closure retains any
previous dark population, as discussed in Sec.~\ref{sec:vtime}.

The optical readout uses two coincidence settings on identically prepared
states. With $S_y=(a^\dagger b-b^\dagger a)/(2i)$, the analyzer
$\mathcal U_{\rm an}=e^{-i\pi S_y/2}e^{i\varphi S_z}$ applies a relative
phase shift followed by a balanced beam splitter. It obeys
$\mathcal U_{\rm an}\ket{D_\varphi}=\ket{1,1}_{ab}$, so
\begin{equation}\label{eq:two-photon-readout}
 \C=1-P_{11}+P_{D_\varphi}
     =1-P_{11}^{\rm bare}+P_{11}^{\rm an}.
\end{equation}
Here the superscripts distinguish detection without and with the analyzer.
Both mean mode occupations remain one throughout this evolution; the
changing complexity is instead resolved by two-photon interference.

\section{Photons in a classical field}\label{sec:photons}
Bosonic optical modes realize the flat and negatively curved members of the
single-ladder coherent-state family.  A classically
forced cavity is governed by the
oscillator algebra and has unbounded
quadratic area growth from the
vacuum, while allowing experimentally prepared nonextremal seeds.  A
signal--idler amplifier is governed by $\mathfrak{su}(1,1)$ and converts the
same distance dependence into elliptic, parabolic, or
hyperbolic pair production.

\subsection{A forced cavity mode}\label{sec:forced}

A single mode driven by a prescribed classical current is the
oscillator member of the family,
\begin{equation}\label{eq:forcedH}
  H(t)=\omega_0a^\dagger a+f(t)a+f^*(t)a^\dagger .
\end{equation}
The resonant interaction lies in the Heisenberg--Weyl displacement
algebra.  A nonzero detuning adds the number rotation $a^\dagger a$, giving
the semidirect oscillator algebra used in the off-resonant solution.
For a resonant drive with fixed rotating-frame phase, the algebra
direction is fixed; detuning or explicit phase modulation rotates it.  The
same displacement Hamiltonian arises from passive
beam-splitter mode conversion, $g\,a^\dagger b+\mathrm{h.c.}$, when one mode
is replaced by an undepleted coherent reference
\cite{CamposSalehTeich1989}.  The conserved total occupation identifies this
as the number-conserving optical realization.  Active signal--idler pair
creation gives the $\mathfrak{su}(1,1)$ realization in
Sec.~\ref{sec:opa}.

For comparison with the atomic Rabi-envelope convention, write a resonant
drive of fixed phase as $f(t)=\Omega(t)e^{i\omega_0t}/2$, with real
$\Omega(t)\ge0$.  Its rotating-frame forcing envelope is then $\Omega(t)/2$,
so $\A(t)=\tfrac12\int_0^t\Omega(t')\,\dd t'$ is the same pulse-area variable
as in Eq.~\eqref{eq:area}.

Starting from the vacuum, the mode remains in the coherent state
$\ket{\zeta(t)}$.  Its amplitude obeys the classical oscillator equation
$i\dot\zeta=\omega_0\zeta+f^*$.  For the exponential drive
$f=f_0e^{-\eta t+i\omega t}$ with $f_0\geq0$ after absorbing its constant
phase, and with real $\eta$, the solution is
\begin{equation}\label{eq:zeta}
  \zeta(t)=-if_0e^{-i\omega_0t}\,
  \frac{1-e^{-[\eta+i(\omega-\omega_0)]t}}{\eta+i(\omega-\omega_0)} ,
\end{equation}
and the complexity is the mean photon number $\C=|\zeta|^2$,
\begin{equation}\label{eq:forcedC}
  \C(t)=\frac{f_0^2}{\eta^2+\Delta^2}
  \left[1+e^{-2\eta t}-2e^{-\eta t}\cos\Delta t\right],
\end{equation}
with $\Delta=\omega_0-\omega$, the same convention as
Eq.~\eqref{eq:rotframe}.
Positive $\eta$ describes decay, negative $\eta$ describes a ramp, and
$\eta=0$ is obtained by the continuous limit.  At the simultaneous corner
$\eta=\Delta=0$, Eqs.~\eqref{eq:zeta} and \eqref{eq:forcedC} mean their joint
limit, $\zeta=-if_0t\,e^{-i\omega_0t}$ and $\C=f_0^2t^2$.
The constant-drive generator in the carrier frame is
$\Delta a^\dagger a+f_0(a+a^\dagger)$.  For $\Delta\ne0$ its coherent
eigenstate has amplitude $\beta=-f_0/\Delta$.  Preparing that displaced vacuum
sets $b_1=0$ and realizes the Heisenberg--Weyl analogue of spin locking.
Independent time dependence in the forcing and detuning changes the algebra
direction and removes the common eigenstate.

\begin{figure}[tb]
\includegraphics[width=\columnwidth]{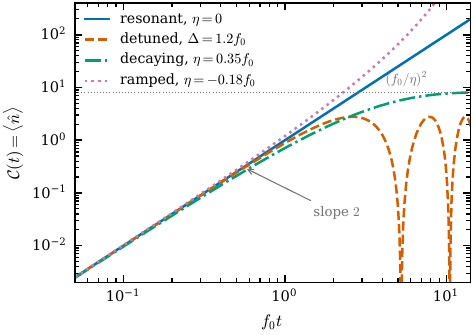}
\caption{\label{fig:forced} Complexity of a forced cavity mode, Eq.~\eqref{eq:forcedC}, which is the mean
photon number. Both axes are logarithmic.  The resonant constant-envelope result
$f_0^2t^2$ is therefore a straight line of slope 2, and each change of slope
directly displays a different growth regime. Detuning makes the mode fill and empty periodically, a
decaying drive freezes the photon number at the finite-area value
$(f_0/\eta)^2$, and a ramped drive bends upward without bound.}
\end{figure}

Figure~\ref{fig:forced} collects four limiting regimes.  A detuned constant
envelope gives
$\C=(4f_0^2/\Delta^2)\sin^2(\Delta t/2)$, with periodic filling and emptying.
On resonance the same envelope gives $\C=f_0^2t^2$, the
flat-orbit relation.
A decaying resonant envelope gives
$\C=(f_0/\eta)^2(1-e^{-\eta t})^2$ and approaches the finite-area value
$(f_0/\eta)^2$.  Reversing the sign of the exponential rate produces a ramp
whose complexity grows at twice that rate.  Detuning changes its prefactor
while preserving the growth exponent.

\subsection{Photon-added seeds}\label{sec:photonadded}

The forced mode makes the initial-state dependence experimentally concrete.
Take the single-photon state $\ket1$, the first photon-added preparation
produced by conditional down-conversion
\cite{AgarwalTara1991,Zavatta2004}.\footnote{Coherent and squeezed states
follow the conventions of Ref.~\cite{WallsMilburn2008}.}
For the resonant fixed
direction $X=a+a^\dagger$, $X\ket0=\ket1$.  Thus changing from the vacuum to
$\ket1$ is a degree-one polynomial jump and the
Christoffel construction applies
\cite{ChowdhuryMahapatra2026}.

For $f_0>0$, let $E$ be the spectral variable of
$H=f_0X$, $X=a+a^\dagger$, and set $\xi=E/(\sqrt2 f_0)$.  The vacuum measure
is proportional to $e^{-\xi^2}\dd\xi$, and
\[
 \ket m=\frac{H_m(X/\sqrt2)}{\sqrt{2^m m!}}\ket0
\]
defines $H_m$ as the physicists' Hermite polynomial.
The single-photon jump therefore multiplies the Gaussian measure by
$\xi^2$. The
orthogonal polynomials of $\xi^2e^{-\xi^2}$ are the generalized Hermite
polynomials, and the chain they define is the parabosonic one,
\begin{equation}\label{eq:paraboseb}
  b_n^2=f_0^2\bigl(n+2[n\ \text{odd}]\bigr),
\end{equation}
where $[n\ {\rm odd}]$ is $1$ for odd $n$ and $0$ for even $n$,
with the even-$n$ coefficients unchanged from the vacuum chain and the odd
ones shifted by two.
Using the pulse area $\A$ defined in Eq.~\eqref{eq:area}, summing the
amplitudes gives
\begin{equation}\label{eq:photonaddedC}
  \C(\A)=\A^2+1-e^{-2\A^2}.
\end{equation}
Thus dependence on the area alone survives the change of seed.

Equations~\eqref{eq:paraboseb} and \eqref{eq:photonaddedC} were obtained for a
fixed generator in Ref.~\cite{ChowdhuryMahapatra2026}.  A resonant envelope
keeps that generator's direction fixed and replaces time by the accumulated
area.  It is therefore the $W(t)=\mathbf 1$ and
$\Delta\tau=\A$ instance of the time-dependent
transfer relation in Eq.~\eqref{eq:dressed-amplitudes}, and the closed complexity holds
for every resonant envelope with a well-defined accumulated area.

A constant detuning is also exactly reducible.  For a nonzero real constant
force $f_0$ and $\Delta\ne0$, write $\hat n=a^\dagger a$.  The carrier-frame
Hamiltonian can then be written
\begin{equation}\label{eq:detuned-fixed-generator}
 \begin{aligned}
 H_{\rm rot}&=\Delta\hat n+f_0(a+a^\dagger)
 =\Delta\bigl(X_{\alpha_0}-\alpha_0^2\bigr),\\
 X_{\alpha_0}&=D(-\alpha_0)\hat nD(\alpha_0),\qquad
 \alpha_0=\frac{f_0}{\Delta}.
 \end{aligned}
\end{equation}
The vacuum-seeded $X_{\alpha_0}$ chain is the displaced-number, or Charlier,
chain, and every finite Fock state $\ket m$ is its $m$th old rung.  Its
seed-adapted chain is therefore fixed exactly by
Eq.~\eqref{eq:old-rung-seed}, although a one-line expression for the full
complexity need not result.  At $f_0=0$ the vacuum chain terminates, and every
$\ket m$ instead generates its own one-site eigenchain.  When
$f(t)/\Delta$ changes, the static displacement in
Eq.~\eqref{eq:detuned-fixed-generator} no longer reduces a fixed laboratory
Fock-seed chain to one time-independent spectral measure.  The vacuum
evolution nevertheless remains the exact coherent-state trajectory of
Sec.~\ref{sec:forced}; a nonextremal fixed-reference chain requires the
time-dependent extended recursion.
Appendix~\ref{app:dunkl} derives Eq.~\eqref{eq:photonaddedC} from the
parabosonic ladder, independently of the connector calculation in
Ref.~\cite{ChowdhuryMahapatra2026}.

\begin{figure}[tb]
\includegraphics[width=\columnwidth]{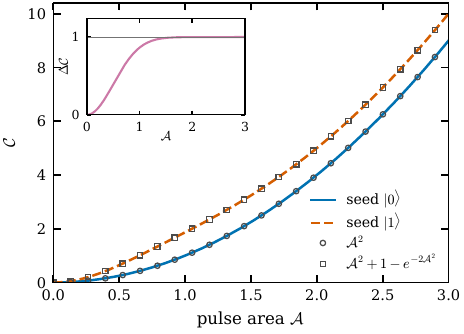}
\caption{\label{fig:photonadded} Complexity of a resonantly forced mode seeded with the vacuum and with the
single-photon-added vacuum, against pulse area. Lines show direct chain computations, and markers
show the closed-form values from $\C=\A^2$ and Eq.~\eqref{eq:photonaddedC}. The inset shows their
difference, which rises on the vacuum scale $\A\sim1$ and approaches one.}
\end{figure}

The optical distinction is most transparent in the difference between the
single-photon and vacuum curves (Fig.~\ref{fig:photonadded}),
\begin{equation}\label{eq:photonadded-offset}
 \C_{\ket1}(\A)-\C_{\ket0}(\A)=1-e^{-2\A^2}.
\end{equation}
It vanishes at the initial time, as every seed-adapted spread complexity must,
and approaches one when $|\A|\gg1$.  The vacuum-scale Gaussian therefore
describes how the transformed chain dynamically resolves the additional node
of the single-photon wavefunction.  The unit offset is an asymptotic property
of the evolution rather than complexity assigned to the seed at $\A=0$.

For a general Fock seed $\ket m$, the spectral measure is multiplied by
$H_m(\xi)^2$ up to normalization and the first hopping remains
in closed form,
\begin{equation}\label{eq:general-fock-b1}
 b_1^2=f_0^2(2m+1),
\end{equation}
the variance of $f_0(a+a^\dagger)$ in $\ket m$.  The case $m=1$ is
especially simple because
$H_1(\xi)^2$ is a pure quadratic power and produces the two-periodic
generalized-Hermite chain summed in Eq.~\eqref{eq:photonaddedC}.  For
$m\geq2$, $H_m$ has $m$ simple real zeros and $H_m^2$ has $m$ distinct double
zeros.  The exact chain is represented by the confluent Christoffel determinant
or the root-free recurrence of Ref.~\cite{ChowdhuryMahapatra2026}.  Thus all finite Fock seeds
belong to the same polynomial-jump hierarchy, while the one-line complexity
in Eq.~\eqref{eq:photonaddedC} reflects
the exceptional two-periodicity of $m=1$.

\subsection{Definite-parity seeds and the chain as an interferometer}
\label{sec:parity}

The photon-added state is the first member of a family with a general
description, and that description connects the Krylov chain to the observable
by which nonclassical states of light are characterized.
Phase-space formulations of state and operator Krylov spreading were
developed by Pal, Pal, and Kim \cite{PalPalKim2026}, who express Krylov-basis
quantities through Weyl--Wigner representations.  In the relation below, the
chain instead samples the seed's Wigner function directly at a point selected
by the drive.

Return to the resonant forced mode, whose generator $H=f(t)(a+a^\dagger)$ keeps
a fixed direction in $\mathfrak{h}(1)$, so that its propagator is the
displacement $U(t)=D(-i\A)$ with $\A$ the accumulated area
defined in Eq.~\eqref{eq:area}. Suppose only that
the seed has definite photon-number parity, $R\ket{K_0}=\pi_0\ket{K_0}$ with
$R=(-1)^{\hat n}$ and $\pi_0=\pm1$. Since $RaR=-a$, the generator is
parity-odd, so each application of $H$ in Eq.~\eqref{eq:lanczos} flips parity and
the $n$th chain vector carries parity $\pi_0(-1)^n$. Chain-site parity
\emph{is} photon-number parity, up to the fixed sign of the seed. The odd-site
occupation of the chain is therefore
$P_{\rm odd}=\tfrac12(1-\pi_0\braket{R})$, and $\braket{R}$ in a displaced state
is, by Royer's identity
$\mathcal W_\psi(\beta)=\tfrac2\pi
\braket{\psi|D(\beta)RD^\dagger(\beta)|\psi}$
\cite{Royer1977}, the seed's own Wigner function evaluated at the displacement,
\begin{equation}\label{eq:wignerscan}
  P_{\rm odd}(\A)=\frac12
  \left[1-\pi_0\,\frac{\pi}{2}\,
  \mathcal W_{K_0}\!\left(i\A\right)\right].
\end{equation}
As the pulse area accumulates, the Krylov chain of a driven mode traces a
straight cut through the Wigner function of its own seed, along the direction in
which the drive displaces it. Equation~\eqref{eq:wignerscan} is exact for every
envelope, with no restriction to small $\A$, and the readout it calls for is
the one already used for cavity-field tomography, photon-number parity after a
controlled displacement \cite{LutterbachDavidovich1997,Deleglise2008}.

The content is easiest to see on the two seeds already in play. For the
vacuum, $\mathcal W_{K_0}$ is a Gaussian and
Eq.~\eqref{eq:wignerscan} returns
$P_{\rm odd}=\tfrac12(1-e^{-2\A^2})$, the odd-photon weight of a coherent
state. For $\ket1$ it returns
$P_{\rm odd}=\tfrac12[1-(1-4\A^2)e^{-2\A^2}]$, the closed form of
Appendix~\ref{app:dunkl}. The two behave quite differently, and what separates
them is the Wigner negativity. The vacuum scan rises monotonically to
$\tfrac12$, while the scan of $\ket1$ starts below $1/2$ while it samples the negative central lobe, crosses
the Wigner zero at $\A=1/2$, and overshoots in the surrounding positive
annulus to $\tfrac12\bigl(1+2e^{-3/2}\bigr)$ at
$\A=\sqrt3/2$, before relaxing to $\tfrac12$. A displaced Gaussian
branch, whose Wigner function is positive everywhere, produces no such
structure.

A superposition produces structure of a different kind. For the even optical cat
$\ket{C_+}\propto\ket\alpha+\ket{-\alpha}$ with $\alpha$ real, the two branches
interfere in $\mathcal W_{K_0}$, and
Eq.~\eqref{eq:wignerscan} returns those
fringes exactly,
\begin{equation}\label{eq:catfringe}
  P_{\rm odd}(\A)=\frac12-\frac{e^{-2\A^2}
  \left[1+e^{2\alpha^2}\cos(4\alpha\A)\right]}{2\left(1+e^{2\alpha^2}\right)} ,
\end{equation}
while the odd cat with $\alpha\ne0$ gives
$P_{\rm odd}=\tfrac12+e^{-2\A^2}[1-e^{2\alpha^2}\cos(4\alpha\A)]
/[2(e^{2\alpha^2}-1)]$, in which the fringe term keeps its sign and the constant offset flips, as
$P_{\rm odd}(0)=0$ requires. Its removable $\alpha\to0$ limit is the
single-photon result above. Throughout, $P_{\rm odd}$ is the occupation of the
odd-numbered \emph{chain} sites, whatever the parity of the seed; the seed
parity enters only through $\pi_0$ in Eq.~\eqref{eq:wignerscan}, which is why
the even and odd cats share a fringe term and differ in offset.

\begin{figure}[tb]
\includegraphics[width=\columnwidth]{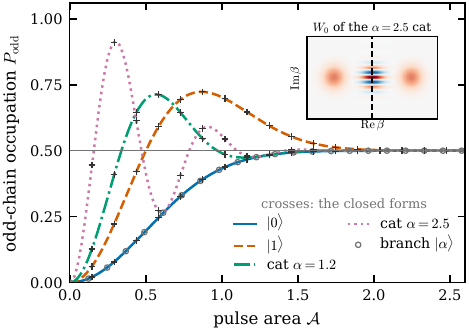}
\caption{\label{fig:parity} Parity interferometry from Eq.~\eqref{eq:wignerscan}, showing odd-site
occupation versus pulse area.  Lines show chain calculations for the
definite-parity seeds, crosses show their closed forms, and grey circles show
the coherent comparator at $\alpha=1.2$.  The photon-added curve crosses
$1/2$ at $\A=1/2$ and peaks at $\tfrac12(1+2e^{-3/2})$ for
$\A=\sqrt3/2$; the cats have the fringes of Eq.~\eqref{eq:catfringe}.  The
inset shows the $\alpha=2.5$ cat Wigner function and the sampled cut.}
\end{figure}

The fringe angular frequency in the scan variable $\A$ is exactly $4\alpha$
(Fig.~\ref{fig:parity}).  For the even cat,
$\bar n=\alpha^2\tanh(\alpha^2)$, so the well-separated-branch regime
$\alpha\gg1$ gives $4\alpha\simeq4\sqrt{\bar n}$; the exact finite-$\alpha$
relation retains the hyperbolic-tangent factor.  These increasingly fine
fringes
are the phase-space interference structure associated with sub-Planck
sensitivity in coherent superpositions,
which occurs in both Heisenberg--Weyl and $SU(2)$ families
\cite{Akhtar}.

The phase $4\alpha\A$ is the Weyl displacement-interference phase for the
quadrature convention used here.  If the two displaced paths are completed
into a closed, oriented phase-space circuit, the same Weyl composition law
identifies it with the signed symplectic area, equivalently twice the
Euclidean oriented area in the complex-$\alpha$ plane, consistently with
Eq.~\eqref{eq:actionangle}.  Equation~\eqref{eq:catfringe} instead measures the
phase along a one-dimensional open scan and therefore needs no auxiliary
closing path.  Its oscillatory envelope is $e^{-2\A^2}$, so the odd-chain
population approaches $1/2$ after one vacuum-scale displacement.

The envelope of Eq.~\eqref{eq:catfringe} is set by the branch width alone,
so the natural way to test that statement is to change the width and leave the
separation fixed, which is what squeezing the branches does. The Gaussian
envelope directly probes the quadrature width of each branch.
To fix the state and quadrature convention, let
$D(\alpha)=\exp(\alpha a^\dagger-\alpha a)$ and
$S(r)=\exp[\tfrac r2(a^2-a^{\dagger2})]$ for real $\alpha$ and $r$, and define
\begin{equation}\label{eq:squeezedcatstate}
 \ket{C_{+,r}}=
 \frac{[D(\alpha)+D(-\alpha)]S(r)\ket0}
 {\sqrt{2[1+e^{-2\alpha^2/w}]}},\quad w=e^{-2r}>0 .
\end{equation}
The branch centres remain at $\pm\sqrt2\,\alpha$ along
$x=(a+a^\dagger)/\sqrt2$, their $x$ variance is $w$ times the vacuum value,
and the parity scan displaces along the conjugate quadrature.  Direct
evaluation of the displaced parity gives
\begin{equation}\label{eq:squeezedcat}
 P_{\rm odd}(\A)=\frac12-e^{-2w\A^2}
 \frac{\cos(4\alpha\A)+e^{-2\alpha^2/w}}
 {2[1+e^{-2\alpha^2/w}]} .
\end{equation}
This result is exact for every real $\alpha$ and $r$ and reduces to
Eq.~\eqref{eq:catfringe} at $w=1$.  Holding the branch centres fixed preserves
the frequency $4\alpha$, while their width sets the envelope
$e^{-2w\A^2}$.  At $\A=1$, for example, squeezing along the branch-separation
axis changes the envelope from $e^{-2}$ to $e^{-2w}$.  This state-geometric
visibility relation is also relevant to squeezed-branch superpositions used in
bosonic cat codes
\cite{SchlegelMingantiSavona2022}.
At fixed branch separation, fitting the Gaussian envelope in
Eq.~\eqref{eq:squeezedcat} directly returns the relative quadrature width
$w=e^{-2r}$.

A single coherent branch at $+\alpha$ gives the monotone vacuum curve, because
its seed-adapted chain is the Fock ladder displaced to $\alpha$ and its site
populations are Poissonian in $\A^2$.\footnote{The coherent branch is included
in Fig.~\ref{fig:parity} as a comparator.  Since it has no definite photon
parity, its curve follows from displacement covariance of the chain rather
than Eq.~\eqref{eq:wignerscan}.}  The oscillatory contribution for the cat
therefore isolates coherence between the two branches.

The atomic analogue has a discrete two-point spectral measure.  For the GHZ
state $(\ket{j,j}+\ket{j,-j})/\sqrt2$ under the longitudinal drive
$H=f(t)S_z$, with $\Theta=\int f(t)\,\dd t$, the cyclic subspace has exactly
two sites for every $j>0$ and
\begin{equation}\label{eq:ghz}
 \C(\Theta)=\sin^2(j\Theta).
\end{equation}
The probability oscillates with angular frequency $2j=N$ and period
$\pi/j$, without an envelope because the spectral measure consists of two sharp lines.  For
integer $j$, equivalently even $N=2j$, the GHZ seed also has definite
excitation parity.  A transverse drive of accumulated area $\A$ then gives
the parity-reduced chain probability
\begin{equation}\label{eq:ghz-transverse-parity}
 P_{\rm odd}(\A)=\frac12\!\left[1-\cos^{2j}(2\A)
            -(-1)^j\sin^{2j}(2\A)\right].
\end{equation}
The transverse features narrow as $j^{-1/2}$, whereas the longitudinal Ramsey
period scales as $j^{-1}$.  The latter is the familiar GHZ phase enhancement
discussed in Ref.~\cite{Bollinger1996}; here the two protocols are displayed in
the same seed-adapted chain language.

The full cat-seed complexity is determined by the orthogonal polynomials of
its quadrature measure.  For real $\alpha$, the even cat has density
proportional to $e^{-\xi^2}\cosh^2(\sqrt2\alpha\xi)$ and becomes doubly peaked
only for $|\alpha|>1/\sqrt2$; the odd cat has density proportional to
$e^{-\xi^2}\sinh^2(\sqrt2\alpha\xi)$ and, for nonzero $\alpha$, is split by its
node at $\xi=0$.  Their recurrences are distinct from
the generalized-Hermite ladder responsible for
Eq.~\eqref{eq:photonaddedC}.  The parity projection remains closed because
Eq.~\eqref{eq:wignerscan} depends only on the exact parity grading of the chain;
it therefore extracts a directly measurable sector of the dynamics even when
the full first moment requires the transformed recurrence.

\subsection{A nonlowest Floquet chain from a periodically forced mode} \label{sec:floquet-forced}

Periodic forcing provides a setting in which time dependence and a
nonextremal seed can be treated together without replacing the stroboscopic
dynamics by a branch-dependent Floquet Hamiltonian.  Let
$f(t+T)=f(t)$ in Eq.~\eqref{eq:forcedH}, write $\hat n=a^\dagger a$, and define
\begin{equation}\label{eq:floquet-monodromy}
 \begin{aligned}
 \Theta_F&=\omega_0T,\\
 \delta_F&=-i e^{-i\Theta_F}\int_0^T
 f^*(s)e^{i\omega_0s}\,\dd s,\\
 F&=U(T,0)=e^{i\chi_F}D(\delta_F)e^{-i\Theta_F\hat n}.
 \end{aligned}
\end{equation}
Here $\chi_F$ is a real phase fixed by the oriented displacement path.  This
exact forced-oscillator monodromy is standard
\cite{FoxVelaArevalo2002}. The rotation angle $\Theta_F$ fixes the folding
studied below, which occurs for constant forcing as well as for modulated
cycles.  Direct Arnoldi evolution of the one-cycle unitary
avoids choosing a logarithmic branch of a Floquet Hamiltonian
\cite{YatesMitra2021,Nizami}.  Related unitary- and operator-Krylov
formulations use orthogonal polynomials on the unit circle
\cite{KolganovTrunin2025,YehMitra2026}.  For the forced oscillator, the
displaced-number spectrum makes the seed dependence analytic and leads to the
cyclic-dimension result below.

When $e^{-i\Theta_F}\ne1$, introduce
\begin{equation}\label{eq:floquet-fixed-generator}
 \begin{aligned}
 \alpha_F&=\frac{\delta_F}{e^{-i\Theta_F}-1},\qquad
 \lambda_F=|\alpha_F|^2,\\
 X_F&=D(-\alpha_F)\hat nD(\alpha_F).
 \end{aligned}
\end{equation}
The Weyl displacement composition identity gives
$F=e^{i(\chi_F+\lambda_F\sin\Theta_F)}e^{-i\Theta_FX_F}$.
Thus the one-period dynamics is projectively a function of one fixed
self-adjoint operator whose phase-space centre is $(-\alpha_F)$.  This is the
displaced-number normal form that also underlies the constant-detuning mode in
Eq.~\eqref{eq:detuned-fixed-generator}.  For
$\alpha_F\ne0$, its vacuum-seeded Lanczos chain is
\begin{equation}\label{eq:floquet-charlier-chain}
 \begin{aligned}
 a_n^{X_F}&=n+\lambda_F,\qquad
 b_{n+1}^{X_F}=\sqrt{\lambda_F(n+1)},\\
 \ket{K_n^{(X_F)}}&=e^{in\arg\alpha_F}\ket n.
 \end{aligned}
\end{equation}
Consequently a Fock preparation $\ket r$ is the $r^{\text{th}}$ old rung of this
Charlier chain, but its seed-adapted stroboscopic ordering must still be
rebuilt.

We therefore apply Arnoldi directly to $F$.  Denote the resulting basis by
$\{\ket{A_n^{(r)}}\}$ and set
\begin{equation}\label{eq:floquet-complexity-definition}
 \begin{aligned}
 \mathcal K_F^+(r)&=\operatorname{span}
 \{\ket r,F\ket r,F^2\ket r,\ldots\},\\
 \C_F^{(r)}(M)&=\sum_{n\ge0}n
 \left|\braket{A_n^{(r)}|F^M|r}\right|^2.
 \end{aligned}
\end{equation}
Here $M\in\mathbb N_0$ counts completed drive periods.
This forward Floquet chain is distinct from both the old
$X_F$-Lanczos chain and the instantaneous extended-generator chain of
Sec.~\ref{sec:framework}.

If $\Theta_F/(2\pi)$ is rational, write it as $p/q_F$ in lowest terms.  For
every $r\in\mathbb N_0$ and $\alpha_F\ne0$, the nontrivial rational case
$q_F\ge2$ and the irrational case give
\begin{equation}\label{eq:floquet-dimension}
 \dim\mathcal K_F^+(r)=
 \begin{cases}
  q_F, & \Theta_F/(2\pi)=p/q_F\in\mathbb Q,\quad q_F\ge2,\\
  \infty, & \Theta_F/(2\pi)\notin\mathbb Q.
 \end{cases}
\end{equation}
The result follows directly from the nonlowest-seed spectral weights.  For
$n\ge r$ they are
\begin{equation}\label{eq:floquet-displaced-number-weight}
 w_n^{(r)}=e^{-\lambda_F}\frac{r!}{n!}\lambda_F^{n-r}
 \left[L_r^{(n-r)}(\lambda_F)\right]^2,
\end{equation}
the displaced-number distribution
\cite{deOliveiraEtAl1990}, where $L_r^{(n-r)}$ is an associated Laguerre
polynomial.  At fixed $r$,
$L_r^{(n-r)}(\lambda_F)$ is a degree-$r$ polynomial in $n$ and therefore
removes only finitely many sites from the large-$n$ tail.  In the rational case,
every residue class modulo $q_F$ retains nonzero weight.  It consequently occupies
all $q_F$ distinct eigenphases, while an irrational angle leaves infinitely
many distinct occupied phases; the cyclic-dimension statement then follows
from the corresponding Vandermonde rank.

The limiting cases complete the classification.  If $\alpha_F=0$ and
$\Theta_F\notin2\pi\mathbb Z$, $\ket r$ is already an eigenstate of the
number rotation and its Floquet chain has one site, for either a rational or
an irrational angle; for $r>0$ it is not an old rung of the terminated vacuum
chain.  When $\Theta_F\in2\pi\mathbb Z$, Eq.~\eqref{eq:floquet-fixed-generator}
is singular and Eq.~\eqref{eq:floquet-monodromy} must be used directly.
Then $\delta_F=0$ gives a one-site chain, whereas $\delta_F\ne0$ gives an
infinite pure-displacement chain for every finite Fock seed.

The half turn, $\Theta_F=\pi$ modulo $2\pi$, is the minimal nontrivial fold.
After removing the scalar phase, its monodromy is the involution
\begin{equation}\label{eq:floquet-half-turn}
 \begin{aligned}
 F_\pi&=D(\delta_F)R,\qquad R=(-1)^{\hat n},\\
 F_\pi^2&=\mathbf1,\qquad \alpha_F=-\frac{\delta_F}{2}.
 \end{aligned}
\end{equation}
For the seed $\ket r$, the return amplitude is
$s_r=(-1)^r e^{-x_F/2}L_r(x_F)$, where $x_F=|\delta_F|^2$.  The return
probability is
$P_{\rm ret}^{(r)}:=|\braket{r|F_\pi|r}|^2=|s_r|^2$.  Hence the exact
 two-site stroboscopic complexity is
\begin{equation}\label{eq:floquet-half-turn-complexity}
 \C_F^{(r)}(M)=\frac{1-(-1)^M}{2}
 \left[1-e^{-x_F}L_r(x_F)^2\right].
\end{equation}
For $r\ge1$, a zero of $L_r(x_F)$ places unit probability on the second site
of the accessible seed-adapted Arnoldi chain at every odd period.  The vacuum
has no such finite-displacement zero.  At $|s_r|=1$ the cyclic space is instead
one-dimensional and a second Arnoldi vector is not defined.  Displaced-Fock
overlaps of precisely this Laguerre form have been measured for motional Fock
states of a trapped ion \cite{WolfEtAl2019}.

\begin{figure}[tb]
\includegraphics[width=\columnwidth]{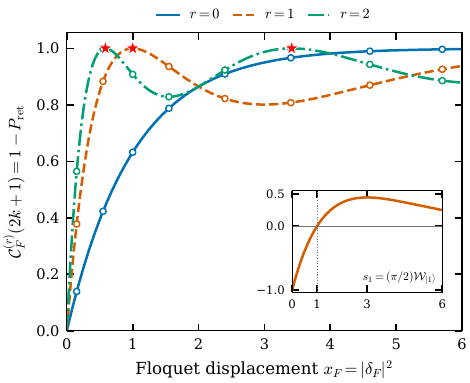}
\caption{\label{fig:floquet-halfturn} Half-turn Floquet readout for
number-state seeds.  Curves give
$\C_F^{(r)}(2k+1)=1-e^{-x_F}L_r(x_F)^2=1-P_{\rm ret}^{(r)}$; open circles are
independent truncated-Fock evolution.  Stars mark unit occupation of the
second Arnoldi site.  The inset gives the signed return amplitude, whose
magnitude, but not sign, follows from the main-panel survival probability.}
\end{figure}

The same involution gives a phase-space interpretation for any normalized
pure seed $\ket\psi$, with $\C_F^\psi$ defined in that seed's own forward
Arnoldi basis.  For $k\in\mathbb N_0$, the displaced-parity identity
$F_\pi=D(\delta_F/2)R D^\dagger(\delta_F/2)$ and Royer's convention
\cite{Royer1977}, also used in Eq.~\eqref{eq:wignerscan}, give
\begin{equation}\label{eq:floquet-half-turn-wigner}
 \C_F^\psi(2k)=0,\qquad
 \C_F^\psi(2k+1)=1-\frac{\pi^2}{4}
 \mathcal W_\psi\!\left(\frac{\delta_F}{2}\right)^2.
\end{equation}
The half-turn complexity therefore detects Wigner zeros and the magnitude
$|\mathcal W_\psi|$ at the fixed centre
$\delta_F/2=-\alpha_F$.  The phase-sensitive return amplitude
$\braket{\psi|F_\pi|\psi}=(\pi/2)\mathcal W_\psi(\delta_F/2)$, rather than the
squared complexity, retains the sign.
This squared Floquet diagnostic complements the continuous-time odd-chain
population in Eq.~\eqref{eq:wignerscan}, which is affine-linear in the Wigner
function and therefore preserves its sign. Figure~\ref{fig:floquet-halfturn}
shows that Fock-state survival gives the complexity without Arnoldi
reconstruction, while an interferometric return amplitude, after calibration
or removal of the known scalar Floquet phase, retains the Wigner sign.

\subsection{Parametric amplification of signal and idler}\label{sec:opa}

Three-wave mixing in an optically nonlinear $\chi^{(2)}$ medium
couples a pump to a signal--idler
pair, $c\,a^\dagger b^\dagger+\mathrm{h.c.}$, where $c$ annihilates a
pump photon and $a,b$ annihilate signal and idler photons, respectively
\cite{mollow1967quantum,YurkeMcCallKlauder1986}.
Replacing the pump by a prescribed, weakly depleted amplitude, whose
envelope $g(t)$ may vary in time, leaves
\begin{equation}\label{eq:opaH}
  H(t)=\omega_aa^\dagger a+\omega_bb^\dagger b
  +\frac{g(t)}{2}\left(e^{-i\omega t}a^\dagger b^\dagger
  +e^{i\omega t}ab\right).
\end{equation}
We choose the pump phase so that $g(t)$ is real; the constant-pump
formulas below use $g\geq0$.

The energy-matching condition is $\omega=\omega_a+\omega_b$, and away from it
the mismatch is
$\Delta=\omega_a+\omega_b-\omega$;
the complexity in
Eq.~\eqref{eq:opaC} is even in $\Delta$, although propagator phases retain its
sign. The conserved
quantity is the number \emph{difference} $\hat n_a-\hat n_b$, not the total,
and in each of its sectors the operators
$K_+=a^\dagger b^\dagger$, $K_-=ab$,
$K_0=\tfrac12(a^\dagger a+b^\dagger b+1)$ close on $\mathfrak{su}(1,1)$ with
$[K_+,K_-]=-2K_0$. The state generated from the pair vacuum is a two-mode squeezed vacuum, whose
Schmidt weights coincide with those of a thermofield double after local phase
rotations and the identification
$\tanh r=e^{-\beta_{\rm th}\omega_a/2}$, where $\beta_{\rm th}$ is the
inverse temperature
\cite{Takahashi1996}. Seeded with the
signal--idler vacuum ($h=1/2$), the chain
is the pair ladder $\ket{n,n}$ and $\C=\langle\hat n_a\rangle$, the number of
generated signal photons, which is the parametric gain minus one. More
generally, a sector with fixed number difference $q_{\rm d}$ has
$h=(|q_{\rm d}|+1)/2$.
Equation~\eqref{eq:opaC} is written for the pair-vacuum sector
$h=1/2$.  For the lowest-weight seed in a general fixed-difference sector,
its right-hand side is multiplied by $2h$.
For a detuned exponential pump, the compact--noncompact continuation in
Sec.~\ref{sec:bessel} gives the corresponding modified-Bessel reduction.

\begin{figure}[tb]
\includegraphics[width=\columnwidth]{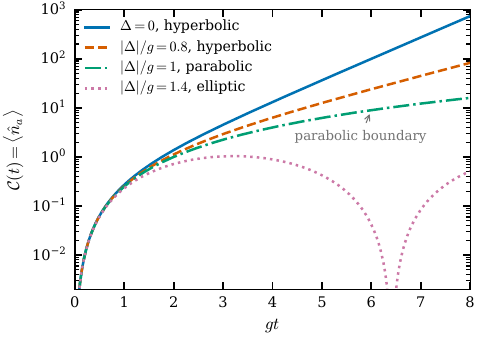}
\caption{\label{fig:opa} Pair-vacuum complexity from Eq.~\eqref{eq:opaC}, equal to the generated
signal count, for $|\Delta|/g=0,0.8,1,$ and $1.4$.  The hyperbolic branch is
exponential, the parabolic boundary quadratic, and the elliptic branch
oscillatory.  These are conjugacy classes of the lossless undepleted-pump
Hamiltonian.}
\end{figure}

For a constant pump the rotating-frame generator is again constant, and its
$SU(1,1)$ conjugacy class fixes the behavior. In the notation of
Appendix~\ref{app:gauss} the generator has $B=\Delta$ and $|AC|=g^2/4$, and the
$\mathfrak{su}(1,1)$ sign change noted there gives $r^2=(\Delta^2-g^2)/4$.
The sign of $r^2$ decides whether the circular functions of that appendix stay
circular or turn hyperbolic, and the three possibilities give
\begin{equation}\label{eq:opaC}
  \C(t)=
  \begin{cases}
   \dfrac{g^2}{\Delta^2-g^2}\sin^2\!\left(\tfrac{t}{2}\sqrt{\Delta^2-g^2}\right),
   & |\Delta|>g,\\[10pt]
   \dfrac{g^2t^2}{4}, & |\Delta|=g,\\[8pt]
   \dfrac{g^2}{g^2-\Delta^2}\sinh^2\!\left(\tfrac{t}{2}\sqrt{g^2-\Delta^2}\right),
   & |\Delta|<g .
  \end{cases}
\end{equation}
Equation~\eqref{eq:opaC} resolves the elliptic, parabolic, and hyperbolic
conjugacy classes of the lossless undepleted-pump Hamiltonian
(Fig.~\ref{fig:opa}).  The pair number oscillates for $|\Delta|>g$, grows as
$g^2t^2/4$ at $|\Delta|=g$, and grows exponentially for $|\Delta|<g$.  The
parabolic expression is the regular fixed-time limit of both neighbouring
branches.  On the elliptic side, the maximum taken over a complete oscillation
diverges as the boundary is approached because the oscillation period diverges
at the same point.  This closed-system result gives the coherent skeleton
of parametric gain.  Damping, cavity escape, and pump replenishment set its experimental
threshold and bandwidth.\footnote{
The hyperbolic branch uses the undepleted-pump approximation.
On resonance, $\C\sim e^{gt}/4$, so pump back-action becomes order unity near
$gt\sim\ln N_{\rm pump}$, after which the full three-wave dynamics is required.}

\subsection{Seeding the amplifier with squeezed light}\label{sec:squeezedseed}

An $SU(1,1)$ coherent input is obtained by squeezing the lowest weight in
a fixed number-difference sector.  The pair-vacuum case has $h=1/2$;
conjugation maps the chain in any sector to one based at that sector's lowest
weight.  The general $SL(2,\mathbb R)$
coherent-seed complexity is derived in
Ref.~\cite{BalasubramanianCaputaSimon2026}.  A complementary
information-geometric analysis in which an operator-Krylov wavefunction is
represented by a two-mode squeezed state is given in
Ref.~\cite{ZhaiLiuZhang2026}.  For a resonant two-mode amplifier seeded by
$z=e^{i\phi}\tanh r_0$, the chain gives the phase-sensitive prefactors
below and a
seed-independent late-time exponent.  The hyperbolic picture makes those two
features geometrically visible.\footnote{Squeezing the signal mode alone prepares a superposition of conserved
number-difference sectors. The same pair Hamiltonian then acts in several
$SU(1,1)$ representations, so the chain must be reconstructed for that
preparation rather than obtained from one fixed lowest-weight orbit.}
\par A resonant
pump acts as a hyperbolic translation, carrying the vacuum a distance $L=gt$
along a geodesic through it. A seeded amplifier starts off that
geodesic, and the displacement produced by the translation depends on that
offset. For a point at perpendicular distance $\rho_\perp$ from the translation axis the
displacement $d$ obeys
\begin{equation}\label{eq:offgeodesic}
  \sinh\frac d2=\cosh\rho_\perp\,\sinh\frac L2 .
\end{equation}
This is the displacement
formula for a hyperbolic translation.  The displacement grows with the offset
and reduces to $d=L$ on the axis itself.
\par Among fixed $SU(1,1)$ coherent seeds, the preserved invariant fixes the
smallest seed-relative spread of a constant detuned amplifier.  In the
hyperbolic regime let $\Gamma=\sqrt{g^2-\Delta^2}$.
Equation~\eqref{eq:displaced-extremal-controls}
gives
\begin{align}\label{eq:hyperbolic-seed-bound}
 \C_{\rm s}(t)&=
 \frac{8h|q_{\rm s}'|^2}{\Gamma^2}
 \sinh^2\!\left(\frac{\Gamma t}{2}\right),\nonumber\\
 \C_{\rm s}(t)&\ge 2h\sinh^2\!\left(\frac{\Gamma t}{2}\right),
 \qquad |\Delta|<g .
\end{align}
Indeed,
$B_{\rm s}'{}^2-4|q_{\rm s}'|^2=\Delta^2-g^2=-\Gamma^2$.
For $t\ne0$, equality holds precisely for a seed on the translation axis, for which
$B_{\rm s}'=0$; one real representative is
$z_{\rm s}=-\Delta/(g+\Gamma)$.  The hyperbolic conjugacy class therefore
sets both the exponent and a coherent-seed-independent floor, while the input position
can only increase the prefactor.  In the elliptic class a coherent fixed-point
seed instead gives $q_{\rm s}'=0$ and zero spread.  At the nonzero parabolic
boundary the infimum is approached only as $|z_{\rm s}|\to1$, so no finite
squeezing attains it.

For the real-pump
convention of Eq.~\eqref{eq:opaH}, the vacuum moves along the imaginary
diameter of the disk.  Thus $z=e^{i\phi}\tanh r_0$ is axial for
$\phi=\pi/2\pmod\pi$ and lies on the perpendicular diameter for
$\phi=0\pmod\pi$; a pump-phase rotation shifts both conditions together.
A seed on the pump-translation axis has $\rho_\perp=0$, which is why it returns the vacuum answer.
Such a squeezed lowest-weight seed of
parameter $r_0$ on the perpendicular diameter has
$|z|=\tanh r_0$ and hence $\rho_\perp=2r_0$ by
Eq.~\eqref{eq:stereo}, and
Eq.~\eqref{eq:distancelaw} then gives for
a seed squeezed by
$r_0$,
\begin{equation}\label{eq:squeezedseed}
  \C(t)=2h\sinh^2\!\left(\frac{gt}{2}\right)\cosh^2(2r_0)
\end{equation}
when the seed lies on the perpendicular disk diameter, and
\begin{equation}\label{eq:squeezedseed-aligned}
  \C(t)=2h\sinh^2\!\left(\frac{gt}{2}\right)
\end{equation}
when the seed lies on the translation axis,
which is exactly the vacuum result for any $r_0$.

\begin{figure}[tb]
\includegraphics[width=\columnwidth]{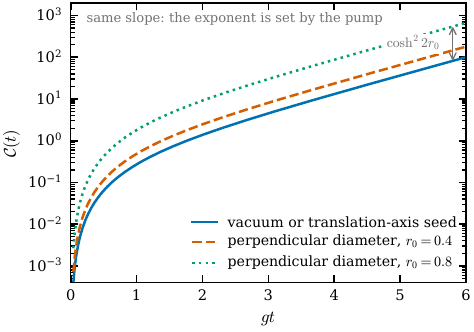}
\caption{\label{fig:squeezedseed} Seed-relative spread for a resonant amplifier with a two-mode squeezed
input, Eqs.~\eqref{eq:squeezedseed} and
\eqref{eq:squeezedseed-aligned}.  A seed on the pump-translation axis
reproduces the vacuum curve; a perpendicular seed multiplies it by
$\cosh^2(2r_0)$.  The common log-scale slope shows that the pump fixes the
growth exponent and the seed fixes the prefactor.}
\end{figure}

Figure~\ref{fig:squeezedseed} shows the seed-relative spread from
Eqs.~\eqref{eq:squeezedseed} and \eqref{eq:squeezedseed-aligned}.  Photon
number retains input population and pump-relative phase, so equal spread can
accompany unequal photon counts.  The pump fixes the spread exponent and the
seed orientation its prefactor, which reaches $\cosh^2(2r_0)$ on the
perpendicular diameter.

\subsection{A frequency quench}\label{sec:quench}

A frequency quench is a nonstationary protocol within the same
$\mathfrak{su}(1,1)$ algebra
as Sec.~\ref{sec:opa}, now in its single-mode sectors.  Here
$K_+=a^{\dagger 2}/2$, $K_-=a^2/2$, and
$K_0=(a^\dagger a+1/2)/2$.  We take the initial state to be the
$\omega_0$-vacuum, the $h=1/4$ lowest weight, with $\omega_0>0$.  At $t=0$ the frequency is switched suddenly from $\omega_0$ to
$\omega_1>0$ for a time $\tau$ and at $t=\tau$ is switched suddenly back to $\omega_0$
\cite{Tibaduiza,PalEtAl2023}.  Writing $\eta_0=(\omega_1^2-\omega_0^2)/(2\omega_0)$ for the quench strength, the
Hamiltonian during the first interval is the $h=1/4$ (even-parity) $\mathfrak{su}(1,1)$
element $H_1=2(\omega_0+\eta_0)K_0+\eta_0(K_++K_-)$; after the
quench, the restored oscillator Hamiltonian is
$H_0=\omega_0(a^\dagger a+\tfrac12)$.
The chain runs over the even Fock states $\ket{2n}$ with
$b_{n+1}=|\eta_0|\sqrt{(n+1)(n+\tfrac12)}$, and
\begin{equation}\label{eq:quenchC}
  \C(t)=\frac{\eta_0^2}{2\omega_1^2}\sin^2(\omega_1t),
  \qquad 0<t<\tau ,
\end{equation}
so the complexity never exceeds
$\eta_0^2/(2\omega_1^2)$. \par
\begin{figure}[tb]
\includegraphics[width=\columnwidth]{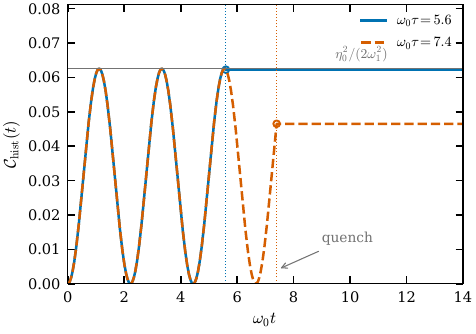}
\caption{\label{fig:quench} History-continued pair-ladder complexity for the sudden quench in
Eq.~\eqref{eq:quenchC}, with $\eta_0/\omega_0=1/2$, hence $\omega_1/\omega_0=\sqrt2$,
and switching times $\omega_0\tau=5.6$ (solid) and $7.4$ (dashed).  Restoring $\omega_0$ freezes the number distribution and
$\C_{\rm hist}$ in the continued even-Fock frame at a switch-selected value
between zero and $\eta_0^2/(2\omega_1^2)$.}
\end{figure}

After the switch, $H_0$ is diagonal in the even-Fock basis.  The photon-number
probabilities generated during the first interval therefore freeze exactly.
The quenched state admits two Krylov descriptions
with different reference frames.
Keeping the pre-switch pair-ladder labels gives the
history-dependent diagnostic
\begin{equation}\label{eq:quench-history}
 \C_{\rm hist}(t)=\C(\tau),\qquad t\geq\tau,
\end{equation}
plotted in Fig.~\ref{fig:quench}.  Restarting the recursion at $t=\tau$ from
$\ket{\psi(\tau)}$ instead probes the subsequent phase evolution relative to
the switched state.

A vacuum-anchored $H_0$ recursion has $b_1=0$ and represents the quenched state only if the switch occurred at a vacuum return.
Equation~\eqref{eq:quench-history} therefore describes
the latched distribution
in a declared history-continued frame, while the frozen Fock probabilities
provide the frame-independent experimental statement.

\section{Bright--dark Krylov geometry in three-level atoms}
\label{sec:three}

The three single-ladder examples align the Krylov
and weight bases for an extremal seed.
A three-level atom adds a branching transition graph, so even the physical
ground-state seed generates coherent superpositions of bare levels.  The
Lanczos recursion then reconstructs the optically natural bright--dark basis
and turns its decoupling, geometric motion, and transfer conditions into
properties of a three-site chain.

\subsection{\texorpdfstring{The $V$ system and its bright--dark chain}{The V system and its bright--dark chain}}\label{sec:vchain}

Take a single atom with one ground state $\ket g$ and two excited states
$\ket{e_1},\ket{e_2}$, each driven from the ground state by a coherent field.
This is the $V$ configuration \cite{DongTang2002} and the first ground-seeded model considered
here in which the bare-level and Krylov bases differ.  The single-Cartan
readout of Eq.~\eqref{eq:dictionary} therefore no longer applies.
In a frame where both drive phases are stationary, and with bare-state phases chosen so
that the couplings are real,
\begin{equation}\label{eq:vH}
  H_V=\begin{pmatrix}0&g_1&g_2\\ g_1&\delta_1&0\\ g_2&0&\delta_2\end{pmatrix}
  \quad\text{in }(\ket g,\ket{e_1},\ket{e_2}),
\end{equation}
with $\delta_{1,2}$ the rotating-frame detunings. The two transitions share the
ground state, so the corresponding $\mathfrak{su}(2)$ subalgebras are not
independent; their commutators generate the rest of $\mathfrak{su}(3)$.
The first excited Krylov vector is therefore a coupling-weighted
superposition of bare levels, and the next vector is its orthogonal
complement.

We assume $G>0$.  If $G=0$, $\ket g$ is an eigenstate and its Krylov
chain has one site.
Starting from the physical ground state $\ket{K_0}=\ket g$, apply
Eq.~\eqref{eq:lanczos}. The first step already does the work. Since
$\braket{g|H_V|g}=0$ we have $a_0=0$, and
$H_V\ket g=g_1\ket{e_1}+g_2\ket{e_2}$, whose norm is
$G=\sqrt{g_1^2+g_2^2}$; normalizing gives $\ket{K_1}$. For $b_2>0$, the second step
subtracts $a_1\ket{K_1}+b_1\ket{K_0}$ from $H_V\ket{K_1}$, and what is left in
the excited manifold is the combination orthogonal to $\ket{K_1}$. Carried
through,
\begin{align}\label{eq:vchain}
 a_0&=0,\qquad b_1=G,\qquad
 \ket{K_1}=\frac{g_1\ket{e_1}+g_2\ket{e_2}}{G},\nonumber\\
 a_1&=\frac{g_1^2\delta_1+g_2^2\delta_2}{G^2},\qquad
 b_2=\frac{|g_1g_2(\delta_1-\delta_2)|}{G^2},\\
 \sigma_D&=\operatorname{sgn}[g_1g_2(\delta_1-\delta_2)],\qquad
 \ket{K_2}=\sigma_D\,\frac{g_2\ket{e_1}-g_1\ket{e_2}}{G},\nonumber\\
 a_2&=\frac{g_2^2\delta_1+g_1^2\delta_2}{G^2}.\nonumber
\end{align}
For $b_2>0$ the recursion has
directly produced the \emph{bright} and \emph{dark}
superpositions of the two excited states, which are the Morris--Shore basis of
a driven multilevel system \cite{MorrisShore1983}. The ground state couples only
to the bright combination, with strength equal to the
total coupling amplitude $G$; the dark combination is
reached only at second order, through $b_2$.  At
$b_2=0$ the recursion terminates at $\ket{K_1}$.  The normalized combination
orthogonal to $\ket B$ then defines an excluded dark direction but is no longer a
Lanczos rung generated by the algorithm. The
Krylov chain of this driven $V$
system is the bright--dark hierarchy.
Its appearance directly from the cyclic subspace shows how the
recursion identifies the optically active and inactive combinations.

\subsection{Time-dependent couplings and a geometric dark rung}
\label{sec:vtime}

The bright--dark chain has a particularly direct extension when the two
couplings vary in amplitude and phase.  Write the rotating-frame Hamiltonian
as
\begin{align}\label{eq:vtime-hamiltonian}
 H_V(t)&=\ket{v(t)}\!\bra g+\ket g\!\bra{v(t)}+H_e(t),\nonumber\\
 \ket{v(t)}&=g_1(t)\ket{e_1}+g_2(t)\ket{e_2}.
\end{align}
Here $H_e(t)$ is the Hermitian excited-manifold block.  Define
$G(t)=\sqrt{|g_1|^2+|g_2|^2}$,
$\ket{B(t)}=\ket v/G$, and
$P_D(t)=\mathbf 1_e-\ket{B}\!\bra{B}$, where $\mathbf 1_e$ is the identity on
the excited doublet.
These quantities are defined on intervals where $G(t)>0$.  An isolated
common zero is a patch boundary.  The chain has a continuous extension through
it when the projective coupling ratio $[g_1:g_2]$ has a common limiting ray;
otherwise the two adjacent pulse intervals define separate bright-state
gauges.

With the physical ground state as the fixed first site, the extended recursion
gives
\begin{align}\label{eq:vtime-chain}
 \ket{K_0}&=\ket g,\qquad a_0=0,\qquad
 b_1=G,\qquad \ket{K_1}=\ket B,\nonumber\\
 a_1&=\bra B(H_e-i\partial_t)\ket B,\\
 b_2^2&=\left\|P_D(H_e-i\partial_t)\ket B\right\|^2.\nonumber
\end{align}
The first hop is controlled by the total coupling amplitude.  The second resolves
two coherent contributions of pulse design,
the action of $H_e$ away from the bright direction and the motion of that
direction itself.

For a degenerate excited doublet, $H_e=\delta(t)\mathbf 1_e$, the dynamical
piece drops out after projection.  Parameterizing
\begin{equation}\label{eq:bright-sphere}
 \ket B=\cos\frac{\vartheta}{2}\ket{e_1}
       +e^{i\varphi_B}\sin\frac{\vartheta}{2}\ket{e_2}
\end{equation}
gives the exact geometric hopping
\begin{equation}\label{eq:v-fs-speed}
 b_2^2=\frac14\left(\dot\vartheta^2+
              \sin^2\!\vartheta\,\dot\varphi_B^2\right).
\end{equation}
Thus $b_2$ is the Fubini--Study speed
of the normalized coupling spinor.  A common
envelope, $\ket v=f(t)\ket{v_0}$, changes $b_1$ while leaving $b_2=0$.
Changing the ratio or relative phase of the two fields instead opens the dark
rung even at equal instantaneous detunings.  Thus the total coupling amplitude and motion
on the bright-state Bloch sphere address the first two Krylov hoppings
separately.

When $H_e$ splits or mixes the excited states, define
$\ket{d_{\rm dyn}}=P_DH_e\ket B$ and
$\ket{d_{\rm geo}}=-iP_D\ket{\dot B}$.  Then
\begin{equation}\label{eq:v-interference}
 b_2^2=\|d_{\rm dyn}\|^2+\|d_{\rm geo}\|^2
 +2\operatorname{Re}\braket{d_{\rm dyn}|d_{\rm geo}}.
\end{equation}
The last term is a coherent, gauge-invariant interference between spectral
mixing and pulse geometry.  It allows the dark-rung coupling to be enhanced or
suppressed without changing $G$.  Equations~\eqref{eq:vtime-chain}--\eqref{eq:v-interference}
identify that coupling with a Krylov hopping and make its projective geometry
explicit. Time-dependent Morris--Shore transformations and their bright--dark
gauge couplings are developed in Ref.~\cite{ZlatanovRangelovVitanov2022}. Appendix~\ref{app:moving-seeds}
derives the result directly from the extended recursion.

The moving ground--bright sector decouples from the dark direction
throughout a pulse interval precisely when
\begin{equation}\label{eq:v-moving-termination}
 P_D(H_e-i\partial_t)\ket B=0 .
\end{equation}
This condition permits two distinct realizations.  A fixed bright direction
with a degenerate excited block makes both projected terms vanish separately.
More generally, a nonzero spectral coupling
$P_DH_e\ket B$ can cancel the geometric coupling
$-iP_D\ket{\dot B}$ coherently.  The relative optical phase and the
excited-manifold control then act as an interference knob that keeps the first
hop $b_1=G$ open while closing $b_2$ along a moving pulse path.
A state initially in the ground--bright sector then remains there.
If earlier driving has populated the dark rung, closing $b_2$ preserves
that population rather than discarding its contribution to $\C$.
This is the same continuation convention used at zero-speed intervals in
Sec.~\ref{sec:dictionary}.

\subsection{Morris--Shore termination and the exact spread bound}\label{sec:cpt}

The second hopping in Eq.~\eqref{eq:vchain} is
$b_2=|g_1g_2(\delta_1-\delta_2)|/G^2$.  Equal detunings therefore close the
bright--dark hopping exactly.  For the ground-state preparation, the cyclic
subspace becomes
\begin{equation}\label{eq:v-reachable-subspace}
 \mathcal K(H_V,\ket g)=\operatorname{span}\{\ket g,\ket B\},
 \quad \ket B=\frac{g_1\ket{e_1}+g_2\ket{e_2}}{G},
\end{equation}
while the orthogonal excited superposition
$\ket D=(g_2\ket{e_1}-g_1\ket{e_2})/G$ in the real-coupling gauge
is the excluded
Morris--Shore sector.  The chain has two sites and hence $0\leq\C(t)\leq1$.

\begin{figure}[tb]
\includegraphics[width=\columnwidth]{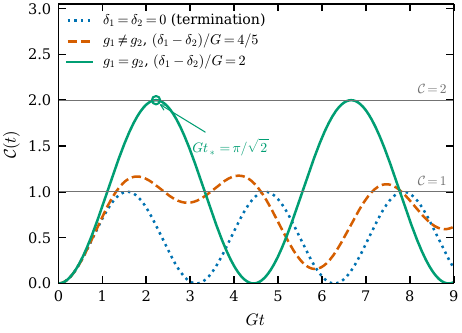}
\caption{\label{fig:vsystem} Ground-seeded $V$-system complexity versus $Gt$, using each curve's
$G=\sqrt{g_1^2+g_2^2}$ as the frequency scale.  The dotted and dashed curves
have $(g_1/G,g_2/G)=(3/5,4/5)$.  The dotted curve has
$\delta_1=\delta_2=0$ and reaches $\C=1$ within the two-site Morris--Shore
sector.  The dashed curve opens the dark rung at
$(\delta_1/G,\delta_2/G)=(2/5,-2/5)$.  The solid uniform-chain point has
$g_1/G=g_2/G=1/\sqrt2$ and $\delta_1/G=-\delta_2/G=1$, reaching $\C=2$ at
$Gt_\ast=\pi/\sqrt2$, Eqs.~\eqref{eq:pstamp}--\eqref{eq:psttime}.}
\end{figure}

Within the reachable sector, the Hamiltonian is an ordinary two-level
generator with coupling $G$ and common detuning
$\bar\delta=\delta_1=\delta_2$.  Its exact maximum bright-state population is
\begin{equation}\label{eq:cptmax}
 \C_{\max}=\frac{4G^2}{4G^2+\bar\delta^2}\leq1,
 \qquad \delta_1=\delta_2=\bar\delta .
\end{equation}
The upper bound is saturated when the common detuning also vanishes
(Fig.~\ref{fig:vsystem}).  If one optical coupling is zero, the same algebraic
termination removes the unused bare excited level.  With both couplings
present, it isolates a coherent bright--dark decomposition and makes the
dimension reduction directly tunable by the detuning difference.

This is the closed-$V$ realization of Morris--Shore decoupling.\footnote{In a $\Lambda$ system, coherent population trapping and STIRAP use a
dark superposition of the two long-lived states, which can itself carry
population \cite{Vitanov2017}.  In the apparent $V$-type two-electron system
of Ref.~\cite{SinghNatarajan2015}, electron--electron interaction converts the
effective level structure into a $\Lambda$ system and thereby enables the
observed dark state.  In the ground-seeded $V$ system here,
the dark vector lies in the excited manifold and remains outside the reachable
cyclic subspace.  Absorption and electromagnetically induced transparency lineshapes
additionally require an open-system probe model with radiative decay and
dephasing.}  The termination
also differs physically from the $b_1=0$ case in Sec.~\ref{sec:tipping}.  There
the seed is an eigenstate of the generator; here $b_1=G$ remains finite and the
level structure removes the next rung.  The two coefficients therefore
separate state-induced localization from spectrum-induced termination.

\subsection{Complete transfer into the
coupling-dark superposition}\label{sec:pst}

The three-site chain bounds the spread by $\C\leq2$, with equality precisely
when the population reaches $\ket{K_2}=\ket D$.  Here \emph{dark} means
uncoupled directly from $\ket g$; at unequal detunings $H_e$ couples this
superposition back to $\ket B$. An exact sufficient route
to that endpoint is to realize a uniform chain.  Engineered perfect-transfer chains exploit
mirror symmetry and spectral phase commensurability
\cite{Christandl2004}; the uniform construction is especially useful here
because its conditions translate into closed optical parameters.

Impose $a_0=a_1=a_2$ and $b_1=b_2$ in Eq.~\eqref{eq:vchain}.  Since $a_0=0$
and a three-site chain requires $\delta_1\neq\delta_2$, the equality
$a_1=a_2$ gives $|g_1|=|g_2|$.  Absorbing their relative phase into the excited
states, set $g_1=g_2=g>0$.  The diagonal condition then requires
$\delta_1+\delta_2=0$, and hopping equality fixes
$|\delta_1-\delta_2|=2\sqrt2\,g$.  One convenient orientation is therefore
\begin{equation}\label{eq:v-uniform-point}
 g_1=g_2=g,\qquad \delta_1=-\delta_2=\sqrt2\,g .
\end{equation}
Both the zero common detuning and the specified detuning difference are part
of the uniform-chain condition.

At Eq.~\eqref{eq:v-uniform-point}, the Krylov matrix has zero diagonal and
$b_1=b_2=G=\sqrt2\,g$.  Starting from $\ket{K_0}$, its far-site amplitude is
\begin{equation}\label{eq:pstamp}
 \braket{K_2|\psi(t)}=\frac12\!\left[\cos(\sqrt2\,Gt)-1\right],
\end{equation}
and complete transfer follows at
\begin{equation}\label{eq:psttime}
 t_\ast=\frac{\pi}{\sqrt2\,G},\qquad \C(t_\ast)=2 .
\end{equation}
The equal-detuning point of Sec.~\ref{sec:cpt} closes the dark rung, whereas
Eq.~\eqref{eq:v-uniform-point} impedance-matches the two open hoppings.
Together they give two experimentally distinct controls of the same
bright--dark chain, one an exact
exclusion and the other an exact population transfer.

\subsection{The time-averaged complexity
profile}\label{sec:lineshape}

For the closed Hamiltonian considered here, the natural stationary diagnostic
is the infinite-time coherent average of the chain spread.  It is obtained by
infinite-time phase averaging in the energy basis. The
equal-detuning point of Sec.~\ref{sec:cpt} and the uniform-chain
point of
Sec.~\ref{sec:pst} are two points of the profile computed here.  A continuous-wave absorption profile is
the corresponding open-system observable and can be built from the same
Hamiltonian after specifying a probe, radiative decay, and dephasing.
\par Writing the chain
matrix as $M$ with eigenvalues $E_r$
and real orthonormal eigenvectors $v^r$, the amplitude on site $n$ is
$\braket{K_n|\psi(t)}=\sum_r e^{-iE_rt}v^r_nv^r_0$, so
$|\braket{K_n|\psi}|^2$ carries phases $e^{-i(E_r-E_s)t}$. When the spectrum is
nondegenerate every $r\neq s$ term averages to zero, leaving
$\overline{|\braket{K_n|\psi}|^2}=\sum_r(v^r_n)^2(v^r_0)^2$ and hence a closed
form with no time integration,
\begin{equation}\label{eq:cbar}
  \bar\C=\sum_r|v^r_0|^2\left(|v^r_1|^2+2|v^r_2|^2\right).
\end{equation}
\par Equation~\eqref{eq:cbar} is evaluated algebraically from the
three-dimensional chain matrix.  A common shift of the two optical detunings
moves the ground-site energy relative to both excited chain sites, so the
average depends on both the detuning difference and the common detuning.  We
isolate the former by choosing the symmetric scan
$\delta_{1,2}=\pm\delta/2$.
For an analytic signed-$\delta$ parametrization we keep the continuous
dark completion $\widetilde K_2=(g_2\ket{e_1}-g_1\ket{e_2})/G$ fixed.  The
off-diagonal entry is then $\mu\delta$; for $\delta\ne0$, the rephasing
$\widetilde K_2\mapsto\operatorname{sgn}(\mu\delta)\widetilde K_2$ restores
the canonical Lanczos coefficient $b_2=|\mu\delta|$ without changing site
probabilities or $\bar\C$.
The chain matrix is then
\begin{equation}\label{eq:chainmatrix}
  M=\begin{pmatrix}0&G&0\\ G&\varepsilon\delta&\mu\delta\\
                   0&\mu\delta&-\varepsilon\delta\end{pmatrix},\quad
  \varepsilon=\frac{g_1^2-g_2^2}{2G^2},\quad \mu=\frac{g_1g_2}{G^2}.
\end{equation}

\begin{figure}[tb]
\includegraphics[width=\columnwidth]{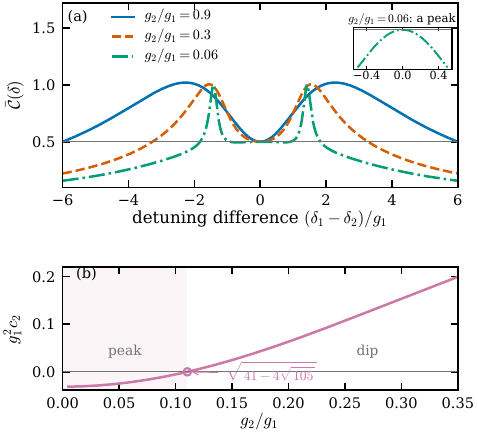}
\caption{\label{fig:lineshape} (a) Closed-unitary time-averaged $V$-system complexity,
Eq.~\eqref{eq:cbar}, versus $\delta/g_1$ for
$\delta_{1,2}=\pm\delta/2$.  The equal-detuning point is a local minimum at
$\bar\C=1/2$ for the main curves and a maximum for the inset ratio
$g_2/g_1=0.06$.  (b) The dimensionless curvature $g_1^2c_2$ from
Eq.~\eqref{eq:curvature}; its zero in Eq.~\eqref{eq:inversion} marks the
inversion, and the shaded band is the local-maximum regime.}
\end{figure}

At $\delta=0$ its eigenvectors are $(1,\pm1,0)/\sqrt2$ and $(0,0,1)$, for
which the sum above gives $\bar\C(0)=\tfrac14+\tfrac14=\tfrac12$, the two-site
value. A normalized Rayleigh--Schr\"odinger expansion of the three eigenvectors
through second order in $\delta$, substituted into Eq.~\eqref{eq:cbar}, gives
\begin{equation}\label{eq:curvature}
  \bar\C(\delta)=\frac12
  +\frac{\tfrac52g_1^2g_2^2-\tfrac1{32}(g_1^2-g_2^2)^2}{G^6}\,\delta^2
  +O(\delta^4),
\end{equation}
with $\delta=\delta_1-\delta_2$. We denote the displayed quadratic
coefficient by $c_2$, so $\bar\C=1/2+c_2\delta^2+O(\delta^4)$.
The equal-detuning termination point therefore
appears in the time-averaged complexity as an extremum of computable curvature
(Fig.~\ref{fig:lineshape}).

The coefficient in Eq.~\eqref{eq:curvature} changes sign at
\begin{equation}\label{eq:inversion}
 \begin{aligned}
 \left|\frac{g_2}{g_1}\right|
 &=\sqrt{41\pm4\sqrt{105}}\,,\\
 \frac{\min(|g_1|,|g_2|)}{\max(|g_1|,|g_2|)}
 &=\sqrt{41-4\sqrt{105}}\simeq0.1104\, .
 \end{aligned}
\end{equation}
The two roots are interchanged by relabelling the optical transitions.  For
the ordering $0<|g_2/g_1|\leq1$ used in Fig.~\ref{fig:lineshape}, the local
minimum becomes a local maximum below
$\sqrt{41-4\sqrt{105}}\simeq0.1104$.  The curvature coefficient takes the
chain form
\begin{equation}\label{eq:curvature-chain}
 \frac{1}{G^2}\left(\frac52\mu^2-\frac18\varepsilon^2\right).
\end{equation}
The first contribution measures the detuning-induced
coupling to the dark rung,
whereas the second measures the differential shift of the
two excited chain
sites.  Their competition therefore determines whether equal detuning
minimizes or maximizes the long-time spread.  Repeated state tomography or calibrated projective readout in the
bright--dark basis can reconstruct this coherent average.  With a weak probe and
calibrated decay rates, the same scan also permits a point-by-point comparison
with the absorption response.

For equal couplings, $\varepsilon=0$, $\mu=\tfrac12$, and the chain matrix has
$b_1=G$, signed second off-diagonal entry $\delta/2$
(equivalently $b_2=|\delta|/2$ in positive Lanczos gauge), and vanishing
diagonal entries. Its eigenvalues are
$0$ and $\pm\Lambda_V$, with
$\Lambda_V^2=G^2+\delta^2/4$. Substitution in
Eq.~\eqref{eq:cbar} gives the complete profile
\begin{equation}\label{eq:vlineshape}
  \bar\C(\delta)=
  \frac{2G^2\left(4G^2+7\delta^2\right)}
       {\left(4G^2+\delta^2\right)^2}.
\end{equation}
\par This rational function of the detuning difference reproduces
Eq.~\eqref{eq:curvature} at small
$\delta$. Differentiation places the two
flanking maxima at
\begin{equation}\label{eq:flank}
  |\delta^\ast|=G\sqrt{\frac{20}{7}},
  \qquad \bar\C(\delta^\ast)=\frac{49}{48},
\end{equation}
while $\bar\C(\delta)\to0$ as $|\delta|\to\infty$. The maxima are
coherent-spread extrema generated by competition between the first and second
Krylov hoppings.  With specified probe and decay parameters, an open-system
calculation can determine how their positions compare with extrema of the
spectroscopic response.

\subsection{Adiabatic passage in the instantaneous-dark-state frame}\label{sec:stirap}

The $\Lambda$ configuration with time-dependent couplings is the setting of
stimulated Raman adiabatic passage \cite{Vitanov2017}, in which a Stokes pulse
precedes a pump pulse and population is transferred between two stable states
through an excited state whose population is suppressed in the adiabatic regime.

A complementary operator-Krylov treatment by Takahashi and del~Campo
expands the adiabatic gauge potential in nested commutators and includes
STIRAP as an application \cite{TakahashiDelCampo2024}.  Here the Krylov basis
is built from states in the moving dark-state frame.  The resulting state-chain
populations resolve diabatic leakage, while the first two hoppings encode the
dark--bright connection and the bright--excited coupling rather than the
operator expansion of the counterdiabatic term.

The general moving-reference state-Krylov construction identifies spread from
an instantaneous eigenstate with nonadiabaticity
\cite{Takahashi:2024hex}.  In STIRAP, the standard adiabatic frame identifies
$\dot\theta$, the rate of change of the mixing angle $\theta$ defined by
$\tan\theta=\Omega_P/\Omega_S$ below, as the dark--bright connection
\cite{Vitanov2017}.  We now specialize these structures
to the optical $\Lambda$ system and show that the connection is exactly the
first Krylov hopping, so the chain resolves where the diabatic leakage resides.

Let $\ket1$ and $\ket3$ be the stable states and let $\ket2\equiv\ket e$ be
the optically excited state.  In the resonant rotating frame,
\begin{equation}\label{eq:stirap-hamiltonian}
 H_\Lambda(t)=\Omega_P(t)(\ket{e}\!\bra{1}+\ket{1}\!\bra{e})
 +\Omega_S(t)(\ket{e}\!\bra{3}+\ket{3}\!\bra{e}).
\end{equation}
Choose fixed optical phases for which the pulse envelopes
$\Omega_P(t),\Omega_S(t)\geq0$, and write
\begin{equation}
 \tan\theta(t)=\frac{\Omega_P(t)}{\Omega_S(t)},\qquad
 G(t)=\sqrt{\Omega_P^2(t)+\Omega_S^2(t)}.
\end{equation}
Here $\Omega_P$ and $\Omega_S$ denote the off-diagonal Hamiltonian matrix
elements; they are half the conventional Rabi frequencies when the interaction
is written with an explicit factor $1/2$.  The instantaneous basis is
\begin{align}
  \ket D&=\cos\theta\ket1-\sin\theta\ket3,\nonumber\\
  \ket B&=\sin\theta\ket1+\cos\theta\ket3,\\
  \ket e&=\ket2 .\nonumber
\end{align}
For any state in this subsection,
$P_\nu(t)=|\braket{\nu|\psi(t)}|^2$ denotes the population of the named bare
or adiabatic state $\ket\nu$.

For the exact one- and two-photon-resonant
Hamiltonian in Eq.~\eqref{eq:stirap-hamiltonian},
$\ket D$ is an exact null eigenvector of $H$.  Its
motion generates the dark--bright coupling through the adiabatic connection,
the same geometric mechanism that opens the second rung of the time-dependent
$V$ chain in Sec.~\ref{sec:vtime}.
In the triple $(\ket D,\ket B,\ket e)$ the
Hamiltonian itself is $H=G(\ket B\bra e+\ket e\bra B)$, since
$H\ket D=0$ and $H\ket B=G\ket e$, while $\partial_t\ket D=-\dot\theta\ket B$
carries the only connection element. The extended generator
in Eq.~\eqref{eq:Gmatrix} is therefore
\begin{equation}\label{eq:stirapmatrix}
  \mathcal G=\begin{pmatrix}0&-i\dot\theta&0\\
                             i\dot\theta&0&G\\ 0&G&0\end{pmatrix}.
\end{equation}
The signs follow directly from $-i\braket{D|\dot B}=-i\dot\theta$ and
$-i\braket{B|\dot D}=i\dot\theta$. On an interval of fixed
$\sigma_\theta=\operatorname{sgn}\dot\theta$, the rephased Lanczos triple
$\ket{K_0}=\ket D$, $\ket{K_1}=i\sigma_\theta\ket B$, and
$\ket{K_2}=i\sigma_\theta\ket e$ makes both
off-diagonal coefficients positive,
\begin{equation}\label{eq:stirapchain}
  b_1=|\dot\theta|,\qquad b_2=G(t).
\end{equation}
The first hopping is generated entirely by the connection; its magnitude
in the positive Lanczos gauge is $b_1=|\dot\theta|$. At isolated zeros of $\dot\theta$
this phase convention is continued patchwise; the moving reference remains
well defined although the instantaneous first coupling vanishes.
Seeded in $\ket D$, which tends to the physical initial state $\ket 1$ in the
asymptotic initial frame $\theta\to0$, the leakage probability is
$P_{\rm leak}=P_B+P_e$, whereas
$\C=P_B+2P_e=P_{\rm leak}+P_e$ additionally weights the optically
excited-state population.

A statically anchored chain would instead measure spread from the
initial laboratory state.  Here the moving frame is retained to resolve
diabatic error into bright and excited components.

For the usual slowly scaled STIRAP pulses, the local adiabaticity
criterion $|\dot\theta|\ll G$ becomes $b_1\ll b_2$ over the pulse-overlap
interval. Localization near the first site additionally requires small
accumulated nonadiabatic amplitudes, which depend on the gaps and the full
pulse history \cite{Vitanov2017}.

\begin{figure}[tb]
\includegraphics[width=\columnwidth]{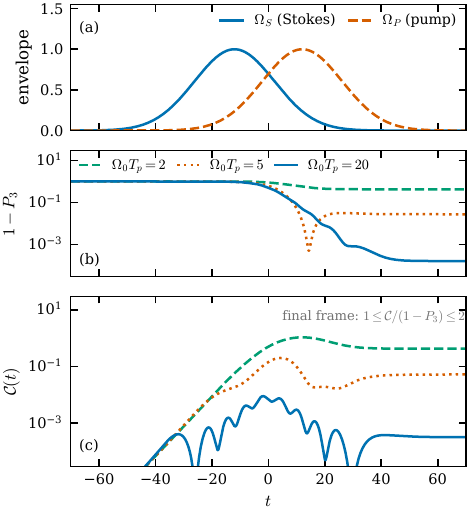}
\caption{\label{fig:stirap} STIRAP for $\Omega_0T_p=2,5,20$.  (a) Counterintuitive Gaussian pulses
$\Omega_{P,S}=\Omega_0e^{-[(t\mp\tau)/T_p]^2}$ with $T_p=20$ and
$\tau/T_p=0.6$.  (b) Transfer error $1-P_3$.  (c) Complexity in the moving
$(\ket D,\ket B,\ket e)$ chain.  Their asymptotic values decrease with
$\Omega_0T_p$ but are unequal.  Asymptotically, $\C\to P_1+2P_2$, whereas
$1-P_3=P_1+P_2$.}
\end{figure}

In the adiabatic limit, transfer carries the
entire population from $\ket1$ to $\ket3$ while the complexity remains near
zero, because the population does not spread along the chain but
is transported with
the moving seed (Fig.~\ref{fig:stirap}).
To leading nonzero order in the nonadiabatic correction,
$P_{\rm leak}\simeq2|c_+|^2$ and
$\C\simeq2|c_+|^2(1+\sin^2\phi_{\rm leak})$ with
$c_+(t)=\tfrac1{\sqrt2}\int_{t_i}^{t}\!\dd t'\,
\dot\theta(t')\exp[-i\int_{t'}^{t}G(s)\,\dd s]$ and
$c_-(t)=c_+^*(t)$ for real pulse envelopes in the chosen gauge, with
$\phi_{\rm leak}=\arg c_+$, so
both $P_{\rm leak}$ and $\C$ are quadratic in the first-order
nonadiabatic amplitudes $c_\pm$, while the factor
$1+\sin^2\phi_{\rm leak}$
oscillates between $1$ and $2$.

For real $\dot\theta$,
the amplitudes on $\ket{\pm}=(\ket B\pm\ket e)/\sqrt2$
are complex conjugates, and
$\C=2(\operatorname{Re}c_+)^2+4(\operatorname{Im}c_+)^2$, which is the form quoted above. At
$\phi_{\rm leak}=0$ the
leakage lies in $\ket B$, whereas at
$\phi_{\rm leak}=\pi/2$ it lies in $\ket e$; the
adiabatic eigenstates $\ket\pm$
are the
balanced
intermediate basis.  For the plotted pulse family the numerical result
converges to this first-order relation
as the overlap-region nonadiabatic coupling is reduced, which is the
perturbative regime used for quantitative comparison.
In the asymptotic final frame
$\theta\to\pi/2$, where $\ket B\to\ket1$, the bound itself does not require
perturbation theory. Since $1-P_3=P_B+P_e$ and $\C=P_B+2P_e$,
\begin{equation}\label{eq:stirap-exact-ratio}
 \frac{\C}{1-P_3}=1+\frac{P_e}{P_B+P_e}\in[1,2]
\end{equation}
whenever the asymptotic transfer error is nonzero.  First order resolves the fraction
through the phase $\phi_{\rm leak}$ of the leaked adiabatic
amplitude.

STIRAP can therefore move the full laboratory population while keeping
the moving-reference spread arbitrarily small.  In this protocol, the site populations resolve diabatic
error, while $\C$ measures departure from the pulse-selected adiabatic frame
and obeys the exact bound in Eq.~\eqref{eq:stirap-exact-ratio}.

\FloatBarrier

\section{Discussion}\label{sec:discussion}
The optical meaning of spread complexity is sharpest when the preparation,
chain, and measurement basis are specified together.  For extremal spin,
oscillator, and amplifier preparations, the Krylov index is the excitation
number in Eq.~\eqref{eq:dictionary}.  Resonant fixed-direction pulses give
$\C=2j\sin^2\A$, $\C=\A^2$, and $\C=2h\sinh^2\A$, while the full populations
obey Eq.~\eqref{eq:eos}.  These distribution-level tests diagnose motion on
the intended coherent-state orbit.

That orbit also carries metric and symplectic information.
Equation~\eqref{eq:distancelaw} relates spread to its geodesic separation,
while Eq.~\eqref{eq:complexity-geometric-phase} gives the geometric phase from
the measured $(\C,\Phi)$ loop.  For displaced Fock states the squared first
hopping and metric acquire the factor $2m+1$, whereas the Berry curvature is
independent of $m$.  Local susceptibility and geometric phase therefore probe
different structures, while the full spread retains further seed dependence.

Table~\ref{tab:seed-control-map} separates local information from complete
dynamics.  Arbitrary controls in one oscillator, $SU(2)$, or $SU(1,1)$ algebra
are completely reduced for $\ket{\ell,0}$ and, by constant conjugation, for a
fixed group-displaced extremal seed $W_{\rm s}\ket{\ell,0}$.  The compact
family also includes $W_{\rm s}\ket{j,j}$ after a fixed $\pi$ rotation and
ladder reversal.
Equations~\eqref{eq:displaced-extremal-controls}--
\eqref{eq:displaced-extremal-complexity} give the transformed controls, chain,
and scalar Gauss equation.  Undoing $W_{\rm s}$ before excitation counting
then gives Eq.~\eqref{eq:displaced-extremal-readout}.  For $r>0$ in the
oscillator and $SU(1,1)$ ladders, or $0<r<2j$ in $SU(2)$, conjugation and $b_1$
remain exact, but the later arbitrary-control recurrence does not follow from
them.
For the spin-one middle weight, Sec.~\ref{sec:tipping} completes this
recurrence for arbitrary controls. Its two-photon realization measures the
full spread through two coincidence settings, making the distinction from
mean excitation directly observable.

Three further reductions determine nonextremal dynamics in broader seed
classes. For fixed $X$, an old
rung uses $Q=p_r$ and starts its Christoffel-weighted chain; a general
admissible $Q$ instead prepares a finite old-rung superposition.  For a
constant force with nonzero detuning, the seed-adapted chain of every Fock
state $\ket r$, with $r\in\mathbb N_0$, is thereby determined exactly.  The
affine family in Eq.~\eqref{eq:dressed-fixed-generator} transports admissible
$Q(X)\ket\chi$ and its positive measure, even for noncommuting laboratory
Hamiltonians; Eq.~\eqref{eq:displacement-dressed-photon-added} is its
one-photon example.  Stroboscopic sampling instead builds an Arnoldi chain from the one-cycle
unitary. Its quasienergy folding depends on the sampled rotation and is
already present for a constant force.  A rational rotation with nonzero displacement and
$q_F\ge2$ has dimension $q_F$ for every Fock seed $\ket r$, whereas an
irrational rotation has infinite dimension.  The half-turn fold has at most
two sites and is solved by Eq.~\eqref{eq:floquet-half-turn-complexity}.

For a constant detuned amplifier, the fixed coherent seed also obeys the
global bound $\C_{\rm s}\ge2h\sinh^2(\Gamma t/2)$ in the hyperbolic class, as
shown in Eq.~\eqref{eq:hyperbolic-seed-bound}.  Equality for $t\ne0$ selects the
translation-axis seed, whereas an elliptic generator admits a normalizable
coherent fixed-point seed with zero spread.  The bound concerns seed-relative
Krylov spread rather than seeded-amplifier photon number.

The cavity results turn these distinctions into measurements.
Equation~\eqref{eq:wignerscan} makes odd-site occupation a displaced-parity
cut through a definite-parity seed, exposing the negative lobe of a
photon-added state or the fringes of a cat.  At the Floquet half turn, a return
measurement gives $\C_F^{(r)}(2k+1)=1-P_{\rm ret}^{(r)}$ without reconstructing the
Arnoldi basis.  Figure~\ref{fig:floquet-halfturn} shows this survival deficit
and its Laguerre zeros; after calibration or removal of the known scalar
Floquet phase, an interferometric return additionally recovers the Wigner sign.
For $q_F>2$, the full complexity still requires the finite
seed-adapted Arnoldi basis.

The three-level atom illustrates a chain whose physical and bare-state bases
differ.  Static ground-seeded recursion gives the bright--dark hierarchy,
termination at equal detunings, complete transfer at the uniform-chain point,
and the flanking maxima in Eq.~\eqref{eq:flank}.  With time-dependent complex
couplings, $b_1=G$, while $b_2$ in Eq.~\eqref{eq:vtime-chain} measures motion
away from the bright direction.  Excited-manifold dynamics can reinforce or
cancel its geometric part through Eq.~\eqref{eq:v-interference}, and
Eq.~\eqref{eq:v-moving-termination} gives the condition for moving-sector decoupling.  In
STIRAP, $|\dot\theta|$ is the first hopping from the instantaneous dark state,
so the chain populations resolve diabatic leakage while the complexity weights
the optically excited component twice.

The scope is closed unitary dynamics with prescribed classical controls.
Spread complexity depends on the Hamiltonian, seed, and reference path.  The
distance relation holds only within a coherent orbit and is not a distance on
the full state space \cite{Aguilar}; operator growth instead tridiagonalizes a
Liouvillian \cite{Parker}.  Pump depletion and loss require, respectively, a
quantized pump and a Liouvillian construction \cite{Bhattacharya:2023zqt}.
Equation~\eqref{eq:qgt-b1} assumes tangency to the declared orbit; the normal
contribution is derived in Appendix~\ref{app:moving-seeds}.

Extending arbitrary-control solutions to general interior weights,
especially the infinite oscillator and $SU(1,1)$ ladders, calls for higher
recurrences beyond the closures developed here. A changing algebra direction
need not preserve the fixed measure used by the Christoffel construction. Floquet
extensions to $SU(2)$ and $SU(1,1)$ add representation edges
and infinite discrete-series support.  Degenerate dark manifolds and open or
nonlinear optics instead call for block Lanczos, Liouvillian, or
seed-dependent recurrences.

The division of roles is concrete.  The algebra organizes the geometry, the
preparation fixes the chain, and the controls determine its path.  In the
solved regimes, excitation counting, parity, return probability, and
interferometric phase provide the optical readouts.

\begin{acknowledgments}
A.C. thanks the organizers of Strings 2024 at CERN, String--Math 2024, and the
School and Workshop on Number Theory and Physics at ICTP for their hospitality
while part of this work was carried out, and acknowledges support from IIT
Bhubaneswar Seed Grant SP--103. Aryabrat Mahapatra started this work as part of his
M.Sc.\ thesis at IIT Bhubaneswar during the 2022--2023 academic year, and
thanks A.C. for supervision, Ankit Gill for useful discussions, and the
Ministry of Education, Government of India, for financial support.
\end{acknowledgments}

\appendix

\section{Gauss decomposition}\label{app:gauss}

For constant coefficients the Riccati equation for $x$ also has constant
coefficients.  When $C\ne0$, the substitution $x=(i/C)\dot u/u$ linearizes it to
$\ddot u+iB\dot u+ACu=0$, whose solutions oscillate at
$r=\sqrt{B^2/4+AC}$. Undoing the substitution and integrating the remaining two
equations by quadrature, the system
in Eq.~\eqref{eq:weinorman} is solved by
\begin{align}
  x(t)&=\frac{-2iA\sin rt}{2r\cos rt+iB\sin rt},\nonumber\\
  y(t)&=-2\ln\!\left[\cos rt+\frac{iB}{2r}\sin rt\right],\\
  z(t)&=\frac{-2iC\sin rt}{2r\cos rt+iB\sin rt},\nonumber
\end{align}
with $r=\sqrt{B^2/4+AC}$ and continuous limits at $r=0$; the small-time
behavior $x=-iAt+O(t^2)$ fixes the signs.
For $C=0$, the Riccati equation is triangular and can be integrated
directly; the displayed group solution is its continuous $C\to0$ limit.
For $\mathfrak{su}(1,1)$, where $[K_+,K_-]=-2K_0$, the commutator sign
replaces the compact invariant by
$r_{1,1}^2=B^2/4-AC$.  Positive, zero, and negative $r_{1,1}^2$ give the
elliptic, parabolic, and hyperbolic branches, respectively.  The last branch
is obtained by analytic continuation of the circular functions and gives
the hyperbolic functions in Eq.~\eqref{eq:opaC}.

\section{Unitarity relation}\label{app:unitarity}

In the spin-$\tfrac12$ representation the factorization
in Eq.~\eqref{eq:gaussfact}
reads
\begin{equation}
  e^{xJ_+}e^{yJ_0}e^{zJ_-}=e^{-y/2}
  \begin{pmatrix} e^{y}+xz & x\\ z& 1\end{pmatrix}.
\end{equation}
Comparing with a general $SU(2)$ element
$\left(\begin{smallmatrix}a&b\\-\bar b&\bar a\end{smallmatrix}\right)$ gives
$e^{-y/2}=\bar a$ and $x=b/\bar a$, so $|a|^2(1+|x|^2)=1$ and
\begin{equation}
  \operatorname{Re}\,y=\ln\left(1+|x|^2\right).
\end{equation}
The relation is a group identity and therefore representation independent. In
a positive discrete-series representation of $\mathfrak{su}(1,1)$, on the unit-disc chart $|x|<1$, the
same argument gives $\operatorname{Re}\,y=\ln(1-|x|^2)$. This is what allows the
complexity in Eq.~\eqref{eq:threesums} to be evaluated from $|x|$ alone.

For the Heisenberg--Weyl family, use the normal ordering
$U_H=e^{xa^\dagger}e^{y_H\mathbf1}e^{za}$, where $y_H$ multiplies the
central identity and $[a,a^\dagger]=\mathbf1$. Unitarity requires
$z=-\bar x$. Vacuum normalization gives
$\|U_H\ket0\|^2=e^{2\operatorname{Re}y_H+|x|^2}=1$, hence
$\operatorname{Re}y_H=-|x|^2/2$. Equivalently, the
Baker--Campbell--Hausdorff identity gives
$U_H=e^{y_H+|x|^2/2}D(x)$, whose scalar prefactor has unit modulus.
The vacuum probabilities are therefore the Poisson member of
Eq.~\eqref{eq:amplitudes}. A number rotation to the right of this
displacement leaves the vacuum unchanged.

\section{The photon-added chain}\label{app:dunkl}

This appendix derives Eq.~\eqref{eq:photonaddedC} of
Sec.~\ref{sec:photonadded} by a route independent of the connector argument
used there. For the resonant generator $H=f_0(a+a^\dagger)$ the vacuum spectral
measure is Gaussian, and the seed
obeys
$\ket1=(a+a^\dagger)\ket0$ and therefore multiplies it by $\xi^2$,
where
$\xi=E/(\sqrt2 f_0)$ is the scaled spectral coordinate defined in
Sec.~\ref{sec:photonadded}. The polynomials orthogonal with respect to
$\xi^2e^{-\xi^2}$ are generalized Hermite polynomials, whose recurrence gives
Eq.~\eqref{eq:paraboseb}, so the chain is the parabosonic oscillator with Dunkl deformation $\mu_{\rm D}=1$, or conventional paraboson
order $p=2\mu_{\rm D}+1=3$ \cite{Rosenblum1994}. Its ladder operators are built from
the Dunkl derivative $T=\partial_\xi+(1-R)/\xi$, with $R$ the parity
reflection. Writing $p_n$ for the normalized generalized-Hermite
polynomial of degree $n$, they satisfy
$a_D^\dagger a_D\,p_n=(n+2[n\ \text{odd}])p_n$. The chain number operator is
therefore
\begin{equation}
  \hat N_{\rm K}=a_D^\dagger a_D-2\Pi_{\rm odd},
\end{equation}
with $\Pi_{\rm odd}$ the projector on odd sites. Both expectation values in
the evolved state are Gaussian integrals,
\begin{align}\label{eq:dunklpieces}
  \langle a_D^\dagger a_D\rangle&=
  \A^2+2\left(1-e^{-2\A^2}\right)+4\A^2e^{-2\A^2},\nonumber\\
  P_{\rm odd}&=\tfrac12-\tfrac12e^{-2\A^2}+2\A^2e^{-2\A^2},
\end{align}
and their combination is Eq.~\eqref{eq:photonaddedC}.

The second line of Eq.~\eqref{eq:dunklpieces} also follows from the
$\ket1$ instance of the parity relation
in Eq.~\eqref{eq:wignerscan}.  Indeed,
\begin{equation}
 \mathcal W_{\ket1}(\beta)=-\frac2\pi
 \left(1-4|\beta|^2\right)e^{-2|\beta|^2}
\end{equation}
and $\pi_0=-1$ reproduce the same $P_{\rm odd}$.  This equality provides a
cross-check between the Dunkl number operator and the displaced-parity
representation of the Wigner function.  The amplitude connector for the
corresponding Christoffel transform gives
a third derivation
\cite{ChowdhuryMahapatra2026}.

The two-periodic parabosonic ladder is special to the seed $\ket1$, for which the
spectral polynomial is proportional to the single-root Hermite polynomial
$H_1(\xi)$.  For $\ket m$ with $m\ge2$, the normalized measure is
$\dd\mu_m(\xi)=H_m(\xi)^2e^{-\xi^2}\dd\xi/
(2^m m!\sqrt\pi)$ and contains all roots of $H_m$.  Its recurrence
is still fixed exactly by the Christoffel-transformed measure, although it no
longer reduces to the one-parameter generalized-Hermite ladder above.  The
parity projection remains closed for every $m$, with
$\pi_0=(-1)^m$ in Eq.~\eqref{eq:wignerscan}.

\section{\texorpdfstring{Moving seeds, dressed generators, and the
time-dependent $V$ chain}{Moving seeds, dressed generators, and the
time-dependent V chain}}
\label{app:moving-seeds}

This appendix collects the derivations behind
Eqs.~\eqref{eq:moving-seed-b1}, \eqref{eq:dressed-amplitudes}, and
\eqref{eq:vtime-chain}.  The moving-seed and $V$-chain identities use unitarity and orthogonal
projection.  The dressed spectral measure additionally uses the spectral
theorem for the self-adjoint operator $X$.

For $\ket{K_0}=W\ket\chi$, the first extended Lanczos residual is
\begin{equation}
 \ket{A_1}=(1-P_0)(H-i\partial_t)W\ket\chi
 =W(1-\ket\chi\!\bra\chi)G_W\ket\chi .
\end{equation}
Taking its norm proves Eq.~\eqref{eq:moving-seed-b1}.  Multiplication by
$i$ writes the same residual as the component of
$(\partial_t+iH)\ket{K_0}$ orthogonal to the ray.  Its squared norm is the
Fubini--Study metric applied to the difference between the reference velocity
and the Hamiltonian velocity, which proves Eq.~\eqref{eq:qgt-b1} whenever both
velocities are tangent to the declared group orbit.  If the Hamiltonian has a
component normal to that orbit, its squared normal component must be added;
Eq.~\eqref{eq:moving-seed-b1} remains the general identity.

The coefficients in Eq.~\eqref{eq:nonextremal-first-hop} follow from
\begin{align}
 \bra m(a+a^\dagger)^2\ket m&=2m+1,\\
 \bra{j,m_z}J_x^2\ket{j,m_z}
 &=\bra{j,m_z}J_y^2\ket{j,m_z}
 =\frac{j(j+1)-m_z^2}{2},\nonumber\\
 \bra{h,n}K_1^2\ket{h,n}
 &=\bra{h,n}K_2^2\ket{h,n}
 =\frac{n^2+2hn+h}{2}, \nonumber
\end{align}
with vanishing mixed symmetrized transverse moments in the stated weight
states.  The last line follows from
$K_+K_-\ket{h,n}=n(2h+n-1)\ket{h,n}$ and
$K_-K_+\ket{h,n}=(n+1)(2h+n)\ket{h,n}$.

To verify the dressed evolution, differentiate Eq.~\eqref{eq:dressed-propagator}.
Direct multiplication on the right by $U^\dagger$ gives
\begin{equation}
 i\dot U U^\dagger
 =\dot\gamma\,\mathbf 1+i\dot W W^\dagger
  +\dot\tau\,WXW^\dagger=H(t).
\end{equation}
Let the fixed polynomial-seed recurrence be
\begin{equation}
 X\ket{K_n^{Q,X}}=
 b_{n+1}^{Q,X}\ket{K_{n+1}^{Q,X}}+
 a_n^{Q,X}\ket{K_n^{Q,X}}+
 b_n^{Q,X}\ket{K_{n-1}^{Q,X}},
\end{equation}
with $b_n^{Q,X}\geq0$.  In the frame transported by $W(t)$, the extended generator
is $G_W=\dot\tau X+\dot\gamma\,\mathbf 1$.  Its diagonal element is therefore
$\dot\gamma+\dot\tau a_n^{Q,X}$, while its two off-diagonal elements are
$\dot\tau b_n^{Q,X}$ and $\dot\tau b_{n+1}^{Q,X}$.  For $\dot\tau>0$ this is already
the positive Lanczos gauge.  For $\dot\tau<0$, rephasing the transported basis
by $\ket{K_n(t)}=(-1)^nW(t)\ket{K_n^{Q,X}}$ changes each off-diagonal sign and
gives $b_n(t)=|\dot\tau|b_n^{Q,X}$.  At an isolated zero of $\dot\tau$ all
instantaneous hoppings vanish, and either neighboring phase convention
continues the probabilities.  This proves
Eq.~\eqref{eq:dressed-lanczos-coefficients} and its patchwise qualification.
Projecting $U\ket{\psi_Q(0)}$ on
$W(t)\ket{K_n^{Q,X}}$ then gives Eq.~\eqref{eq:dressed-amplitudes}.  If
$\dd\mu(E)=\bra\chi\dd P_X(E)\ket\chi$ is the spectral measure of $X$ at
$\ket\chi$, the spectral theorem gives
\begin{align}
 \dd\mu_Q(E)
 &=\frac{1}{\mathcal N_Q}
   \bra\chi Q^\dagger(X)\dd P_X(E)Q(X)\ket\chi\nonumber\\
 &=\frac{|Q(E)|^2}{\mathcal N_Q}\dd\mu(E).
\end{align}
This proves Eq.~\eqref{eq:time-dependent-christoffel}, including complex
polynomials.

For the displacement-dressed optical example, set
$X=a+a^\dagger$, $\ket\chi=\ket0$, and $Q(X)=X$.  Since
$X\ket0=\ket1$ and $\mathcal N_Q=1$, Eq.~\eqref{eq:dressed-amplitudes}
reduces to the photon-added fixed-generator problem at the effective time
$\Delta\tau$.  Inserting Eq.~\eqref{eq:photonaddedC} proves
Eq.~\eqref{eq:displacement-dressed-photon-added}.  The displacement path
$\alpha(t)$ is retained in the transported basis and in the laboratory
Hamiltonian, while the transported-chain probabilities depend only on
$\Delta\tau$.

Finally, on an interval with $G(t)>0$,
Eq.~\eqref{eq:vtime-hamiltonian} gives
$(H_V-i\partial_t)\ket g=G\ket B$.  Hence $a_0=0$, $b_1=G$, and
$\ket{K_1}=\ket B$.  The next residual is
\begin{align}
 \ket{A_2}
 &=(H_V-i\partial_t-a_1)\ket B-G\ket g\nonumber\\
 &=P_D(H_e-i\partial_t)\ket B,
\end{align}
which proves Eq.~\eqref{eq:vtime-chain}.  Substitution of
Eq.~\eqref{eq:bright-sphere} gives
$\|P_D\ket{\dot B}\|^2=
(\dot\vartheta^2+\sin^2\vartheta\,\dot\varphi_B^2)/4$.
Splitting the same residual into $d_{\rm dyn}+d_{\rm geo}$ gives
Eq.~\eqref{eq:v-interference}; both vectors acquire the same phase under
$\ket B\mapsto e^{i\lambda_B(t)}\ket B$, so their interference is gauge
invariant.  Moreover, $b_2(t)=0$ if and only if the residual itself vanishes
at that time.  When it vanishes throughout an interval, the Hermitian amplitude
generator decouples the ground--bright sector from the dark direction.
This proves Eq.~\eqref{eq:v-moving-termination} with the preparation and
continuation conditions stated in Sec.~\ref{sec:vtime}.

\renewcommand{\bibfont}{\fontsize{8}{9}\selectfont}

\bibliographystyle{apsrev4-2}
\bibliography{Krylov}

\begin{thebibliography}{74}%
\makeatletter
\providecommand \@ifxundefined [1]{%
 \@ifx{#1\undefined}
}%
\providecommand \@ifnum [1]{%
 \ifnum #1\expandafter \@firstoftwo
 \else \expandafter \@secondoftwo
 \fi
}%
\providecommand \@ifx [1]{%
 \ifx #1\expandafter \@firstoftwo
 \else \expandafter \@secondoftwo
 \fi
}%
\providecommand \natexlab [1]{#1}%
\providecommand \enquote  [1]{``#1''}%
\providecommand \bibnamefont  [1]{#1}%
\providecommand \bibfnamefont [1]{#1}%
\providecommand \citenamefont [1]{#1}%
\providecommand \href@noop [0]{\@secondoftwo}%
\providecommand \href [0]{\begingroup \@sanitize@url \@href}%
\providecommand \@href[1]{\@@startlink{#1}\@@href}%
\providecommand \@@href[1]{\endgroup#1\@@endlink}%
\providecommand \@sanitize@url [0]{\catcode `\\12\catcode `\$12\catcode
  `\&12\catcode `\#12\catcode `\^12\catcode `\_12\catcode `\%12\relax}%
\providecommand \@@startlink[1]{}%
\providecommand \@@endlink[0]{}%
\providecommand \url  [0]{\begingroup\@sanitize@url \@url }%
\providecommand \@url [1]{\endgroup\@href {#1}{\urlprefix }}%
\providecommand \urlprefix  [0]{URL }%
\providecommand \Eprint [0]{\href }%
\providecommand \doibase [0]{https://doi.org/}%
\providecommand \selectlanguage [0]{\@gobble}%
\providecommand \bibinfo  [0]{\@secondoftwo}%
\providecommand \bibfield  [0]{\@secondoftwo}%
\providecommand \translation [1]{[#1]}%
\providecommand \BibitemOpen [0]{}%
\providecommand \bibitemStop [0]{}%
\providecommand \bibitemNoStop [0]{.\EOS\space}%
\providecommand \EOS [0]{\spacefactor3000\relax}%
\providecommand \BibitemShut  [1]{\csname bibitem#1\endcsname}%
\let\auto@bib@innerbib\@empty
\bibitem [{\citenamefont {Balasubramanian}\ \emph {et~al.}(2022)\citenamefont
  {Balasubramanian}, \citenamefont {Caputa}, \citenamefont {Magan},\ and\
  \citenamefont {Wu}}]{Balasubramanian:2022tpr}%
  \BibitemOpen
  \bibfield  {author} {\bibinfo {author} {\bibfnamefont {V.}~\bibnamefont
  {Balasubramanian}}, \bibinfo {author} {\bibfnamefont {P.}~\bibnamefont
  {Caputa}}, \bibinfo {author} {\bibfnamefont {J.~M.}\ \bibnamefont {Magan}},\
  and\ \bibinfo {author} {\bibfnamefont {Q.}~\bibnamefont {Wu}},\ }\href
  {https://doi.org/10.1103/PhysRevD.106.046007} {\bibfield  {journal} {\bibinfo
   {journal} {Phys. Rev. D}\ }\textbf {\bibinfo {volume} {106}},\ \bibinfo
  {pages} {046007} (\bibinfo {year} {2022})}\BibitemShut {NoStop}%
\bibitem [{\citenamefont {Nandy}\ \emph {et~al.}(2025)\citenamefont {Nandy},
  \citenamefont {Matsoukas-Roubeas}, \citenamefont {Mart{\'\i}nez-Azcona},
  \citenamefont {Dymarsky},\ and\ \citenamefont {del Campo}}]{Nandy:2024htc}%
  \BibitemOpen
  \bibfield  {author} {\bibinfo {author} {\bibfnamefont {P.}~\bibnamefont
  {Nandy}}, \bibinfo {author} {\bibfnamefont {A.~S.}\ \bibnamefont
  {Matsoukas-Roubeas}}, \bibinfo {author} {\bibfnamefont {P.}~\bibnamefont
  {Mart{\'\i}nez-Azcona}}, \bibinfo {author} {\bibfnamefont {A.}~\bibnamefont
  {Dymarsky}},\ and\ \bibinfo {author} {\bibfnamefont {A.}~\bibnamefont {del
  Campo}},\ }\href {https://doi.org/10.1016/j.physrep.2025.05.001} {\bibfield
  {journal} {\bibinfo  {journal} {Phys. Rep.}\ }\textbf {\bibinfo {volume}
  {1125--1128}},\ \bibinfo {pages} {1} (\bibinfo {year} {2025})}\BibitemShut
  {NoStop}%
\bibitem [{\citenamefont {Takahashi}\ and\ \citenamefont {del
  Campo}(2025)}]{Takahashi:2024hex}%
  \BibitemOpen
  \bibfield  {author} {\bibinfo {author} {\bibfnamefont {K.}~\bibnamefont
  {Takahashi}}\ and\ \bibinfo {author} {\bibfnamefont {A.}~\bibnamefont {del
  Campo}},\ }\href {https://doi.org/10.1103/PhysRevLett.134.030401} {\bibfield
  {journal} {\bibinfo  {journal} {Phys. Rev. Lett.}\ }\textbf {\bibinfo
  {volume} {134}},\ \bibinfo {pages} {030401} (\bibinfo {year}
  {2025})}\BibitemShut {NoStop}%
\bibitem [{\citenamefont {Grabarits}\ \emph {et~al.}(2026)\citenamefont
  {Grabarits}, \citenamefont {Medina-Guerra},\ and\ \citenamefont {del
  Campo}}]{GrabaritsMedinaGuerraDelCampo2026}%
  \BibitemOpen
  \bibfield  {author} {\bibinfo {author} {\bibfnamefont {A.}~\bibnamefont
  {Grabarits}}, \bibinfo {author} {\bibfnamefont {E.}~\bibnamefont
  {Medina-Guerra}},\ and\ \bibinfo {author} {\bibfnamefont {A.}~\bibnamefont
  {del Campo}},\ }\href@noop {} {\bibinfo {title} {Krylov dynamics and operator
  growth in time-dependent systems via {L}ie algebras}} (\bibinfo {year}
  {2026}),\ \Eprint {https://arxiv.org/abs/2605.05290} {arXiv:2605.05290
  [quant-ph]} \BibitemShut {NoStop}%
\bibitem [{\citenamefont {Nizami}\ and\ \citenamefont
  {Shrestha}(2023)}]{Nizami}%
  \BibitemOpen
  \bibfield  {author} {\bibinfo {author} {\bibfnamefont {A.~A.}\ \bibnamefont
  {Nizami}}\ and\ \bibinfo {author} {\bibfnamefont {A.~W.}\ \bibnamefont
  {Shrestha}},\ }\href {https://doi.org/10.1103/PhysRevE.108.054222} {\bibfield
   {journal} {\bibinfo  {journal} {Phys. Rev. E}\ }\textbf {\bibinfo {volume}
  {108}},\ \bibinfo {pages} {054222} (\bibinfo {year} {2023})}\BibitemShut
  {NoStop}%
\bibitem [{\citenamefont {Faraji~Astaneh}\ and\ \citenamefont
  {Kafashi}(2026)}]{FarajiAstanehKafashi2026}%
  \BibitemOpen
  \bibfield  {author} {\bibinfo {author} {\bibfnamefont {A.}~\bibnamefont
  {Faraji~Astaneh}}\ and\ \bibinfo {author} {\bibfnamefont {P.}~\bibnamefont
  {Kafashi}},\ }\href@noop {} {\bibinfo {title} {Krylov complexity for
  time-dependent {H}amiltonians}} (\bibinfo {year} {2026}),\ \Eprint
  {https://arxiv.org/abs/2607.10454} {arXiv:2607.10454 [hep-th]} \BibitemShut
  {NoStop}%
\bibitem [{\citenamefont {Tavis}\ and\ \citenamefont
  {Cummings}(1968)}]{TavisCummings1968}%
  \BibitemOpen
  \bibfield  {author} {\bibinfo {author} {\bibfnamefont {M.}~\bibnamefont
  {Tavis}}\ and\ \bibinfo {author} {\bibfnamefont {F.~W.}\ \bibnamefont
  {Cummings}},\ }\href {https://doi.org/10.1103/PhysRev.170.379} {\bibfield
  {journal} {\bibinfo  {journal} {Phys. Rev.}\ }\textbf {\bibinfo {volume}
  {170}},\ \bibinfo {pages} {379} (\bibinfo {year} {1968})}\BibitemShut
  {NoStop}%
\bibitem [{\citenamefont {Walls}\ and\ \citenamefont
  {Milburn}(2008)}]{WallsMilburn2008}%
  \BibitemOpen
  \bibfield  {author} {\bibinfo {author} {\bibfnamefont {D.~F.}\ \bibnamefont
  {Walls}}\ and\ \bibinfo {author} {\bibfnamefont {G.~J.}\ \bibnamefont
  {Milburn}},\ }\href {https://doi.org/10.1007/978-3-540-28574-8} {\emph
  {\bibinfo {title} {Quantum Optics}}},\ \bibinfo {edition} {2nd}\ ed.\
  (\bibinfo  {publisher} {Springer},\ \bibinfo {address} {Berlin},\ \bibinfo
  {year} {2008})\BibitemShut {NoStop}%
\bibitem [{\citenamefont {Mollow}\ and\ \citenamefont
  {Glauber}(1967)}]{mollow1967quantum}%
  \BibitemOpen
  \bibfield  {author} {\bibinfo {author} {\bibfnamefont {B.~R.}\ \bibnamefont
  {Mollow}}\ and\ \bibinfo {author} {\bibfnamefont {R.~J.}\ \bibnamefont
  {Glauber}},\ }\href {https://doi.org/10.1103/PhysRev.160.1076} {\bibfield
  {journal} {\bibinfo  {journal} {Phys. Rev.}\ }\textbf {\bibinfo {volume}
  {160}},\ \bibinfo {pages} {1076} (\bibinfo {year} {1967})}\BibitemShut
  {NoStop}%
\bibitem [{\citenamefont {Yurke}\ \emph {et~al.}(1986)\citenamefont {Yurke},
  \citenamefont {McCall},\ and\ \citenamefont
  {Klauder}}]{YurkeMcCallKlauder1986}%
  \BibitemOpen
  \bibfield  {author} {\bibinfo {author} {\bibfnamefont {B.}~\bibnamefont
  {Yurke}}, \bibinfo {author} {\bibfnamefont {S.~L.}\ \bibnamefont {McCall}},\
  and\ \bibinfo {author} {\bibfnamefont {J.~R.}\ \bibnamefont {Klauder}},\
  }\href {https://doi.org/10.1103/PhysRevA.33.4033} {\bibfield  {journal}
  {\bibinfo  {journal} {Phys. Rev. A}\ }\textbf {\bibinfo {volume} {33}},\
  \bibinfo {pages} {4033} (\bibinfo {year} {1986})}\BibitemShut {NoStop}%
\bibitem [{\citenamefont {Morris}\ and\ \citenamefont
  {Shore}(1983)}]{MorrisShore1983}%
  \BibitemOpen
  \bibfield  {author} {\bibinfo {author} {\bibfnamefont {J.~R.}\ \bibnamefont
  {Morris}}\ and\ \bibinfo {author} {\bibfnamefont {B.~W.}\ \bibnamefont
  {Shore}},\ }\href {https://doi.org/10.1103/PhysRevA.27.906} {\bibfield
  {journal} {\bibinfo  {journal} {Phys. Rev. A}\ }\textbf {\bibinfo {volume}
  {27}},\ \bibinfo {pages} {906} (\bibinfo {year} {1983})}\BibitemShut
  {NoStop}%
\bibitem [{\citenamefont {Dong}\ and\ \citenamefont
  {Tang}(2002)}]{DongTang2002}%
  \BibitemOpen
  \bibfield  {author} {\bibinfo {author} {\bibfnamefont {P.}~\bibnamefont
  {Dong}}\ and\ \bibinfo {author} {\bibfnamefont {S.~H.}\ \bibnamefont
  {Tang}},\ }\href {https://doi.org/10.1103/PhysRevA.65.033816} {\bibfield
  {journal} {\bibinfo  {journal} {Phys. Rev. A}\ }\textbf {\bibinfo {volume}
  {65}},\ \bibinfo {pages} {033816} (\bibinfo {year} {2002})}\BibitemShut
  {NoStop}%
\bibitem [{\citenamefont {Espa\~nol}\ and\ \citenamefont
  {Wisniacki}(2023)}]{Espanol:2022cqr}%
  \BibitemOpen
  \bibfield  {author} {\bibinfo {author} {\bibfnamefont {B.~L.}\ \bibnamefont
  {Espa\~nol}}\ and\ \bibinfo {author} {\bibfnamefont {D.~A.}\ \bibnamefont
  {Wisniacki}},\ }\href {https://doi.org/10.1103/PhysRevE.107.024217}
  {\bibfield  {journal} {\bibinfo  {journal} {Phys. Rev. E}\ }\textbf {\bibinfo
  {volume} {107}},\ \bibinfo {pages} {024217} (\bibinfo {year}
  {2023})}\BibitemShut {NoStop}%
\bibitem [{\citenamefont {Chowdhury}\ and\ \citenamefont
  {Mahapatra}(2026)}]{ChowdhuryMahapatra2026}%
  \BibitemOpen
  \bibfield  {author} {\bibinfo {author} {\bibfnamefont {A.}~\bibnamefont
  {Chowdhury}}\ and\ \bibinfo {author} {\bibfnamefont {A.~P.}\ \bibnamefont
  {Mahapatra}},\ }\href@noop {} {\bibinfo {title} {Polynomial initial-state
  jumps and {C}hristoffel transforms in {K}rylov complexity}} (\bibinfo {year}
  {2026}),\ \Eprint {https://arxiv.org/abs/2607.05294} {arXiv:2607.05294
  [hep-th]} \BibitemShut {NoStop}%
\bibitem [{\citenamefont {Caputa}\ \emph {et~al.}(2022)\citenamefont {Caputa},
  \citenamefont {Magan},\ and\ \citenamefont {Patramanis}}]{Caputa:2021sib}%
  \BibitemOpen
  \bibfield  {author} {\bibinfo {author} {\bibfnamefont {P.}~\bibnamefont
  {Caputa}}, \bibinfo {author} {\bibfnamefont {J.~M.}\ \bibnamefont {Magan}},\
  and\ \bibinfo {author} {\bibfnamefont {D.}~\bibnamefont {Patramanis}},\
  }\href {https://doi.org/10.1103/PhysRevResearch.4.013041} {\bibfield
  {journal} {\bibinfo  {journal} {Phys. Rev. Res.}\ }\textbf {\bibinfo {volume}
  {4}},\ \bibinfo {pages} {013041} (\bibinfo {year} {2022})}\BibitemShut
  {NoStop}%
\bibitem [{\citenamefont {Berry}(1984)}]{Berry1984}%
  \BibitemOpen
  \bibfield  {author} {\bibinfo {author} {\bibfnamefont {M.~V.}\ \bibnamefont
  {Berry}},\ }\href {https://doi.org/10.1098/rspa.1984.0023} {\bibfield
  {journal} {\bibinfo  {journal} {Proc. R. Soc. Lond. A}\ }\textbf {\bibinfo
  {volume} {392}},\ \bibinfo {pages} {45} (\bibinfo {year} {1984})}\BibitemShut
  {NoStop}%
\bibitem [{\citenamefont {Aharonov}\ and\ \citenamefont
  {Anandan}(1987)}]{AharonovAnandan1987}%
  \BibitemOpen
  \bibfield  {author} {\bibinfo {author} {\bibfnamefont {Y.}~\bibnamefont
  {Aharonov}}\ and\ \bibinfo {author} {\bibfnamefont {J.}~\bibnamefont
  {Anandan}},\ }\href {https://doi.org/10.1103/PhysRevLett.58.1593} {\bibfield
  {journal} {\bibinfo  {journal} {Phys. Rev. Lett.}\ }\textbf {\bibinfo
  {volume} {58}},\ \bibinfo {pages} {1593} (\bibinfo {year}
  {1987})}\BibitemShut {NoStop}%
\bibitem [{\citenamefont {Zheng}\ \emph {et~al.}(2026)\citenamefont {Zheng},
  \citenamefont {He}, \citenamefont {Liu},\ and\ \citenamefont
  {Zhang}}]{ZhengHeLiuZhang2026}%
  \BibitemOpen
  \bibfield  {author} {\bibinfo {author} {\bibfnamefont {H.-Q.}\ \bibnamefont
  {Zheng}}, \bibinfo {author} {\bibfnamefont {P.-Z.}\ \bibnamefont {He}},
  \bibinfo {author} {\bibfnamefont {L.-H.}\ \bibnamefont {Liu}},\ and\ \bibinfo
  {author} {\bibfnamefont {H.-Q.}\ \bibnamefont {Zhang}},\ }\href@noop {}
  {\bibinfo {title} {Krylov complexity and {B}erry phase in quantum adiabatic
  dynamics}} (\bibinfo {year} {2026}),\ \Eprint
  {https://arxiv.org/abs/2608.14325} {arXiv:2608.14325 [quant-ph]} \BibitemShut
  {NoStop}%
\bibitem [{\citenamefont {Lutterbach}\ and\ \citenamefont
  {Davidovich}(1997)}]{LutterbachDavidovich1997}%
  \BibitemOpen
  \bibfield  {author} {\bibinfo {author} {\bibfnamefont {L.~G.}\ \bibnamefont
  {Lutterbach}}\ and\ \bibinfo {author} {\bibfnamefont {L.}~\bibnamefont
  {Davidovich}},\ }\href {https://doi.org/10.1103/PhysRevLett.78.2547}
  {\bibfield  {journal} {\bibinfo  {journal} {Phys. Rev. Lett.}\ }\textbf
  {\bibinfo {volume} {78}},\ \bibinfo {pages} {2547} (\bibinfo {year}
  {1997})}\BibitemShut {NoStop}%
\bibitem [{\citenamefont {Del\'eglise}\ \emph {et~al.}(2008)\citenamefont
  {Del\'eglise}, \citenamefont {Dotsenko}, \citenamefont {Sayrin},
  \citenamefont {Bernu}, \citenamefont {Brune}, \citenamefont {Raimond},\ and\
  \citenamefont {Haroche}}]{Deleglise2008}%
  \BibitemOpen
  \bibfield  {author} {\bibinfo {author} {\bibfnamefont {S.}~\bibnamefont
  {Del\'eglise}}, \bibinfo {author} {\bibfnamefont {I.}~\bibnamefont
  {Dotsenko}}, \bibinfo {author} {\bibfnamefont {C.}~\bibnamefont {Sayrin}},
  \bibinfo {author} {\bibfnamefont {J.}~\bibnamefont {Bernu}}, \bibinfo
  {author} {\bibfnamefont {M.}~\bibnamefont {Brune}}, \bibinfo {author}
  {\bibfnamefont {J.-M.}\ \bibnamefont {Raimond}},\ and\ \bibinfo {author}
  {\bibfnamefont {S.}~\bibnamefont {Haroche}},\ }\href
  {https://doi.org/10.1038/nature07288} {\bibfield  {journal} {\bibinfo
  {journal} {Nature}\ }\textbf {\bibinfo {volume} {455}},\ \bibinfo {pages}
  {510} (\bibinfo {year} {2008})}\BibitemShut {NoStop}%
\bibitem [{\citenamefont {Zlatanov}\ \emph {et~al.}(2022)\citenamefont
  {Zlatanov}, \citenamefont {Rangelov},\ and\ \citenamefont
  {Vitanov}}]{ZlatanovRangelovVitanov2022}%
  \BibitemOpen
  \bibfield  {author} {\bibinfo {author} {\bibfnamefont {K.~N.}\ \bibnamefont
  {Zlatanov}}, \bibinfo {author} {\bibfnamefont {A.~A.}\ \bibnamefont
  {Rangelov}},\ and\ \bibinfo {author} {\bibfnamefont {N.~V.}\ \bibnamefont
  {Vitanov}},\ }\href {https://doi.org/10.1088/1361-6455/ac8d3f} {\bibfield
  {journal} {\bibinfo  {journal} {J. Phys. B: At. Mol. Opt. Phys.}\ }\textbf
  {\bibinfo {volume} {55}},\ \bibinfo {pages} {204001} (\bibinfo {year}
  {2022})}\BibitemShut {NoStop}%
\bibitem [{\citenamefont {Vitanov}\ \emph {et~al.}(2017)\citenamefont
  {Vitanov}, \citenamefont {Rangelov}, \citenamefont {Shore},\ and\
  \citenamefont {Bergmann}}]{Vitanov2017}%
  \BibitemOpen
  \bibfield  {author} {\bibinfo {author} {\bibfnamefont {N.~V.}\ \bibnamefont
  {Vitanov}}, \bibinfo {author} {\bibfnamefont {A.~A.}\ \bibnamefont
  {Rangelov}}, \bibinfo {author} {\bibfnamefont {B.~W.}\ \bibnamefont
  {Shore}},\ and\ \bibinfo {author} {\bibfnamefont {K.}~\bibnamefont
  {Bergmann}},\ }\href {https://doi.org/10.1103/RevModPhys.89.015006}
  {\bibfield  {journal} {\bibinfo  {journal} {Rev. Mod. Phys.}\ }\textbf
  {\bibinfo {volume} {89}},\ \bibinfo {pages} {015006} (\bibinfo {year}
  {2017})}\BibitemShut {NoStop}%
\bibitem [{\citenamefont {Lanczos}(1950)}]{Lanczos1950}%
  \BibitemOpen
  \bibfield  {author} {\bibinfo {author} {\bibfnamefont {C.}~\bibnamefont
  {Lanczos}},\ }\href {https://doi.org/10.6028/jres.045.026} {\bibfield
  {journal} {\bibinfo  {journal} {J. Res. Natl. Bur. Stand.}\ }\textbf
  {\bibinfo {volume} {45}},\ \bibinfo {pages} {255} (\bibinfo {year}
  {1950})}\BibitemShut {NoStop}%
\bibitem [{\citenamefont {Howland}(1974)}]{Howland1974}%
  \BibitemOpen
  \bibfield  {author} {\bibinfo {author} {\bibfnamefont {J.~S.}\ \bibnamefont
  {Howland}},\ }\href {https://doi.org/10.1007/BF01351346} {\bibfield
  {journal} {\bibinfo  {journal} {Math. Ann.}\ }\textbf {\bibinfo {volume}
  {207}},\ \bibinfo {pages} {315} (\bibinfo {year} {1974})}\BibitemShut
  {NoStop}%
\bibitem [{\citenamefont {Peierls}(1933)}]{Peierls1933}%
  \BibitemOpen
  \bibfield  {author} {\bibinfo {author} {\bibfnamefont {R.}~\bibnamefont
  {Peierls}},\ }\href {https://doi.org/10.1007/BF01342591} {\bibfield
  {journal} {\bibinfo  {journal} {Z. Phys.}\ }\textbf {\bibinfo {volume}
  {80}},\ \bibinfo {pages} {763} (\bibinfo {year} {1933})}\BibitemShut
  {NoStop}%
\bibitem [{\citenamefont {Wei}\ and\ \citenamefont
  {Norman}(1963)}]{WeiNorman1963}%
  \BibitemOpen
  \bibfield  {author} {\bibinfo {author} {\bibfnamefont {J.}~\bibnamefont
  {Wei}}\ and\ \bibinfo {author} {\bibfnamefont {E.}~\bibnamefont {Norman}},\
  }\href {https://doi.org/10.1063/1.1703993} {\bibfield  {journal} {\bibinfo
  {journal} {J. Math. Phys.}\ }\textbf {\bibinfo {volume} {4}},\ \bibinfo
  {pages} {575} (\bibinfo {year} {1963})}\BibitemShut {NoStop}%
\bibitem [{\citenamefont {Ban}(1993)}]{Ban:93}%
  \BibitemOpen
  \bibfield  {author} {\bibinfo {author} {\bibfnamefont {M.}~\bibnamefont
  {Ban}},\ }\href {https://doi.org/10.1364/JOSAB.10.001347} {\bibfield
  {journal} {\bibinfo  {journal} {J. Opt. Soc. Am. B}\ }\textbf {\bibinfo
  {volume} {10}},\ \bibinfo {pages} {1347} (\bibinfo {year}
  {1993})}\BibitemShut {NoStop}%
\bibitem [{\citenamefont {Perelomov}(1972)}]{perelomov1972coherent}%
  \BibitemOpen
  \bibfield  {author} {\bibinfo {author} {\bibfnamefont {A.~M.}\ \bibnamefont
  {Perelomov}},\ }\href {https://doi.org/10.1007/BF01645091} {\bibfield
  {journal} {\bibinfo  {journal} {Communications in Mathematical Physics}\
  }\textbf {\bibinfo {volume} {26}},\ \bibinfo {pages} {222} (\bibinfo {year}
  {1972})}\BibitemShut {NoStop}%
\bibitem [{\citenamefont {Glauber}(1963)}]{Glauber1963}%
  \BibitemOpen
  \bibfield  {author} {\bibinfo {author} {\bibfnamefont {R.~J.}\ \bibnamefont
  {Glauber}},\ }\href {https://doi.org/10.1103/PhysRev.131.2766} {\bibfield
  {journal} {\bibinfo  {journal} {Phys. Rev.}\ }\textbf {\bibinfo {volume}
  {131}},\ \bibinfo {pages} {2766} (\bibinfo {year} {1963})}\BibitemShut
  {NoStop}%
\bibitem [{\citenamefont {Seetharaman}\ \emph {et~al.}(2025)\citenamefont
  {Seetharaman}, \citenamefont {Singh},\ and\ \citenamefont
  {Nath}}]{SeetharamanSinghNath2025}%
  \BibitemOpen
  \bibfield  {author} {\bibinfo {author} {\bibfnamefont {S.}~\bibnamefont
  {Seetharaman}}, \bibinfo {author} {\bibfnamefont {C.}~\bibnamefont {Singh}},\
  and\ \bibinfo {author} {\bibfnamefont {R.}~\bibnamefont {Nath}},\ }\href
  {https://doi.org/10.1103/PhysRevD.111.076014} {\bibfield  {journal} {\bibinfo
   {journal} {Phys. Rev. D}\ }\textbf {\bibinfo {volume} {111}},\ \bibinfo
  {pages} {076014} (\bibinfo {year} {2025})}\BibitemShut {NoStop}%
\bibitem [{\citenamefont {Cafaro}\ \emph {et~al.}(2026)\citenamefont {Cafaro},
  \citenamefont {Clements},\ and\ \citenamefont
  {Anuboyina}}]{CafaroClementsAnuboyina2026}%
  \BibitemOpen
  \bibfield  {author} {\bibinfo {author} {\bibfnamefont {C.}~\bibnamefont
  {Cafaro}}, \bibinfo {author} {\bibfnamefont {E.}~\bibnamefont {Clements}},\
  and\ \bibinfo {author} {\bibfnamefont {V.~V.}\ \bibnamefont {Anuboyina}},\
  }\href {https://doi.org/10.1140/epjp/s13360-026-07857-5} {\bibfield
  {journal} {\bibinfo  {journal} {Eur. Phys. J. Plus}\ }\textbf {\bibinfo
  {volume} {141}},\ \bibinfo {pages} {639} (\bibinfo {year}
  {2026})}\BibitemShut {NoStop}%
\bibitem [{\citenamefont {Fu}\ \emph {et~al.}(2025)\citenamefont {Fu},
  \citenamefont {Kim}, \citenamefont {Pal},\ and\ \citenamefont
  {Pal}}]{FuKimPalPal2024}%
  \BibitemOpen
  \bibfield  {author} {\bibinfo {author} {\bibfnamefont {Y.}~\bibnamefont
  {Fu}}, \bibinfo {author} {\bibfnamefont {K.-Y.}\ \bibnamefont {Kim}},
  \bibinfo {author} {\bibfnamefont {K.}~\bibnamefont {Pal}},\ and\ \bibinfo
  {author} {\bibfnamefont {K.}~\bibnamefont {Pal}},\ }\href
  {https://doi.org/10.1007/JHEP06(2025)139} {\bibfield  {journal} {\bibinfo
  {journal} {JHEP}\ }\textbf {\bibinfo {volume} {2025}}\bibinfo  {number} {
  (06)},\ \bibinfo {pages} {139}}\BibitemShut {NoStop}%
\bibitem [{\citenamefont {Chowdhury}(2026)}]{SChowdhury2026}%
  \BibitemOpen
\bibfield  {number} {  }\bibfield  {author} {\bibinfo {author} {\bibfnamefont
  {S.}~\bibnamefont {Chowdhury}},\ }\href
  {https://doi.org/10.1007/JHEP06(2026)091} {\bibfield  {journal} {\bibinfo
  {journal} {JHEP}\ }\textbf {\bibinfo {volume} {2026}}\bibinfo  {number} {
  (06)},\ \bibinfo {pages} {091}}\BibitemShut {NoStop}%
\bibitem [{\citenamefont {Haque}\ \emph {et~al.}(2024)\citenamefont {Haque},
  \citenamefont {Murugan}, \citenamefont {Tladi},\ and\ \citenamefont
  {Van~Zyl}}]{HaqueJacobi2024}%
  \BibitemOpen
\bibfield  {number} {  }\bibfield  {author} {\bibinfo {author} {\bibfnamefont
  {S.~S.}\ \bibnamefont {Haque}}, \bibinfo {author} {\bibfnamefont
  {J.}~\bibnamefont {Murugan}}, \bibinfo {author} {\bibfnamefont
  {M.}~\bibnamefont {Tladi}},\ and\ \bibinfo {author} {\bibfnamefont
  {H.~J.~R.}\ \bibnamefont {Van~Zyl}},\ }\href
  {https://doi.org/10.1007/JHEP05(2024)220} {\bibfield  {journal} {\bibinfo
  {journal} {JHEP}\ }\textbf {\bibinfo {volume} {2024}}\bibinfo  {number} {
  (05)},\ \bibinfo {pages} {220}}\BibitemShut {NoStop}%
\bibitem [{\citenamefont {Mandel}(1979)}]{Mandel1979}%
  \BibitemOpen
\bibfield  {number} {  }\bibfield  {author} {\bibinfo {author} {\bibfnamefont
  {L.}~\bibnamefont {Mandel}},\ }\href {https://doi.org/10.1364/OL.4.000205}
  {\bibfield  {journal} {\bibinfo  {journal} {Opt. Lett.}\ }\textbf {\bibinfo
  {volume} {4}},\ \bibinfo {pages} {205} (\bibinfo {year} {1979})}\BibitemShut
  {NoStop}%
\bibitem [{\citenamefont {Chakraborty}(2026)}]{Chakraborty2026}%
  \BibitemOpen
  \bibfield  {author} {\bibinfo {author} {\bibfnamefont {D.}~\bibnamefont
  {Chakraborty}},\ }\href@noop {} {\bibinfo {title} {Krylov complexity from
  {L}oschmidt amplitude}} (\bibinfo {year} {2026}),\ \Eprint
  {https://arxiv.org/abs/2607.10921} {arXiv:2607.10921 [hep-th]} \BibitemShut
  {NoStop}%
\bibitem [{\citenamefont {Gerry}(1989)}]{Gerry1989}%
  \BibitemOpen
  \bibfield  {author} {\bibinfo {author} {\bibfnamefont {C.~C.}\ \bibnamefont
  {Gerry}},\ }\href {https://doi.org/10.1103/PhysRevA.39.3204} {\bibfield
  {journal} {\bibinfo  {journal} {Phys. Rev. A}\ }\textbf {\bibinfo {volume}
  {39}},\ \bibinfo {pages} {3204} (\bibinfo {year} {1989})}\BibitemShut
  {NoStop}%
\bibitem [{\citenamefont {Vega}\ \emph {et~al.}(2024)\citenamefont {Vega},
  \citenamefont {Chore{\~n}o}, \citenamefont {Ojeda-Guill{\'e}n},\ and\
  \citenamefont {Mota}}]{VegaChorenoOjedaGuillenMota2024}%
  \BibitemOpen
  \bibfield  {author} {\bibinfo {author} {\bibfnamefont {J.~C.}\ \bibnamefont
  {Vega}}, \bibinfo {author} {\bibfnamefont {E.}~\bibnamefont {Chore{\~n}o}},
  \bibinfo {author} {\bibfnamefont {D.}~\bibnamefont {Ojeda-Guill{\'e}n}},\
  and\ \bibinfo {author} {\bibfnamefont {R.~D.}\ \bibnamefont {Mota}},\ }\href
  {https://doi.org/10.1364/JOSAB.517533} {\bibfield  {journal} {\bibinfo
  {journal} {J. Opt. Soc. Am. B}\ }\textbf {\bibinfo {volume} {41}},\ \bibinfo
  {pages} {1084} (\bibinfo {year} {2024})}\BibitemShut {NoStop}%
\bibitem [{\citenamefont {Aguilar-Gutierrez}\ \emph {et~al.}(2025)\citenamefont
  {Aguilar-Gutierrez}, \citenamefont {Fu}, \citenamefont {Pal},\ and\
  \citenamefont {Parmentier}}]{AguilarGutierrez2025}%
  \BibitemOpen
  \bibfield  {author} {\bibinfo {author} {\bibfnamefont {S.~E.}\ \bibnamefont
  {Aguilar-Gutierrez}}, \bibinfo {author} {\bibfnamefont {Y.}~\bibnamefont
  {Fu}}, \bibinfo {author} {\bibfnamefont {K.}~\bibnamefont {Pal}},\ and\
  \bibinfo {author} {\bibfnamefont {K.}~\bibnamefont {Parmentier}},\ }\href
  {https://doi.org/10.1007/JHEP09(2025)039} {\bibfield  {journal} {\bibinfo
  {journal} {JHEP}\ }\textbf {\bibinfo {volume} {2025}}\bibinfo  {number} {
  (09)},\ \bibinfo {pages} {039}}\BibitemShut {NoStop}%
\bibitem [{\citenamefont {Szeg\H{o}}(1939)}]{Szego1939}%
  \BibitemOpen
\bibfield  {number} {  }\bibfield  {author} {\bibinfo {author} {\bibfnamefont
  {G.}~\bibnamefont {Szeg\H{o}}},\ }\href@noop {} {\emph {\bibinfo {title}
  {Orthogonal Polynomials}}},\ \bibinfo {series} {American Mathematical Society
  Colloquium Publications}, Vol.~\bibinfo {volume} {23}\ (\bibinfo  {publisher}
  {American Mathematical Society},\ \bibinfo {address} {Providence, RI},\
  \bibinfo {year} {1939})\BibitemShut {NoStop}%
\bibitem [{\citenamefont {Balasubramanian}\ \emph {et~al.}(2026)\citenamefont
  {Balasubramanian}, \citenamefont {Caputa},\ and\ \citenamefont
  {Sim\'on}}]{BalasubramanianCaputaSimon2026}%
  \BibitemOpen
  \bibfield  {author} {\bibinfo {author} {\bibfnamefont {V.}~\bibnamefont
  {Balasubramanian}}, \bibinfo {author} {\bibfnamefont {P.}~\bibnamefont
  {Caputa}},\ and\ \bibinfo {author} {\bibfnamefont {J.}~\bibnamefont
  {Sim\'on}},\ }\href {https://doi.org/10.1007/JHEP04(2026)172} {\bibfield
  {journal} {\bibinfo  {journal} {JHEP}\ }\textbf {\bibinfo {volume}
  {2026}}\bibinfo  {number} { (04)},\ \bibinfo {pages} {172}}\BibitemShut
  {NoStop}%
\bibitem [{\citenamefont {McCall}\ and\ \citenamefont
  {Hahn}(1969)}]{McCallHahn1969}%
  \BibitemOpen
\bibfield  {number} {  }\bibfield  {author} {\bibinfo {author} {\bibfnamefont
  {S.~L.}\ \bibnamefont {McCall}}\ and\ \bibinfo {author} {\bibfnamefont
  {E.~L.}\ \bibnamefont {Hahn}},\ }\href
  {https://doi.org/10.1103/PhysRev.183.457} {\bibfield  {journal} {\bibinfo
  {journal} {Phys. Rev.}\ }\textbf {\bibinfo {volume} {183}},\ \bibinfo {pages}
  {457} (\bibinfo {year} {1969})}\BibitemShut {NoStop}%
\bibitem [{\citenamefont {Robertson}(1929)}]{Robertson1929}%
  \BibitemOpen
  \bibfield  {author} {\bibinfo {author} {\bibfnamefont {H.~P.}\ \bibnamefont
  {Robertson}},\ }\href {https://doi.org/10.1103/PhysRev.34.163} {\bibfield
  {journal} {\bibinfo  {journal} {Phys. Rev.}\ }\textbf {\bibinfo {volume}
  {34}},\ \bibinfo {pages} {163} (\bibinfo {year} {1929})}\BibitemShut
  {NoStop}%
\bibitem [{\citenamefont {Mandelstam}\ and\ \citenamefont
  {Tamm}(1945)}]{MandelstamTamm1945}%
  \BibitemOpen
  \bibfield  {author} {\bibinfo {author} {\bibfnamefont {L.}~\bibnamefont
  {Mandelstam}}\ and\ \bibinfo {author} {\bibfnamefont {I.}~\bibnamefont
  {Tamm}},\ }\href@noop {} {\bibfield  {journal} {\bibinfo  {journal} {J. Phys.
  (USSR)}\ }\textbf {\bibinfo {volume} {9}},\ \bibinfo {pages} {249} (\bibinfo
  {year} {1945})}\BibitemShut {NoStop}%
\bibitem [{\citenamefont {H\"ornedal}\ \emph {et~al.}(2022)\citenamefont
  {H\"ornedal}, \citenamefont {Carabba}, \citenamefont {Matsoukas-Roubeas},\
  and\ \citenamefont {del Campo}}]{HornedalCarabbaMatsoukasDelCampo2022}%
  \BibitemOpen
  \bibfield  {author} {\bibinfo {author} {\bibfnamefont {N.}~\bibnamefont
  {H\"ornedal}}, \bibinfo {author} {\bibfnamefont {N.}~\bibnamefont {Carabba}},
  \bibinfo {author} {\bibfnamefont {A.~S.}\ \bibnamefont {Matsoukas-Roubeas}},\
  and\ \bibinfo {author} {\bibfnamefont {A.}~\bibnamefont {del Campo}},\ }\href
  {https://doi.org/10.1038/s42005-022-00985-1} {\bibfield  {journal} {\bibinfo
  {journal} {Commun. Phys.}\ }\textbf {\bibinfo {volume} {5}},\ \bibinfo
  {pages} {207} (\bibinfo {year} {2022})}\BibitemShut {NoStop}%
\bibitem [{\citenamefont {Demkov}(1964)}]{Demkov1964}%
  \BibitemOpen
  \bibfield  {author} {\bibinfo {author} {\bibfnamefont {Y.~N.}\ \bibnamefont
  {Demkov}},\ }\href@noop {} {\bibfield  {journal} {\bibinfo  {journal} {Sov.
  Phys. JETP}\ }\textbf {\bibinfo {volume} {18}},\ \bibinfo {pages} {138}
  (\bibinfo {year} {1964})}\BibitemShut {NoStop}%
\bibitem [{\citenamefont {Redfield}(1955)}]{Redfield1955}%
  \BibitemOpen
  \bibfield  {author} {\bibinfo {author} {\bibfnamefont {A.~G.}\ \bibnamefont
  {Redfield}},\ }\href {https://doi.org/10.1103/PhysRev.98.1787} {\bibfield
  {journal} {\bibinfo  {journal} {Phys. Rev.}\ }\textbf {\bibinfo {volume}
  {98}},\ \bibinfo {pages} {1787} (\bibinfo {year} {1955})}\BibitemShut
  {NoStop}%
\bibitem [{\citenamefont {Zener}(1932)}]{Zener1932}%
  \BibitemOpen
  \bibfield  {author} {\bibinfo {author} {\bibfnamefont {C.}~\bibnamefont
  {Zener}},\ }\href {https://doi.org/10.1098/rspa.1932.0165} {\bibfield
  {journal} {\bibinfo  {journal} {Proc. R. Soc. Lond. A}\ }\textbf {\bibinfo
  {volume} {137}},\ \bibinfo {pages} {696} (\bibinfo {year}
  {1932})}\BibitemShut {NoStop}%
\bibitem [{\citenamefont {Grabarits}\ and\ \citenamefont {del
  Campo}(2025)}]{GrabaritsDelCampo2025}%
  \BibitemOpen
  \bibfield  {author} {\bibinfo {author} {\bibfnamefont {A.}~\bibnamefont
  {Grabarits}}\ and\ \bibinfo {author} {\bibfnamefont {A.}~\bibnamefont {del
  Campo}},\ }\href@noop {} {\bibinfo {title} {Universal growth of {K}rylov
  complexity across a quantum phase transition}} (\bibinfo {year} {2025}),\
  \Eprint {https://arxiv.org/abs/2510.13947} {arXiv:2510.13947 [quant-ph]}
  \BibitemShut {NoStop}%
\bibitem [{\citenamefont {Campos}\ \emph {et~al.}(1989)\citenamefont {Campos},
  \citenamefont {Saleh},\ and\ \citenamefont {Teich}}]{CamposSalehTeich1989}%
  \BibitemOpen
  \bibfield  {author} {\bibinfo {author} {\bibfnamefont {R.~A.}\ \bibnamefont
  {Campos}}, \bibinfo {author} {\bibfnamefont {B.~E.~A.}\ \bibnamefont
  {Saleh}},\ and\ \bibinfo {author} {\bibfnamefont {M.~C.}\ \bibnamefont
  {Teich}},\ }\href {https://doi.org/10.1103/PhysRevA.40.1371} {\bibfield
  {journal} {\bibinfo  {journal} {Phys. Rev. A}\ }\textbf {\bibinfo {volume}
  {40}},\ \bibinfo {pages} {1371} (\bibinfo {year} {1989})}\BibitemShut
  {NoStop}%
\bibitem [{\citenamefont {Agarwal}\ and\ \citenamefont
  {Tara}(1991)}]{AgarwalTara1991}%
  \BibitemOpen
  \bibfield  {author} {\bibinfo {author} {\bibfnamefont {G.~S.}\ \bibnamefont
  {Agarwal}}\ and\ \bibinfo {author} {\bibfnamefont {K.}~\bibnamefont {Tara}},\
  }\href {https://doi.org/10.1103/PhysRevA.43.492} {\bibfield  {journal}
  {\bibinfo  {journal} {Phys. Rev. A}\ }\textbf {\bibinfo {volume} {43}},\
  \bibinfo {pages} {492} (\bibinfo {year} {1991})}\BibitemShut {NoStop}%
\bibitem [{\citenamefont {Zavatta}\ \emph {et~al.}(2004)\citenamefont
  {Zavatta}, \citenamefont {Viciani},\ and\ \citenamefont
  {Bellini}}]{Zavatta2004}%
  \BibitemOpen
  \bibfield  {author} {\bibinfo {author} {\bibfnamefont {A.}~\bibnamefont
  {Zavatta}}, \bibinfo {author} {\bibfnamefont {S.}~\bibnamefont {Viciani}},\
  and\ \bibinfo {author} {\bibfnamefont {M.}~\bibnamefont {Bellini}},\ }\href
  {https://doi.org/10.1126/science.1103190} {\bibfield  {journal} {\bibinfo
  {journal} {Science}\ }\textbf {\bibinfo {volume} {306}},\ \bibinfo {pages}
  {660} (\bibinfo {year} {2004})}\BibitemShut {NoStop}%
\bibitem [{\citenamefont {Pal}\ \emph {et~al.}(2026)\citenamefont {Pal},
  \citenamefont {Pal},\ and\ \citenamefont {Kim}}]{PalPalKim2026}%
  \BibitemOpen
  \bibfield  {author} {\bibinfo {author} {\bibfnamefont {K.}~\bibnamefont
  {Pal}}, \bibinfo {author} {\bibfnamefont {K.}~\bibnamefont {Pal}},\ and\
  \bibinfo {author} {\bibfnamefont {K.-Y.}\ \bibnamefont {Kim}},\ }\href
  {https://doi.org/10.1007/JHEP08(2026)187} {\bibfield  {journal} {\bibinfo
  {journal} {JHEP}\ }\textbf {\bibinfo {volume} {2026}}\bibinfo  {number} {
  (08)},\ \bibinfo {pages} {187}}\BibitemShut {NoStop}%
\bibitem [{\citenamefont {Royer}(1977)}]{Royer1977}%
  \BibitemOpen
\bibfield  {number} {  }\bibfield  {author} {\bibinfo {author} {\bibfnamefont
  {A.}~\bibnamefont {Royer}},\ }\href {https://doi.org/10.1103/PhysRevA.15.449}
  {\bibfield  {journal} {\bibinfo  {journal} {Phys. Rev. A}\ }\textbf {\bibinfo
  {volume} {15}},\ \bibinfo {pages} {449} (\bibinfo {year} {1977})}\BibitemShut
  {NoStop}%
\bibitem [{\citenamefont {Akhtar}\ \emph {et~al.}(2021)\citenamefont {Akhtar},
  \citenamefont {Sanders},\ and\ \citenamefont {Navarrete-Benlloch}}]{Akhtar}%
  \BibitemOpen
  \bibfield  {author} {\bibinfo {author} {\bibfnamefont {N.}~\bibnamefont
  {Akhtar}}, \bibinfo {author} {\bibfnamefont {B.~C.}\ \bibnamefont
  {Sanders}},\ and\ \bibinfo {author} {\bibfnamefont {C.}~\bibnamefont
  {Navarrete-Benlloch}},\ }\href {https://doi.org/10.1103/PhysRevA.103.053711}
  {\bibfield  {journal} {\bibinfo  {journal} {Phys. Rev. A}\ }\textbf {\bibinfo
  {volume} {103}},\ \bibinfo {pages} {053711} (\bibinfo {year}
  {2021})}\BibitemShut {NoStop}%
\bibitem [{\citenamefont {Schlegel}\ \emph {et~al.}(2022)\citenamefont
  {Schlegel}, \citenamefont {Minganti},\ and\ \citenamefont
  {Savona}}]{SchlegelMingantiSavona2022}%
  \BibitemOpen
  \bibfield  {author} {\bibinfo {author} {\bibfnamefont {D.~S.}\ \bibnamefont
  {Schlegel}}, \bibinfo {author} {\bibfnamefont {F.}~\bibnamefont {Minganti}},\
  and\ \bibinfo {author} {\bibfnamefont {V.}~\bibnamefont {Savona}},\ }\href
  {https://doi.org/10.1103/PhysRevA.106.022431} {\bibfield  {journal} {\bibinfo
   {journal} {Phys. Rev. A}\ }\textbf {\bibinfo {volume} {106}},\ \bibinfo
  {pages} {022431} (\bibinfo {year} {2022})}\BibitemShut {NoStop}%
\bibitem [{\citenamefont {Bollinger}\ \emph {et~al.}(1996)\citenamefont
  {Bollinger}, \citenamefont {Itano}, \citenamefont {Wineland},\ and\
  \citenamefont {Heinzen}}]{Bollinger1996}%
  \BibitemOpen
  \bibfield  {author} {\bibinfo {author} {\bibfnamefont {J.~J.}\ \bibnamefont
  {Bollinger}}, \bibinfo {author} {\bibfnamefont {W.~M.}\ \bibnamefont
  {Itano}}, \bibinfo {author} {\bibfnamefont {D.~J.}\ \bibnamefont
  {Wineland}},\ and\ \bibinfo {author} {\bibfnamefont {D.~J.}\ \bibnamefont
  {Heinzen}},\ }\href {https://doi.org/10.1103/PhysRevA.54.R4649} {\bibfield
  {journal} {\bibinfo  {journal} {Phys. Rev. A}\ }\textbf {\bibinfo {volume}
  {54}},\ \bibinfo {pages} {R4649} (\bibinfo {year} {1996})}\BibitemShut
  {NoStop}%
\bibitem [{\citenamefont {Fox}\ and\ \citenamefont
  {Vela-Arevalo}(2002)}]{FoxVelaArevalo2002}%
  \BibitemOpen
  \bibfield  {author} {\bibinfo {author} {\bibfnamefont {R.~F.}\ \bibnamefont
  {Fox}}\ and\ \bibinfo {author} {\bibfnamefont {L.~V.}\ \bibnamefont
  {Vela-Arevalo}},\ }\href {https://doi.org/10.1103/PhysRevA.66.053402}
  {\bibfield  {journal} {\bibinfo  {journal} {Phys. Rev. A}\ }\textbf {\bibinfo
  {volume} {66}},\ \bibinfo {pages} {053402} (\bibinfo {year}
  {2002})}\BibitemShut {NoStop}%
\bibitem [{\citenamefont {Yates}\ and\ \citenamefont
  {Mitra}(2021)}]{YatesMitra2021}%
  \BibitemOpen
  \bibfield  {author} {\bibinfo {author} {\bibfnamefont {D.~J.}\ \bibnamefont
  {Yates}}\ and\ \bibinfo {author} {\bibfnamefont {A.}~\bibnamefont {Mitra}},\
  }\href {https://doi.org/10.1103/PhysRevB.104.195121} {\bibfield  {journal}
  {\bibinfo  {journal} {Phys. Rev. B}\ }\textbf {\bibinfo {volume} {104}},\
  \bibinfo {pages} {195121} (\bibinfo {year} {2021})}\BibitemShut {NoStop}%
\bibitem [{\citenamefont {Kolganov}\ and\ \citenamefont
  {Trunin}(2025)}]{KolganovTrunin2025}%
  \BibitemOpen
  \bibfield  {author} {\bibinfo {author} {\bibfnamefont {N.}~\bibnamefont
  {Kolganov}}\ and\ \bibinfo {author} {\bibfnamefont {D.~A.}\ \bibnamefont
  {Trunin}},\ }\href {https://doi.org/10.1103/PhysRevE.111.L052202} {\bibfield
  {journal} {\bibinfo  {journal} {Phys. Rev. E}\ }\textbf {\bibinfo {volume}
  {111}},\ \bibinfo {pages} {L052202} (\bibinfo {year} {2025})}\BibitemShut
  {NoStop}%
\bibitem [{\citenamefont {Yeh}\ and\ \citenamefont
  {Mitra}(2026)}]{YehMitra2026}%
  \BibitemOpen
  \bibfield  {author} {\bibinfo {author} {\bibfnamefont {H.-C.}\ \bibnamefont
  {Yeh}}\ and\ \bibinfo {author} {\bibfnamefont {A.}~\bibnamefont {Mitra}},\
  }\href {https://doi.org/10.1103/jp5f-wpkq} {\bibfield  {journal} {\bibinfo
  {journal} {Phys. Rev. B}\ }\textbf {\bibinfo {volume} {113}},\ \bibinfo
  {pages} {024308} (\bibinfo {year} {2026})}\BibitemShut {NoStop}%
\bibitem [{\citenamefont {{de Oliveira}}\ \emph {et~al.}(1990)\citenamefont
  {{de Oliveira}}, \citenamefont {Kim}, \citenamefont {Knight},\ and\
  \citenamefont {Bu{\v z}ek}}]{deOliveiraEtAl1990}%
  \BibitemOpen
  \bibfield  {author} {\bibinfo {author} {\bibfnamefont {F.~A.~M.}\
  \bibnamefont {{de Oliveira}}}, \bibinfo {author} {\bibfnamefont {M.~S.}\
  \bibnamefont {Kim}}, \bibinfo {author} {\bibfnamefont {P.~L.}\ \bibnamefont
  {Knight}},\ and\ \bibinfo {author} {\bibfnamefont {V.}~\bibnamefont {Bu{\v
  z}ek}},\ }\href {https://doi.org/10.1103/PhysRevA.41.2645} {\bibfield
  {journal} {\bibinfo  {journal} {Phys. Rev. A}\ }\textbf {\bibinfo {volume}
  {41}},\ \bibinfo {pages} {2645} (\bibinfo {year} {1990})}\BibitemShut
  {NoStop}%
\bibitem [{\citenamefont {Wolf}\ \emph {et~al.}(2019)\citenamefont {Wolf},
  \citenamefont {Shi}, \citenamefont {Heip}, \citenamefont {Gessner},
  \citenamefont {Pezz\`e}, \citenamefont {Smerzi}, \citenamefont {Schulte},
  \citenamefont {Hammerer},\ and\ \citenamefont {Schmidt}}]{WolfEtAl2019}%
  \BibitemOpen
  \bibfield  {author} {\bibinfo {author} {\bibfnamefont {F.}~\bibnamefont
  {Wolf}}, \bibinfo {author} {\bibfnamefont {C.}~\bibnamefont {Shi}}, \bibinfo
  {author} {\bibfnamefont {J.~C.}\ \bibnamefont {Heip}}, \bibinfo {author}
  {\bibfnamefont {M.}~\bibnamefont {Gessner}}, \bibinfo {author} {\bibfnamefont
  {L.}~\bibnamefont {Pezz\`e}}, \bibinfo {author} {\bibfnamefont
  {A.}~\bibnamefont {Smerzi}}, \bibinfo {author} {\bibfnamefont
  {M.}~\bibnamefont {Schulte}}, \bibinfo {author} {\bibfnamefont
  {K.}~\bibnamefont {Hammerer}},\ and\ \bibinfo {author} {\bibfnamefont
  {P.~O.}\ \bibnamefont {Schmidt}},\ }\href
  {https://doi.org/10.1038/s41467-019-10576-4} {\bibfield  {journal} {\bibinfo
  {journal} {Nat. Commun.}\ }\textbf {\bibinfo {volume} {10}},\ \bibinfo
  {pages} {2929} (\bibinfo {year} {2019})}\BibitemShut {NoStop}%
\bibitem [{\citenamefont {Takahashi}\ and\ \citenamefont
  {Umezawa}(1996)}]{Takahashi1996}%
  \BibitemOpen
  \bibfield  {author} {\bibinfo {author} {\bibfnamefont {Y.}~\bibnamefont
  {Takahashi}}\ and\ \bibinfo {author} {\bibfnamefont {H.}~\bibnamefont
  {Umezawa}},\ }\href {https://doi.org/10.1142/S0217979296000817} {\bibfield
  {journal} {\bibinfo  {journal} {Int. J. Mod. Phys. B}\ }\textbf {\bibinfo
  {volume} {10}},\ \bibinfo {pages} {1755} (\bibinfo {year}
  {1996})}\BibitemShut {NoStop}%
\bibitem [{\citenamefont {Zhai}\ \emph {et~al.}(2026)\citenamefont {Zhai},
  \citenamefont {Liu},\ and\ \citenamefont {Zhang}}]{ZhaiLiuZhang2026}%
  \BibitemOpen
  \bibfield  {author} {\bibinfo {author} {\bibfnamefont {K.-H.}\ \bibnamefont
  {Zhai}}, \bibinfo {author} {\bibfnamefont {L.-H.}\ \bibnamefont {Liu}},\ and\
  \bibinfo {author} {\bibfnamefont {H.-Q.}\ \bibnamefont {Zhang}},\ }\href
  {https://doi.org/10.1016/j.aop.2026.170534} {\bibfield  {journal} {\bibinfo
  {journal} {Ann. Phys.}\ }\textbf {\bibinfo {volume} {491}},\ \bibinfo {pages}
  {170534} (\bibinfo {year} {2026})}\BibitemShut {NoStop}%
\bibitem [{\citenamefont {Tibaduiza}\ \emph {et~al.}(2020)\citenamefont
  {Tibaduiza}, \citenamefont {Pires}, \citenamefont {Szilard}, \citenamefont
  {Zarro}, \citenamefont {Farina},\ and\ \citenamefont {Rego}}]{Tibaduiza}%
  \BibitemOpen
  \bibfield  {author} {\bibinfo {author} {\bibfnamefont {D.~M.}\ \bibnamefont
  {Tibaduiza}}, \bibinfo {author} {\bibfnamefont {L.}~\bibnamefont {Pires}},
  \bibinfo {author} {\bibfnamefont {D.}~\bibnamefont {Szilard}}, \bibinfo
  {author} {\bibfnamefont {C.~A.~D.}\ \bibnamefont {Zarro}}, \bibinfo {author}
  {\bibfnamefont {C.}~\bibnamefont {Farina}},\ and\ \bibinfo {author}
  {\bibfnamefont {A.~L.~C.}\ \bibnamefont {Rego}},\ }\href
  {https://doi.org/10.1007/s13538-020-00770-x} {\bibfield  {journal} {\bibinfo
  {journal} {Brazilian Journal of Physics}\ }\textbf {\bibinfo {volume} {50}},\
  \bibinfo {pages} {634} (\bibinfo {year} {2020})}\BibitemShut {NoStop}%
\bibitem [{\citenamefont {Pal}\ \emph {et~al.}(2023)\citenamefont {Pal},
  \citenamefont {Pal}, \citenamefont {Gill},\ and\ \citenamefont
  {Sarkar}}]{PalEtAl2023}%
  \BibitemOpen
  \bibfield  {author} {\bibinfo {author} {\bibfnamefont {K.}~\bibnamefont
  {Pal}}, \bibinfo {author} {\bibfnamefont {K.}~\bibnamefont {Pal}}, \bibinfo
  {author} {\bibfnamefont {A.}~\bibnamefont {Gill}},\ and\ \bibinfo {author}
  {\bibfnamefont {T.}~\bibnamefont {Sarkar}},\ }\href
  {https://doi.org/10.1103/PhysRevB.108.104311} {\bibfield  {journal} {\bibinfo
   {journal} {Phys. Rev. B}\ }\textbf {\bibinfo {volume} {108}},\ \bibinfo
  {pages} {104311} (\bibinfo {year} {2023})}\BibitemShut {NoStop}%
\bibitem [{\citenamefont {Singh}\ and\ \citenamefont
  {Natarajan}(2015)}]{SinghNatarajan2015}%
  \BibitemOpen
  \bibfield  {author} {\bibinfo {author} {\bibfnamefont {A.~K.}\ \bibnamefont
  {Singh}}\ and\ \bibinfo {author} {\bibfnamefont {V.}~\bibnamefont
  {Natarajan}},\ }\href {https://doi.org/10.1088/1367-2630/17/3/033044}
  {\bibfield  {journal} {\bibinfo  {journal} {New J. Phys.}\ }\textbf {\bibinfo
  {volume} {17}},\ \bibinfo {pages} {033044} (\bibinfo {year}
  {2015})}\BibitemShut {NoStop}%
\bibitem [{\citenamefont {Christandl}\ \emph {et~al.}(2004)\citenamefont
  {Christandl}, \citenamefont {Datta}, \citenamefont {Ekert},\ and\
  \citenamefont {Landahl}}]{Christandl2004}%
  \BibitemOpen
  \bibfield  {author} {\bibinfo {author} {\bibfnamefont {M.}~\bibnamefont
  {Christandl}}, \bibinfo {author} {\bibfnamefont {N.}~\bibnamefont {Datta}},
  \bibinfo {author} {\bibfnamefont {A.}~\bibnamefont {Ekert}},\ and\ \bibinfo
  {author} {\bibfnamefont {A.~J.}\ \bibnamefont {Landahl}},\ }\href
  {https://doi.org/10.1103/PhysRevLett.92.187902} {\bibfield  {journal}
  {\bibinfo  {journal} {Phys. Rev. Lett.}\ }\textbf {\bibinfo {volume} {92}},\
  \bibinfo {pages} {187902} (\bibinfo {year} {2004})}\BibitemShut {NoStop}%
\bibitem [{\citenamefont {Takahashi}\ and\ \citenamefont {del
  Campo}(2024)}]{TakahashiDelCampo2024}%
  \BibitemOpen
  \bibfield  {author} {\bibinfo {author} {\bibfnamefont {K.}~\bibnamefont
  {Takahashi}}\ and\ \bibinfo {author} {\bibfnamefont {A.}~\bibnamefont {del
  Campo}},\ }\href {https://doi.org/10.1103/PhysRevX.14.011032} {\bibfield
  {journal} {\bibinfo  {journal} {Phys. Rev. X}\ }\textbf {\bibinfo {volume}
  {14}},\ \bibinfo {pages} {011032} (\bibinfo {year} {2024})}\BibitemShut
  {NoStop}%
\bibitem [{\citenamefont {Aguilar-Gutierrez}\ and\ \citenamefont
  {Rolph}(2024)}]{Aguilar}%
  \BibitemOpen
  \bibfield  {author} {\bibinfo {author} {\bibfnamefont {S.~E.}\ \bibnamefont
  {Aguilar-Gutierrez}}\ and\ \bibinfo {author} {\bibfnamefont {A.}~\bibnamefont
  {Rolph}},\ }\href {https://doi.org/10.1103/PhysRevD.109.L081701} {\bibfield
  {journal} {\bibinfo  {journal} {Phys. Rev. D}\ }\textbf {\bibinfo {volume}
  {109}},\ \bibinfo {pages} {L081701} (\bibinfo {year} {2024})}\BibitemShut
  {NoStop}%
\bibitem [{\citenamefont {Parker}\ \emph {et~al.}(2019)\citenamefont {Parker},
  \citenamefont {Cao}, \citenamefont {Avdoshkin}, \citenamefont {Scaffidi},\
  and\ \citenamefont {Altman}}]{Parker}%
  \BibitemOpen
  \bibfield  {author} {\bibinfo {author} {\bibfnamefont {D.~E.}\ \bibnamefont
  {Parker}}, \bibinfo {author} {\bibfnamefont {X.}~\bibnamefont {Cao}},
  \bibinfo {author} {\bibfnamefont {A.}~\bibnamefont {Avdoshkin}}, \bibinfo
  {author} {\bibfnamefont {T.}~\bibnamefont {Scaffidi}},\ and\ \bibinfo
  {author} {\bibfnamefont {E.}~\bibnamefont {Altman}},\ }\href
  {https://doi.org/10.1103/PhysRevX.9.041017} {\bibfield  {journal} {\bibinfo
  {journal} {Phys. Rev. X}\ }\textbf {\bibinfo {volume} {9}},\ \bibinfo {pages}
  {041017} (\bibinfo {year} {2019})}\BibitemShut {NoStop}%
\bibitem [{\citenamefont {Bhattacharya}\ \emph {et~al.}(2023)\citenamefont
  {Bhattacharya}, \citenamefont {Nandy}, \citenamefont {Nath},\ and\
  \citenamefont {Sahu}}]{Bhattacharya:2023zqt}%
  \BibitemOpen
  \bibfield  {author} {\bibinfo {author} {\bibfnamefont {A.}~\bibnamefont
  {Bhattacharya}}, \bibinfo {author} {\bibfnamefont {P.}~\bibnamefont {Nandy}},
  \bibinfo {author} {\bibfnamefont {P.~P.}\ \bibnamefont {Nath}},\ and\
  \bibinfo {author} {\bibfnamefont {H.}~\bibnamefont {Sahu}},\ }\href
  {https://doi.org/10.1007/JHEP12(2023)066} {\bibfield  {journal} {\bibinfo
  {journal} {JHEP}\ }\textbf {\bibinfo {volume} {2023}}\bibinfo  {number} {
  (12)},\ \bibinfo {pages} {066}}\BibitemShut {NoStop}%
\bibitem [{\citenamefont {Rosenblum}(1994)}]{Rosenblum1994}%
  \BibitemOpen
\bibfield  {number} {  }\bibfield  {author} {\bibinfo {author} {\bibfnamefont
  {M.}~\bibnamefont {Rosenblum}},\ }in\ \href
  {https://doi.org/10.1007/978-3-0348-8522-5_15} {\emph {\bibinfo {booktitle}
  {Nonselfadjoint Operators and Related Topics}}},\ \bibinfo {editor} {edited
  by\ \bibinfo {editor} {\bibfnamefont {A.}~\bibnamefont {Feintuch}}\ and\
  \bibinfo {editor} {\bibfnamefont {I.}~\bibnamefont {Gohberg}}}\ (\bibinfo
  {publisher} {Birkh{\"a}user},\ \bibinfo {address} {Basel},\ \bibinfo {year}
  {1994})\ pp.\ \bibinfo {pages} {369--396}\BibitemShut {NoStop}%
\end{thebibliography}%

\end{document}